\documentclass[12pt,singlespace,vi,twoside]{mitthesis}

\usepackage{cmap}
\usepackage{antpolt,eulervm}
\usepackage[utf8]{inputenc}
\usepackage[T1]{fontenc}

\usepackage{hyperref}
\hypersetup{colorlinks=true, linkcolor=blue, filecolor=magenta, urlcolor=blue,
			citecolor=red}
\usepackage[nottoc,notlot,notlof]{tocbibind}

\usepackage{amssymb,amsthm,bm,physics,upgreek,epigraph,enumitem}

\newtheorem{theorem}{\textsc{Theorem}}[chapter] 
\newenvironment{ftheorem}
  {\begin{mdframed}\begin{theorem}}
  {\end{theorem}\end{mdframed}}

\newtheorem{cor}{\textsc{Corollary}}[theorem] 
\newtheorem{lem}[theorem]{\textsc{Lemma}} 
\newtheorem{prop}[theorem]{\textsc{Proposition}} 
\newtheorem{define}[theorem]{\textsc{Definition}} 
\newenvironment{fdefine}
  {\begin{mdframed}\begin{define}}
  {\end{define}\end{mdframed}}

\newtheorem{rem}{\textsc{Remark}}
\newtheorem{ex}{\textsc{Example}}
\renewenvironment{proof}{\noindent{\textbf{\textsc{Proof.~}}}}{}
\renewcommand{\qed}{\unskip\nobreak\qquad\blacksquare}

\newtheorem{axiom}{\textsc{Axiom}}

\newtheorem{inneraxm}{\textsc{Axiom}}
\newenvironment{axm}[1]
  {\renewcommand\theinneraxm{#1}\inneraxm}
  {\endinneraxm}

\newcommand{\fra}{{\frak A}}
\newcommand{\frm}{{\frak M}}
\newcommand{\Ad}{{\sf Ad}}
\newcommand{\Dom}{{\sf Dom}}
\newcommand{\Dim}{{\sf dim}}
\newcommand{\Hom}{{\sf Hom}}
\newcommand{\End}{{\sf End}}
\newcommand{\trc}{{\sf Tr}}
\newcommand{\Aut}{{\sf Aut}}

\newcommand{\A}{{\cal A}}
\newcommand{\B}{{\cal B}}

\newcommand{\D}{{\cal D}}
\newcommand{\E}{{\cal E}}
\newcommand{\F}{{\cal F}}
\newcommand{\HH}{{\cal H}}
\newcommand{\J}{{\cal J}}
\newcommand{\M}{{\cal M}}
\newcommand{\Q}{{\cal Q}}
\newcommand{\R}{{\cal R}}
\newcommand{\T}{{\cal T}}
\newcommand{\U}{{\cal U}}
\newcommand{\V}{{\cal V}}
\newcommand{\W}{{\cal W}}
\newcommand{\X}{{\cal X}}

\newcommand{\+}{\oplus}
\newcommand{\x}{\otimes}

\newcommand{\DX}{\eth_X}
\newcommand{\dx}{{\frak D}^X}
\newcommand{\ds}{{\frak D}^{\cal S}}
\newcommand{\gmn}{g_{\mu\nu}}
\newcommand{\gMN}{g^{\mu\nu}}
\newcommand{\dm}{\partial_\mu}
\newcommand{\dn}{\partial_\nu}
\newcommand{\gm}{\gamma^\mu}
\newcommand{\gn}{\gamma^\nu}
\newcommand{\gk}{\gamma^\kappa}
\newcommand{\gl}{\gamma^\lambda}
\newcommand{\Ox}{\bm\omega_X}

\newcommand{\om}{\omega_\mu}
\newcommand{\on}{\omega_\nu}
\newcommand{\oxm}{\omega^X_\mu}
\newcommand{\oxn}{\omega^X_\nu}
\newcommand{\osm}{\omega^{\cal S}_\mu}

\newcommand{\half}{\frac1{2}}

\newcommand{\Id}{\mathbb{I}}
\newcommand{\id}{{\sf id}}

\newcommand{\nxm}{\nabla^X_\mu}
\newcommand{\nxn}{\nabla^X_\nu}
\newcommand{\nsm}{\nabla^{\cal S}_\mu}

\newcommand{\s}{\sigma}
\newcommand{\g}{\gamma}	
\newcommand{\G}{\Gamma}
\newcommand{\gf}{\gamma^5}

\newcommand{\dmu}{{\sf d}\mu}
\newcommand{\intbar}{\text{--}\!\!\!\!\int}

\long\def\comment#1{} 

\usepackage{mdframed}
\usepackage{lipsum}

\begin{document}
%
%
%
%
%
%
%
%
%
%
%


\title{Twisting Noncommutative Geometries\\with Applications to High Energy Physics}
\author{Devashish Singh}


\department{Department of Mathematics}


\degree{Ph.D.~in Mathematics and Applications}


\degreemonth{November}
\degreeyear{2019}
\thesisdate{\today}


\copyrightnoticetext{}


\supervisor{Dr.~Pierre Martinetti}{Researcher}


\chairman{Prof.~Stefano Vigni}{Director of the Graduate Program}


\maketitle




\cleardoublepage


\setcounter{savepage}{\thepage}

\begin{abstractpage}
%
%
%

\bigskip

With the bare essentials of noncommutative geometry (defined by a spectral triple),
we first describe how it naturally gives rise to gauge theories. Then, we quickly
review the notion of twisting (in particular, minimally) noncommutative geometries 
and how it induces a Wick rotation, that is, a transition of the metric signature 
from euclidean to lorentzian. 
We focus on comparatively more tractable examples of spectral triples; such as the
ones corresponding to a closed riemannian spin manifold, $U(1)$ gauge theory, 
and electrodynamics. By minimally twisting these examples and computing their 
associated fermionic actions, we demonstrate how to arrive at physically relevant actions (such as the 
Weyl and Dirac actions) in Lorentz signature, even though starting from 
euclidean spectral triples. In the process, not only do we extract a physical 
interpretation of the twist, but we also capture exactly how the Wick rotation 
takes place at the level of the fermionic action.

\bigskip
\end{abstractpage}


\cleardoublepage

\section*{Acknowledgements}

\bigskip
\bigskip

First and foremost, it is hard to overstate my gratitude to Pierre Martinetti for 
being my advisor, mentor, and collaborator throughout my time as a Ph.D.\ candidate. 
I am deeply indebted for his patient guidance, continuous support, and encouragement;
all of which were indispensably conducive to the fruition of this thesis. I have 
been extremely fortunate to have a supervisor who provided me with an example, 
not only of \emph{how to be a mathematical physicist}
but also of \emph{how to be}.

\bigskip

I extend my thanks to Walter van Suijlekom and his Ph.D.\ students (Teun van Nuland, 
Ruben Stienstra) for useful discussions and hospitality, 
during May 2019, at the Mathematics Department and the Institute for Mathematics, 
Astrophysics and Particle Physics (IMAPP) of Radboud University Nijmegen, 
Netherlands where a part of the research for this thesis was completed.

\bigskip

I am thankful to Latham Boyle for inviting me to Perimeter Institute, during June 
2019, where a part of this thesis was written. The vibrant and inspiring 
environment there provided me with excellent working conditions and 
added to my Ph.D.\ experience. 
Research at Perimeter Institute is supported in part by the Government of Canada 
through the Department of Innovation, Science and Economic Development Canada and 
by the Province of Ontario through the Ministry of Economic Development, Job 
Creation and Trade.

\bigskip

I am grateful for having a spectrum of wonderful colleagues and awesome officemates
at the Department of Mathematics in Genoa, who were always welcoming whenever I 
needed help of any sort. I express my thanks to 
Francesca Bartolucci,
Elisa Biagioli, 
Jo\~ao Braga de G\'oes Vasconcellos, 
Federico Faldino, 
Nicol\`o Ginatta,
Sepehr Jafari, and
Paola Magrone.

\bigskip

A special thanks goes to my family and friends who support me in everything I do.

\bigskip

Last but not least, thanks to Alice, for paying heed to my ramblings on 
mathematics, science, philosophy, life, and everything in between; and for all her
love and support.


\pagestyle{plain}

\tableofcontents

	\chapter*{}

{\it ``The steady progress of physics requires for its theoretical 
formulation a mathematics that gets continually more advanced. This is only natural 
and to be expected. What, however, was not expected by the scientific workers of 
the last century was the particular form that the line of advancement of the 
mathematics would take, namely, it was expected that the mathematics would get more
and more complicated, but would rest on a permanent basis of axioms and definitions,
while actually the modern physical developments have required a mathematics that 
continually shifts its foundations and gets more abstract. Non-euclidean geometry 
and non-commutative algebra, which were at one time considered to be purely 
fictions of the mind and pastimes for logical thinkers, have now been found to be 
very necessary for the description of general facts of the physical world. It seems 
likely that this process of increasing abstraction will continue in the future and 
that advance in physics is to be associated with a continual modification and 
generalisation of the axioms at the base of the mathematics rather than with a 
logical development of any one mathematical scheme on a fixed foundation. There 
are at present fundamental problems in theoretical physics awaiting solution, e.g., 
the relativistic formulation of quantum mechanics and the nature of atomic nuclei 
(to be followed by more difficult ones such as the problem of life), the solution 
of which problems will presumably require a more drastic revision of our 
fundamental concepts than any that have gone before. Quite likely these changes 
will be so great that it will be beyond the power of human intelligence to get the 
necessary new ideas by direct attempts to formulate the experimental data in 
mathematical terms. The theoretical worker in the future will therefore have to 
proceed in a more indirect way. The most powerful method of advance that can be 
suggested at present is to employ all the resources of pure mathematics in attempts 
to perfect and generalise the mathematical formalism that forms the existing basis 
of theoretical physics, and after each success in this direction, to try to 
interpret the new mathematical features in terms of physical entities (by a process 
like Eddington's Principle of Identification).''}

\bigskip
\bigskip

\begin{flushright}
-- P.A.M.\ Dirac, \emph{Quantised singularities in the electromagnetic field},\\
\emph{Proceedings of the Royal Society of London A} {\bf 133}: 821 (1931).
\end{flushright}

\clearpage
\newpage

\chapter*{Introduction}
\addcontentsline{toc}{chapter}{Introduction}
\label{sec:intro}

Understanding the intrinsic nature of spacetime is not only one of the most 
fundamental quests for theoretical physicists, but also a mountainous challenge
for mathematicians. In the light of how the advent of general relativity was 
facilitated and brought forth by Riemann's very broad extension and abstract 
generalization of euclidean differential geometry of surfaces in $\mathbb{R}^3$, 
it is quite likely that accommodating Planck scale physics calls for a revision of 
our notions about geometry -- towards geometric objects that are much more flexible
than differentiable manifolds.

\smallskip

\emph{Noncommutative geometry} (NCG) \cite{C94} is such an approach to generalize 
riemannian geometry by giving a purely spectral/operator-algebraic 
characterization of geometry \cite{C13}; in the same sense as Gel'fand duality 
(\S \ref{app:A}) provides an algebraic characterization of topology. 
The mathematical object encapsulating such a characterization of geometry in a 
generalized sense is a \emph{spectral triple} $(\A,\HH,\D)$ consisting of a 
unital $*$-algebra $\A$ of bounded operators in a Hilbert space $\HH$ and a 
self-adjoint operator $\D$ with compact resolvent on $\HH$ such that the commutator
$[\D,a]$ is bounded for any $a\in\A$. Spectral triples naturally give rise to 
gauge theories.

\smallskip

One of the most spectacular achievements of NCG for particle physics is the full
derivation of its most important gauge theory, i.e.\ the Standard Model (SM) 
lagrangian, along with all its delicacies including the Higgs potential, 
spontaneous symmetry breaking, neutrino mixing, etc.\ \cite{CCM} and the 
Einstein-Hilbert action \cite{CC96, CC97}. In fact, during the early 
developmental stages, SM was regarded as one of the guiding principles behind the 
blueprints of the framework \cite{C95}. NCG provides a purely geometric/gravitational 
description of the SM, where gravity is naturally present
with minimal coupling to matter \cite{C96, CC10, CS}.

\smallskip

NCG offers various ways to build models even beyond the SM, see e.g.\ \cite{CS,DKL} for 
a recent review. One of them involves 
twisting the spectral triple of the SM by an algebra automorphism \cite{DLM1,DLM2,DM}, 
in the sense of Connes and Moscovici \cite{CMo}. This provides a
  mathematical justification to the extra scalar
  field introduced in \cite{CC12} to both fit the mass of the
  Higgs and stabilize the electroweak
  vacuum. A significant difference from the construction based on
  spectral triples without  first-order 
  condition \cite{CCS1,CCS2} is that the twist does
  not only yield an extra scalar field $\sigma$, but also 
  a supplementary real one-form field $X_\mu$,{\footnote{In \cite{DM} this field
was improperly called \emph{vector field}.} whose meaning was rather unclear 
so far.

\pagebreak

Although Connes' work provides a spectral characterization of
compact riemannian spin manifolds \cite{C13} along with the tools for their 
noncommutative generalization \cite{C96}, extending this program to the 
pseudo-riemannian case is notoriously difficult, and there has, so far, been no 
completely convincing model of pseudo-riemannian spectral triples. 
However, several interesting results in this context have been obtained recently, 
see e.g. \cite{BB, BBB, DPR, Fr, FE}, there is nevertheless no reconstruction theorem 
for pseudo-riemannian manifolds in view, and it is still unclear how the spectral 
action should be handled in a pseudo-riemannian signature. 

\smallskip

Quite unexpectedly, the twist of the SM, which has been introduced in a purely 
riemannian context, seems to have some link with Wick rotation. In fact, the inner 
product induced by the twist on the Hilbert space of euclidean spinors on a
four-dimensional manifold $\M$, coincides with the Kre\u{\i}n product of lorentzian
spinors \cite{DFLM}. This is not so surprising, for the twist $\rho$ coincides
with the automorphism that exchanges the two eigenspaces of the grading operator 
(in physicist's words: that exchanges the left and the right components of 
spinors). And this is nothing but the inner automorphism induced by the first Dirac
matrix $\gamma^0=c(dx^0)$. This explains why, by twisting, one is somehow able
to single out the $x_0$ direction among the four riemannian dimensions of
$\M$. However, the promotion of this $x_0$ to a ``time direction'' is not
fully accomplished, at least not in the sense of Wick rotation \cite{D'AKL}.
Indeed, regarding the Dirac matrices, the inner automorphism induced
by $\gamma^0$ does not implement the Wick rotation $\W$ 
(which maps the spatial Dirac matrices as
$\g^j \to \W(\g^j) := i\g^j$)  but actually its square:
\begin{equation}
	\rho(\gamma^j) = \gamma^0\gamma^j\gamma^0 
	= -\gamma^j = \W^2(\gamma^j), \qquad \text{for} \quad j=1,2,3.
\label{eq:1}
 \end{equation}
Nevertheless, a transition from the euclidean to the lorentzian does
occur, and the $x_0$ direction gets promoted to a time
direction, but this happens at the level of the fermionic action. This
is one of the main results of this thesis, summarized in Prop.~\ref{Prop:Weyl} and
Prop.~\ref{prop:eq-Dirac}.

\smallskip

More specifically, starting with the twisting of a \emph{euclidean} manifold, then 
that of a $U(1)$ gauge theory, and finally the twisting of the spectral triple of
electrodynamics in euclidean signature \cite{DS}; we show how the fermionic action 
for twisted spectral triples, proposed in \cite{DFLM}, actually yields the Weyl 
(Prop.~\ref{Prop:Weyl}) and the Dirac (Prop.~\ref{prop:eq-Dirac}) equations in 
\emph{Lorentz signature}. In addition, the zeroth component of the extra one-form 
field $X_\mu$ acquires a clear interpretation as an energy.

\smallskip

The following two aspects of the fermionic action for twisted spectral triples
explain the above-mentioned change of the metric signature from euclidean to
lorentzian:
\begin{enumerate}
	\item First, in order to guarantee that the fermionic action is symmetric when 
evaluated on Gra{\ss}mann variables,\footnote{which is an important requirement for the whole physical interpretation of the action formula, also in the non-twisted case \cite{CCM})} one restricts the 
bilinear form that defines the action to the $+1$-eigenspace $\HH_\R$ of the 
unitary $\cal R$ implementing the twist on the Hilbert space $\HH$; whereas in the 
non-twisted case, one restricts the bilinear form to the $+1$-eigenspace of the 
grading $\g$, in order to solve the fermion doubling problem \cite{LMMS}. This 
different choice of eigenspace had been noticed in \cite{DFLM} but the physical 
consequences were not drawn. As already emphasized above, in the models relevant 
for physics, ${\cal R}=\gamma^0$, and once restricted to $\HH_\R$, the bilinear 
form no longer involves derivative in the $x_0$ direction. In other words, the 
restriction to $\HH_\R$ projects the euclidean fermionic action to what will 
constitute its spatial part in lorentzian signature.
	\item Second, the twisted fluctuations of the Dirac operator of a 
$4$-dimensional riemannian manifold are not necessarily zero \cite{DM,LM1}; in 
contrast with the non-twisted case where those fluctuations are always vanishing. 
and these are parametrized by the above-mentioned real one-form field $X_\mu$. By 
interpreting the zeroth component of this field as an energy, one recovers a 
derivative in the $x_0$ direction -- but now in Lorentz signature.
\end{enumerate}
In addition to the second point above, at least for the spectral action of the 
twisted SM, the contribution of the $X_\mu$ field is 
minimized when the field itself vanishes, i.e.\ there is no twist or 
$\rho = {\sf id}$ \cite{DM}.
Thus, one may view the twist as a vacuum fluctuation around the noncommutative
riemannian geometry or its spontaneous symmetry breaking to a lorentzian
(twisted) noncommutative geometry.

\smallskip

The manuscript has been organized as follows. 

\smallskip

Chapter \ref{chap:axioms} defines the notion of a \emph{noncommutative geometry} in 
terms of spectral triples (Def.\ \ref{def:1.1}). In \S\ref{sec:1.1}, we first list 
out the five axioms a spectral triple with a commutative algebra must obey 
to satisfy Connes' reconstruction theorem (Theorem \ref{thrm1}) for riemannian 
manifolds. 
Two extra axioms added to the list take into account the spin structure and then we
are also able to give a purely spectral characterization to riemannian spin 
geometries (\S\ref{sec:1.2}) and define a \emph{canonical} spectral triple 
associated to them (Def.\ \ref{def:CT}). 
Considering the real structure, in \S\ref{sec:1.3}, four of the above-mentioned 
seven axioms are modified to be more flexible and suitable to accommodate  
noncommutative algebras and, thus, generalizing the geometry defined by the 
spectral triples to a noncommutative setting.

\smallskip

As discussed in chapter \ref{chap:GTST}, spectral triples naturally give rise to 
gauge theories. There exists a more general notion of equivalence than isomorphism 
between algebras known as Morita equivalence (\S\ref{sec:MorEq}). 
This notion of equivalence between algebras when lifted to the level of spectral 
triples -- in a manner consistent with the real structure -- gives rise to 
generalized gauge fields (\S \ref{sec:MorEqSpecTr}). 
These gauge fields obtained as perturbations of the Dirac operator (encoding 
the metric information as it defines the distance formula in noncommutative 
geometry \cite{C96}) are referred to as the \emph{inner fluctuations} (of the 
metric). 

\smallskip

The gauge transformations of generalized gauge fields (or, gauge 
potentials in physical gauge theories) thus obtained are arrived at via
a change of connection on the bimodule (that implements Morita self-equivalence of 
real spectral triples) induced by an adjoint action of the group of unitaries of 
the algebra (\S\ref{sec:GT}). Then, in \S\ref{subsec:gauge_inv}, we defined two of the 
most important gauge-invariant functionals on spectral triples: the spectral action
and the fermionic action. 

\smallskip

In \S \ref{sec:ACG}, we define one of the most important classes of 
noncommutative geometries from the standpoint of physics -- almost-commutative 
geometries, which will be very useful for our purposes later in this thesis. We 
then give a few examples of the physically relevant models describes by such
geometries: such as $U(1)$ gauge theory (\S \ref{subsec:2.4.1}), electrodynamics
(\S \ref{subsec:2.4.2}), the Standard Model of particle physics and its extensions 
(\S \ref{subsec:2.4.3}).

\smallskip

In \S \ref{subsec:twistedsp}, we first review the known material regarding spectral 
triples twisted by using algebra automorphisms and their compatibility with the 
real structure. We then recall the notions of a covariant Dirac operator, inner
fluctuations (\S \ref{subsec:twist-inner-fluc}), and gauge transformations (\S 
\ref{sec:twstgaugetrans}) for twisted real spectral triples. We further discuss how
the twist naturally induces a new inner product on the initial Hilbert space (\S 
\ref{subsec:rho}), which helps to define a corresponding gauge-invariant 
fermionic action in the twisted case (\S \ref{subsec:fermionaction}). 

\smallskip

Furthermore, in \S\ref{sec:2.2}, we outline the construction named `minimal 
twist by grading' that associates a \emph{minimally} twisted counterpart to any 
given graded spectral triple -- meaning the twisted spectral triple has the same 
Hilbert space and Dirac operator as that of the initial one, but the algebra is 
doubled in order to make the twisting possible \cite{LM1}. 

\smallskip

Chapter \ref{chap:examples} is the main and new contribution of the thesis, 
which is primarily concerned with making use of 
the methods summarized in the previous chapter, which entails applying the 
procedure of minimal twist by grading to three very simple yet concrete examples of
spectral triples and computing the corresponding fermionic actions:
\begin{enumerate}
	\item \emph{A closed riemannian spin manifold}. 
We investigate the minimal twist of a four-dimensional manifold in \S\ref{sec:Weyl}
and show that the twisted fluctuations of corresponding Dirac operator are 
parametrized by a real one-form field $X_\mu$ (\S \ref{subsec:Xmu}) -- first 
discovered in \cite{DM}. We further recall how to deal with gauge transformations 
in the twisted case (\S \ref{sec:3.1.2}) -- along the lines of \cite{LM2}; and then
compute the fermionic action showing that it yields a lagrangian density very 
similar to that of the Weyl lagrangian in lorentzian signature, as soon as one 
interprets the zeroth component of $X_\mu$ as the time component of the 
energy-momentum four-vector of fermions (\S \ref{subsec:fermioncactionmanif}). 
However, this lagrangian does not possess enough spinorial degrees of freedom to 
produce Weyl equations.
	\item \emph{A $U(1)$ gauge theory}.
The previous example dictates that one must consider a spectral triple with 
slightly more room for the Hilbert space. So, in \S\ref{sec:doublem}, we double its
size by taking two copies of the background manifold -- which describes a $U(1)$ 
gauge theory \cite{DS}. We then compute the twisted-covariant Dirac operator (\S
\ref{subsec:twistfluc2man}) and obtain the Weyl equations from the associated 
fermionic action (\S\ref{subsec:Weyl}).
	\item \emph{The gauge theory of electrodynamics}. 
In \S\ref{sec:electrody}, we apply the same construction as above to the spectral 
triple of electrodynamics proposed in \cite{DS}. We write down its minimal twist 
and calculate the twisted fluctuations for both the free and the finite parts of 
the Dirac operator in \S\ref{sec:4.3}. The gauge transformations are derived in \S
\ref{subsec:gaugetransformED} and, finally, the Dirac equation in Lorentz signature
is obtained in \S\ref{sec:Dirtaceq}.
\end{enumerate}
This results not only in finding a physical interpretation for the twist, but also 
roots the link between the twist and Wick rotation -- as depicted in \cite{DFLM} --
on a much more firm ground.

\smallskip

Chapter \ref{chap:5} addresses some issues that opened up due to this work and  
are yet to be settled. For instance, although we showed that the $\rho$-inner product 
and, hence, the fermionic action of a minimally twisted manifold are both invariant 
under Lorentz boosts (\S \ref{sec:Lorentzinv}); how exactly Lorentz transformations
arise within the framework of (twisted) noncommutative geometry is however rather 
unclear so far. 

\smallskip

Another relevant question that naturally arises in this context is 
that of the spectral action. In \S \ref{sec:5.2}, we compute the Lichnerowicz 
formula for the twisted-covariant Dirac operator of a closed riemannian spin 
manifold (with non-zero curvature), which is the very first step towards writing 
down the heat-kernel expansion for the spectral action.

\smallskip

Finally, we conclude with some outlook and perspective. 

\smallskip

The appendices contain a brief recollection of Gel'fand duality (\S \ref{app:A}), 
the definitions of and notations for the Clifford algebras and Cliffords algebra 
bundles (\S \ref{app:cliff}), the modular theory of Tomita and Takesaki (\S 
\ref{appb}), and all the required notations for $\g$-matrices (in chiral 
representation) and for the Weyl and Dirac equations -- both in euclidean space and 
minkowskian spacetime (\S \ref{GammaMatrices}).

\chapter{The Axioms of Noncommutative Geometry}
\label{chap:axioms}


{\small{\it \quad ``It is known that geometry assumes, as things given, both the notion
of space and the first principles of constructions in space. She gives definitions
of them which are merely nominal, while the true determinations appear in
the form of axioms. The relation of these assumptions remains consequently
in darkness; we neither perceive whether and how far their connection is
necessary, nor a priori, whether it is possible.''}
\begin{flushright}
-- Bernhard Riemann, 
\emph{On the Hypotheses Which Lie at the Bases of Geometry} (1854)\\
(Original: \emph{\"Uber die Hypothesen, welche der Geometrie zu Grunde liegen})
\end{flushright}}

\smallskip

\noindent In \S\ref{sec:1.1}, we look at the commutative case -- that is, riemannian 
geometry -- and list out the five axioms for Connes' reconstruction theorem 
\cite{C13}. The reconstruction theorem is at the heart of the subject and it lays 
down the foundation for a nontrivial generalization of riemannian geometry by 
giving a way to translate the geometric information on riemannian manifolds into a 
spectral/operator-algebraic language and vice-versa. With two additional axioms 
(\S\ref{sec:1.2}), the theorem also holds for riemannian spin manifolds \cite{C96}. Finally, 
in \S\ref{sec:1.3}, we state a slightly modified version of four out of the seven 
axioms to make them suitable for the said generalization. These seven 
(including the modified ones) will form the set of axioms a spectral triple must 
fulfill to define a noncommutative geometry.

\smallskip

\begin{fdefine}[from \cite{CMa}]
\label{def:1.1}
	A {\bf spectral triple} $(\A,\HH,\D)$ consists of
\begin{enumerate}
\setlength\itemsep{-0.5em}
	\item a unital $*$-algebra $\A$ (see definition in App.~\ref{app:A}),
	\item a Hilbert space $\HH$ on which $\A$ acts as bounded 
	operators, via a representation
	$\pi:\A\to\B(\HH)$,
	\item a (not necessarily bounded) self-adjoint operator $\D:\HH\to\HH$ such 
	that its resolvent $(i+\D)^{-1}$ is compact and its commutator with $\A$ is 
	bounded, that is, $[\D,a]\in\B(\HH),\;\forall a\in\A$.
\end{enumerate}
A spectral triple is said to be {\bf commutative} if $\A$ is commutative.
\medskip
\end{fdefine}

\pagebreak

\section{Connes' reconstruction theorem}
\label{sec:1.1}

Riemannian geometry is described by a commutative spectral
triple $(\A,\HH,\D)$ satisfying the following five axioms:

\begin{axiom}[Dimension] 
\label{ax1}
	$\D$ is an infinitesimal of the order $n \in \mathbb{N}$, i.e.\
	the $k$-th characteristic value of its resolvent $\D^{-1}$ is 
	$O(k^{-\frac 1{n}})$. The spectral triple is said to be of {\bf $KO$-dimension} 
	$n$.
\end{axiom}

Axiom \ref{ax1} fixes the dimension of the riemannian manifold that
the spectral triple describes.

\begin{axiom}[First-order or order-one condition]
\label{ax2}
	Omitting the symbol $\pi$ of the representation of the algebra $\A$,\footnote{For brevity of notation, from now on and wherever applicable, we use $a$ to mean its representation $\pi(a)$. Thus, $a^*$ denotes $\pi(a^*) = \pi(a)^\dag$, where $^*$ is the involution of ${\cal A}$ and $^\dag$ is the hermitian conjugation on $\HH$.} 
	one has that
\begin{equation}
	\big[[\D,a],b\big] = 0, \qquad \forall a,b\in \A.
\end{equation}
\end{axiom}

Axiom \ref{ax2} ensures that $\D$ is a first-order differential operator.

\begin{axiom}[Regularity condition] 
\label{ax3}
	For any $a\in\A$, both $a$ and $[\D,a]$ are in the smooth domain of the 
	derivation $\delta(\cdot) := \big[|\D|,\cdot\big]$, where $|\D|:=\sqrt{\D\D^*}$ denotes
	the absolute value of $\D$. That is, for any $m\in\mathbb{N}$,
\begin{equation}
	a \in \Dom(\delta^m), \quad [\D,a] \in \Dom(\delta^m), 
	\qquad \forall a \in \A,
\end{equation}
	where 
\begin{equation}
	\Dom(\delta) \equiv \Big\{ T\in\B(\HH) \;\Big|\; 
	T\,\Dom\,|\D| \subset \Dom\,|\D|,\; \delta(T) \in \B(\HH) \Big\}.
\end{equation}
In other words, both $\delta^m(a)$ and $\delta^m([\D,a])$ are bounded operators.
\end{axiom}

Axiom \ref{ax3} is an algebraic formulation of the smoothness of coordinates.
For the next axiom we need the following definition.

\begin{define}
	An $n$-dimensional {\bf Hochschild cycle} is a finite sum of the elements of 
	$\A^{\x(n+1)}:=\A\x\A\x\ldots\x\A$ ($n+1$ times), given by
	$c = \sum_j (a_j^0 \x a_j^1 \x \ldots \x a_j^n)$, such that the contraction 
	$bc = 0$, where, by definition, $b$ is linear and satisfies 
\begin{equation} 
\begin{split}
	b(a^0 \x a^1 \x \ldots \x a^n) 
& = (a^0a^1 \x a^2 \x \ldots \x a^n) - (a^0 \x a^1a^2 \x \ldots \x a^n) + \cdots \\
& \quad + (-1)^k(a^0 \x \ldots \x a^ka^{k+1} \x \ldots \x a^n) + \cdots \\
& \quad + (-1)^n(a^na^0 \x \ldots \x a^{n-1}).
\end{split} 
\end{equation} 
\end{define} 
\pagebreak
A Hochschild cycle is the algebraic formulation of a differential form. For a 
commutative algebra $\A$, it can be constructed easily by taking 
any $a^j$ and considering the following sum running over all the permutations $\s$ 
of $\{1,\ldots,n\}:$
\begin{equation}
	c = \sum \varepsilon(\s)(a^0 \x a^{\s(1)} \x a^{\s(2)} \x \ldots \x a^{\s(n)}),
\end{equation}
which corresponds to the familiar differential form $a^0da^1 \land da^2 \land 
\ldots \land da^n$, requiring no previous knowledge of the tangent bundle 
whatsoever, yet providing the volume form.

\begin{axiom}[Orientability condition] 
\label{ax4}
	Let $\pi_\D(c)$ be the representation of a Hochschild cycle $c$ on $\HH$,
	induced by the following linear map:
\begin{equation}
	\pi_\D(a_0 \x a_1 \x \ldots \x a_n) = a_0\big[\D,a_1\big]\ldots\big[\D,a_n\big], 
	\qquad \forall a_j \in \A.  
\end{equation}
	There exists a Hochschild cycle $c \in Z_n(\A,\A)$, 
	such that for odd $n$, we have $\pi_\D(c) = 1$, 
	whereas for even $n$, we have that $\pi_\D(c) = \g$ is a 
	$\mathbb{Z}_2$-grading satisfying:
\begin{equation}
\label{eq:grad}
	\g = \g^*, \quad \g^2 = 1, \quad \g\D = -\D\g,
	\qquad \g a = a\g, \quad \forall a \in \A.
\end{equation}
\end{axiom}

\begin{axiom}[Finiteness and absolute continuity] 
\label{ax5}
	The space $\HH^\infty := \cap_m \Dom(\D^m)$ is a finite projective 
	$\A$-module \eqref{eq:E=pAN}, with a natural hermitian structure $(\cdot,\cdot)$
	\eqref{eq:hermstruct}-\eqref{eq:hermstruct1}, given by
\begin{equation}
	\big\langle \zeta, a\eta \big\rangle = \intbar \, a(\zeta,\eta)|\D|^{-n}, 
	\qquad \forall a \in \A, \; \forall \zeta,\eta \in \HH^\infty,
\end{equation}
	where $\intbar$ denotes the noncommutative integral given by the Dixmier trace.
\end{axiom}


%
%

The following reconstruction theorem (see \cite[\S11]{C13} for the proof) provides 
us a purely spectral/operator-algebraic characterization of riemannian 
geometries and, hence, facilitates their generalization to a noncommutative 
setting.

\begin{ftheorem}
\label{thrm1}
	For a commutative spectral triple $(\A,\HH,\D)$ that respects Axioms \ref{ax1}, 
	\ref{ax2}, \ref{ax4}, \ref{ax5}, a stronger\footnotemark~regularity condition 
	(Axiom \ref{ax3}), and is equipped with an antisymmetric Hochschild cycle 
	$c$; there exists a compact oriented smooth manifold $\M$ such that $\A$ is the
	algebra $C^\infty(\M)$ of smooth functions on $\M$. 
\medskip
\end{ftheorem}

\footnotetext{All the elements of the endomorphism algebra $\End_\A(\HH^\infty)$ are regular.\label{note1}}

Moreover, the converse of the above statement also holds. That is, given any 
compact oriented smooth manifold $\M$, one can associate to it a strongly regular\textsuperscript{\ref{note1}}
commutative spectral triple $(C^\infty(\M),\HH,\D)$ respecting the Axioms 
\ref{ax1}, \ref{ax2}, \ref{ax4}, \ref{ax5}, and equipped with an antisymmetric 
Hochschild cycle $c$. 

\pagebreak

\section{Riemannian spin geometry}
\label{sec:1.2}

With two more axioms, Theorem~\ref{thrm1} can be extended to incorporate spin 
structures and, thus, gives a spectral characterization of 
riemannian spin manifolds, which are of importance for our purposes.

\smallskip

Before moving on to the axioms and the extension of the theorem, we quickly 
review some standard definitions (from \cite[\S4.2]{Su}) of spin geometry \cite{LM}.
For the definitions and notations related to Clifford algebras see App.~\ref{app:cliff}.

\begin{define}
	A {\bf spin$^c$ structure} $(\M,\cal S)$ on a riemannian manifold $(\M,g)$ 
	consists of a vector bundle ${\cal S} \to \M$ such that 
\begin{equation}
	\End({\cal S}) \simeq
	\begin{cases}
		\mathbb{C}l({\cal TM}), & \qquad \M \text{ even-dimensional} \\
		\mathbb{C}l({\cal TM})^0, & \qquad \M \text{ odd-dimensional}
	\end{cases},
\end{equation}
	as algebra bundles, where $\mathbb{C}l({\cal TM})$ is the complex 
	Clifford algebra bundle over $\M$
	and $\mathbb{C}l({\cal TM})^0$ its even part. A riemannian manifold with a 
	spin$^c$ structure is called a {\bf spin$^c$ manifold}.
\end{define}

\begin{define}
	If there exists a spin$^c$ structure $(\M,\cal S)$, then ${\cal S}\to\M$ is 
	called a {\bf spinor bundle} and the continuous sections $\psi\in\G({\cal S})$ 
	of the spinor bundle are called {\bf spinors}. The completion of the space
	$\G({\cal S})$ in the norm with respect to the inner product
\begin{equation}
	\big\langle \psi_1,\psi_2 \big\rangle := \int_\M dx\sqrt{g}\,(\psi_1,\psi_2)(x),
\end{equation}
	with $dx\sqrt{g}$ being the riemannian volume form; is called the {\bf Hilbert 
	space $L^2(\M,{\cal S})$ of square integrable spinors}.
\end{define}

\begin{define}
	Given a spin$^c$ structure $(\M,{\cal S})$. If there exists an anti-unitary 
	operator $\J : \G({\cal S}) \to \G({\cal S})$ commuting with:
\begin{enumerate}
	\item the action of real-valued continuous functions on $\G(\cal S)$, and
	\item the algebra $Cliff^{-1}(\M)$ \eqref{alg-Cliff} or its even part 
	$Cliff^{-1}(\M)^0$ if $\M$ is odd-dimensional,
\end{enumerate}	
	then, $\J$ is called the {\bf charge conjugation} operator. Also,
	$({\cal S},\J)$ defines a {\bf spin structure} and $\M$ is, then, referred to 
	as a {\bf spin manifold}.
\end{define}

The following Axiom \ref{ax6} is a $K$-theoretic reformulation (i.e.\ in terms of 
$K$-homologies) of the Poincar\'e duality, which will not be used within the scope
of this thesis. So we state it here without explanation and refer to \cite{C94} for
the details.

\begin{axiom}[Poincar\'e duality]
\label{ax6}
	The intersection form $K_*(\A) \times K_*(\A) \to \mathbb{Z}$ is invertible.
\end{axiom}

\begin{axiom}[Real structure] 
\label{ax7}
	There exists an antilinear isometry $J : \HH \to \HH$ such that 
\begin{equation}
\label{eq:RealStruct}
	J a J^{-1} = a^*, \qquad \forall a \in \A,
\end{equation}
	and
\begin{equation}
\label{eq:Reality}
	J^2 = \epsilon\Id_\HH, \qquad J\D = \epsilon'\D J, 
	\qquad J\g = \epsilon''\g J,
\end{equation}
	where the signs $\epsilon,\epsilon',\epsilon''\in\{\pm1\}$ are determined by 
	the $KO$-dimension $n$ modulo $8$, as per Table \ref{tab:KO-dim}, which 
	come from the classification of the irreducible representations of Clifford 
	algebras (see \cite[\S4.1]{Su}). 
\begin{table}[h]
\centering
\begin{tabular}{|c|c|c|c|c|c|c|c|c|}
\hline
$n$ & $0$ & $1$ & $2$ & $3$ & $4$ & $5$ & $6$ & $7$ \\
\hline
$\epsilon$ & $1$ & $1$ & $-1$ & $-1$ & $-1$ & $-1$ & $1$ & $1$ \\
$\epsilon'$ & $1$ & $-1$ & $1$ & $1$ & $1$ & $-1$ & $1$ & $1$ \\
$\epsilon''$ & $1$ & & $-1$ & & $1$ & & $-1$ & \\
\hline
\end{tabular}
\caption{The signs of $\epsilon,\epsilon',\epsilon''$ for a spectral triple of 
KO-dimension $n$ (mod $8$).}
\label{tab:KO-dim}
\end{table}
\end{axiom}

\begin{define}
	A spectral triple is said to be {\bf real} if it is endowed with a real 
	structure $J$, as defined in the Axiom \ref{ax7}.
\end{define}

\begin{ftheorem}
	For a commutative real spectral triple $(\A,\HH,\D)$ satisfying the Axioms 
	\ref{ax1}--\ref{ax5} and a weaker\footnotemark~Poincar\'e duality (Axiom 
	\ref{ax6}), there exists a smooth oriented compact spin\footnotemark~manifold 
	$\M$ such that $\A=C^\infty(\M)$. 
\medskip
\end{ftheorem}

\addtocounter{footnote}{-2}
\stepcounter{footnote}
\footnotetext{The multiplicity of the action of the bicommutant $\A''$ of $\A$ in 
$\HH$ is $2^{n/2}$, see Theorem 11.5 of \cite{C13}.}
\stepcounter{footnote}
\footnotetext{Without the real structure, the theorem continues to hold but rather 
for a spin$^c$ manifold; see the Remark 5 about Theorem 1 in \cite{C96}.}

\begin{define}
\label{def:CT}
	A {\bf canonical triple} is the real spectral triple 
\begin{equation}
\label{eq:CT}
	\big( \A := C^\infty(\M),\; \HH := L^2(\M,{\cal S}),\; \D := \eth \big),
\end{equation}
	associated to an $n$-dimensional closed spin manifold $\M$. The algebra 
	$C^\infty(\M)$ acts on the Hilbert space $L^2(\M,{\cal S})$ by point-wise 
	multiplication
\begin{equation}
\label{eq:pim}
	\big(\pi_\M(f)\big)(x) := f(x)\psi(x).
\end{equation}
	$\eth$ is the 
	Dirac operator
\begin{equation}
	\eth := -i\sum_{\mu=1}^n \gm\nsm, \qquad \nsm := \dm + \osm,
\end{equation}
	where $\g$'s are the self-adjoint Euclidean Dirac matrices (see \S\ref{GammaMatrices}) 
	and $\nabla^{\cal S}$ is the covariant derivative associated to the 
	spin connection $\omega^{\cal S}$ on the spinor bundle. The real structure
	$J$ is given by the charge conjugation operator $\J$.
\end{define}

One checks, from the Table \ref{tab:KO-dim}, 
that for the canonical triple \eqref{eq:CT},
\begin{equation}
	KO\text{-}\Dim(\M) = \Dim(\M) = n.
\end{equation}

\section{Generalization to the noncommutative case}
\label{sec:1.3}

In general, the algebra $\A$ of a spectral triple $(\A,\HH,\D)$ may not necessarily
be commutative, in which case the Axioms \ref{ax1}, \ref{ax3} and \ref{ax5} stay 
unchanged while the remaining axioms are modified. 

\begin{define}
\label{def:OppAlg}
	The {\bf opposite algebra} $\A^\circ$ associated to an algebra $\A$ is 
	isomorphic to $\A$ (as a vector space), but it is endowed with an 
	{\bf opposite product} $\bullet$ given by 
\begin{equation}
\label{eq:OppProd}
	a^\circ\bullet b^\circ := (ba)^\circ, \qquad 
	\forall a^\circ,b^\circ\in\A^\circ,
\end{equation}
where $a\mapsto a^\circ$ is the isomorphism between the vector spaces underlying
$\A$ and $\A^\circ$.
\end{define}

Tomita-Takesaki theory (App.~\ref{appb}) asserts, 
for a weakly closed $*$-algebra $\A$ of bounded operators on a Hilbert space $\HH$ 
(endowed with a cyclic and separating vector), 
the existence of a canonical \emph{modular involution} $J:\HH\to\HH$, 
which is an antilinear isometry 
\begin{equation}
	J:\A\to\A' \simeq J\A J^{-1}, \qquad \A\ni a \mapsto Ja^*J^{-1}\in\A',
\end{equation}
defining a $\mathbb{C}$-linear anti-isomorphism from the algebra to its commutant
\eqref{eq:b4}. One identifies this with the antilinear isometry
in Axiom \ref{ax7}, which is the real structure $J$, and requires it to induce an
isomorphism between $\A$ and $\A^\circ$ given by the map
\begin{equation}
\label{eq:OppAlg}
	\A \ni a \mapsto a^\circ := J a^*J^{-1} \in \A^\circ,
\end{equation}
	which defines a right representation of $\A$ or, equivalently a left 
	representation of $\A^\circ$, on $\HH$, thus
\begin{equation}
	\psi a := a^\circ\psi = Ja^*J^{-1}\psi, \qquad \psi \in \HH.
\end{equation}

The opposite product $\bullet$ \eqref{eq:OppProd} in the opposite algebra 
$\A^\circ$ is `opposite' to the product $\cdot$ in the algebra $\A$ in the 
following sense. For $a,b\in\A$, the map \eqref{eq:OppAlg} sends $a\cdot b\in\A$ 
to
\begin{equation}
	(a\cdot b)^\circ = J(a\cdot b)^*J^{-1} = Jb^*a^*J^{-1} = 
	(Jb^*J^{-1})(Ja^*J^{-1}) = b^\circ \bullet a^\circ.
\end{equation}

Consequently, the Axioms \ref{ax7} and \ref{ax2} are adapted accordingly as
below and will be used extensively later.

\begin{axm}{7'}[Commutant property, zeroth-order or order-zero condition]
\label{axm7'}
	One has that
\begin{equation}
\label{eq:commprop}
	\big[a,b^\circ\big] = ab^\circ - b^\circ a = 0, \qquad \forall a,b \in \A.
\end{equation}
\end{axm}

\begin{axm}{2'}[First-order or order-one condition]
\label{axm2'}
	One has that
\begin{equation}
	\big[[\D,a],b^\circ\big] = 0 \quad \text{or, equivalently,} \quad 
	\big[[\D,b^\circ],a\big] = 0, \qquad \forall a, b \in \A,
\end{equation}
where the equivalence follows from Axiom \ref{axm7'}.
\end{axm}

The commutativity in \eqref{eq:commprop} provides $\HH$ with the structure of an 
$(\A\x\A^\circ)$-module or, equivalently, an $\A$-bimodule:
\begin{equation}
	(a\x b^\circ)\psi = aJb^*J^{-1}\psi, 
	\qquad \forall a, b \in \A,\, \forall \psi \in \HH.
\end{equation}

Similar to Axiom \ref{ax6}, we shall not make use
of the corresponding modification, i.e.\ the following Axiom \ref{axm6'}, which 
reformulates Poincar\'e duality in the framework of Atiyah's $KR$-theory. 
Nevertheless, for completeness, we give the statement without explanation and refer
to \cite{C95} for the details.

\begin{axm}{6'}
\label{axm6'}
	The Kasparov cup product with the class $\mu \in KR^n(\A\x\A^\circ)$ 
	of the Fredholm module associated to the spectral triple provides an isomorphism:
\begin{equation}
	K_*(\A) \xrightarrow{\cap_\mu} K^*(\A)
\end{equation}
	from $K$-cohomology to $K$-homology.
\end{axm}

\begin{axm}{4'}
\label{axm4'}
	There exists a Hochschild cycle $c \in Z_n(\A,\A\x\A^\circ)$, such that for odd
	$n$, we have $\pi_\D(c) = 1$, whereas for even $n$, we have that $\pi_\D(c) = 
	\g$ is a $\mathbb{Z}_2$-grading with
\begin{equation}
	\g = \g^*, \quad \g^2 = 1, \quad \g\D = -\D\g, 
	\qquad \g a = a\g, \quad \forall a \in \A.
\end{equation}
\end{axm}

\bigskip

Thus, an $n$-dimensional (metric) {\bf noncommutative geometry} is defined by a 
spectral triple $(\A,\HH,\D)$ respecting the Axioms \ref{ax1}, \ref{axm2'}, 
\ref{ax3}, \ref{axm4'}, \ref{ax5}, \ref{axm6'} and \ref{axm7'}. 
Such spectral triples will be the key objects that we will work with.

\clearpage
\newpage 

%

%

\chapter{Gauge Theories from Spectral Triples}
\label{chap:GTST}

Morita equivalence is a generalized notion of isomorphisms between algebras 
(\S\ref{sec:MorEq}). The existence of Morita equivalence generates \emph{inner
fluctuations} in noncommutative geometry, which are interpreted as generalized 
gauge fields. In other words, Morita equivalence of algebras when lifted to
spectral triples (\S\ref{sec:MorEqSpecTr}) gives rise to gauge theories on 
noncommutative geometries. More concretely, exporting the geometry of a spectral
triple $(\A,\HH,\D)$ on to an algebra $\B$ Morita equivalent to $\A$ in a manner
consistent with the real structure gives rise to the fluctuation of the metric
(\S\ref{subsec:innerfluc}).

\smallskip

Since in Def.~\ref{def:1.1}, we defined spectral triples for unital $*$-algebras;
from here onwards we will work with unital algebras. The material presented in 
this section is well-known and has been mostly adapted from 
\cite{C94,C96,CMa,LM2,Su}. 

\section{Morita equivalence of algebras}
\label{sec:MorEq}

\begin{define}
	A {\bf left $A$-module} ${_A\E}$ is a vector space $\E$ with a left representation 
of the algebra $A$ given by a bilinear map
\begin{equation}
	A \times \E \to \E, \quad (a,\zeta) \mapsto a\zeta,
	\qquad \forall a \in A, \, \forall \zeta \in \E,
\end{equation}
such that
\begin{equation}
	(a_1a_2)\zeta = a_1(a_2\zeta),
	\qquad \forall a_1,a_2 \in A, \, \forall \zeta \in \E.
\end{equation}
A {\bf right $B$-module} $\E_B$ is a vector space $\E$ with a right 
representation of the algebra $B$ given by a bilinear map
\begin{equation}
	\E \times B \to \E, \quad (\zeta,b) \mapsto \zeta b, 
	\qquad \forall b \in B, \, \forall \zeta \in \E,
\end{equation}
such that
\begin{equation}
	\zeta(b_1b_2) = (\zeta b_1)b_2, 
	\qquad \forall b_1,b_2 \in B, \, \forall \zeta \in \E.
\end{equation}
An $A$-$B$-{\bf bimodule} $_A\E_B$ is both a left $A$-module $_A\E$ and a 
right $B$-module $\E_B$ with the actions of the left and the right representations 
commuting with each other:
\begin{equation}
	a(\zeta b) = (a\zeta)b, 
	\qquad \forall a \in A, \, \forall b \in B, \, \forall \zeta \in \E. 
\end{equation}
\end{define}

\begin{define}
A {\bf module homomorphism} is a linear map $\varphi : \E \to \E'$ respecting 
the algebra representation on the module $\E$, in the following manner
\begin{align}
	\text{for } {_A\E} : 
	&& \varphi(a\zeta) & = a\varphi(\zeta),  		 
	&& \forall a \in A, \, \forall \zeta \in \E, \nonumber \\
	\text{for } \E_B : 
	&& \varphi(\zeta b) & = \varphi(\zeta)b, 
	&& \forall b \in B, \, \forall \zeta \in \E, \\
	\text{for } {_A\E_B} :
	&& \varphi(a\zeta b) & = a\varphi(\zeta)b, 
	&& \forall a \in A, \, \forall b \in B, \, \forall \zeta \in \E. \nonumber
\end{align}
Let ${\sf Hom}_A(\E,\E') := \{ \varphi : \E \to \E' \}$ denote the 
{\bf space of $A$-linear module homomorphisms}. Then, the 
{\bf algebra of $A$-linear endomorphisms} of $\E$ is given by
${\sf Hom}_A(\E,\E) =: {\sf End}_A(\E)$.
\end{define}

\begin{define}
A {\bf balanced tensor product} of modules $\E_A$ and $_A\E'$ is defined as
\begin{equation}
	\E \x_A \E' := \E \x \E' / \left\{ 
	\textstyle\sum_j (\zeta_ja_j\x\zeta_j' - \zeta_j\x a_j\zeta_j');
	\; \forall a_j\in A,\,\forall\zeta_j\in\E,\,\forall\zeta_j'\in\E' \right\}\!,
\end{equation}
where the quotient ensures the $A$-linearity of the tensor product:
\begin{equation}
\label{eq:Alinrty}
	\zeta a \x_A \zeta' = \zeta \x_A a\zeta', 
	\qquad \forall a\in A,\,\forall\zeta\in\E,\,\forall\zeta'\in\E'.
\end{equation}
\end{define}

\begin{mdframed}
\begin{define}
	The algebras $A$ and $B$ are called {\bf Morita equivalent} to each other if 
	there exist bimodules $_A\E_B$ and $_B\E'_A$ such that
\begin{equation}
	\E \x_B \E' \simeq A \qquad \text{and} \qquad \E' \x_A \E \simeq B,
\end{equation}
	as $A$-bimodule and $B$-bimodule, respectively. 
\end{define}
\medskip
\end{mdframed}

\begin{ex}
	For any $n\in\mathbb{N}$, $B = M_n(A)$ is Morita equivalent to $A$, where both
	of the bimodules $\E$ and $\E'$, implementing Morita equivalence, are given by 
\begin{equation}
\label{eq:AN}
	A^n := \underbrace{A\+\cdots\+A}_{n \text{ times}}\,,
\end{equation}
	viewed as an $A$-$M_n(A)$-bimodule and an $M_n(A)$-$A$-bimodule, respectively, 
	so that
\begin{equation}
	A^n \x_{M_n(A)} A^n \simeq A \qquad \text{and} \qquad
	A^n \x_A A^n \simeq M_n(A).
\end{equation}
	In particular, for $n=1$, $B = M_1(A) = A$. That is, an algebra $A$ is Morita 
	equivalent to itself, with the bimodules implementing Morita equivalence being 
	$\E = \E' = A$. 
\end{ex}

\begin{define}
	An $A$-module $\E$ is called {\bf finite projective} (or, finitely 
	generated projective), if there exists an idempotent matrix 
	$\wp = \wp^2 \in M_n(A)$, for some $n\in\mathbb{N}$, such that 
\begin{equation}
\label{eq:E=pAN}
	\E_A \simeq \wp A^n \qquad \text{or} \qquad 
	{_A\E} \simeq A^n\wp,
\end{equation}
	where $A^n$ is the bimodule given by \eqref{eq:AN}. 
\end{define}

Further, it follows that $\E_A$ is a finite projective module iff
\begin{equation}
\label{eq:End=ExHom}
	{\sf End}_A(\E) \simeq \E \x_A {\sf Hom}_A(\E,A),
\end{equation}
with ${\sf Hom}_A(\E,A)$ as a left $A$-module.

\smallskip

Morita equivalence of algebras can be characterized in terms of endomorphism 
algebras of finite projective modules as follows. 
\begin{mdframed}
\smallskip
Two unital algebras $A$ and $B$ are Morita equivalent iff
\begin{equation}
\label{eq:Morita}
	B \simeq {\sf End}_A(\E),
\end{equation}
for a finite projective module $\E$.
\smallskip
\end{mdframed}

Thus, \eqref{eq:Morita}, with \eqref{eq:End=ExHom}, implies that all the algebras 
Morita equivalent to a unital algebra $A$ are of the form $\E\x_A{\sf Hom}(\E,A)$ 
for some finite projective module $\E$. We notice, in particular, if $B = A$, 
then $\E = A$.

\subsection{$*$-algebras and Hilbert bimodules}

The above discussion on algebras and modules specializes to $*$-algebras
\eqref{def:*-alg}, with
additionally requiring $\wp$ in \eqref{eq:E=pAN} to be an \emph{orthogonal 
projection}, that is, $\wp^* = \wp$.

\begin{define}
For a right $\A$-module $\E_\A$, the {\bf conjugate module} $_\A\E^\circ$ is a left
$\A$-module
\begin{equation}
	\E^\circ := \left\{ \overline\zeta\;\vert\;\zeta\in\E \right\}, \qquad \text{with} 
	\quad a\overline\zeta = \overline{\zeta a^*}, \qquad \forall a\in \A.
\end{equation}
\end{define}

In case $\E_\A$ is a finite projective module, then it follows that 
$\E^\circ \simeq {\sf Hom}_\A(\E,\A)$, as a left $\A$-module, 
and \eqref{eq:End=ExHom} gives 
\begin{equation}
	{\sf End}_\A(\E) \simeq \E \x_\A \E^\circ\!,
\end{equation}
which, following \eqref{eq:Morita}, implies that all the $*$-algebras Morita 
equivalent to a unital $*$-algebra $\A$ are of the form $\E \x_\A \E^\circ$. 

\pagebreak

\begin{define} 
On a finite projective right $\A$-module $\E_\A$, there is a 
sesquilinear map\footnote{antilinear in the first entry and linear in the second} 
$\big\langle\cdot,\cdot\big\rangle_\E: \E\times\E \to \A$ 
called the {\bf hermitian structure}, which defines an $\A$-valued inner product 
on $\E$ satisfying
\begin{align}
\label{eq:hermstruct}
	\big\langle \zeta_1 a_1, \zeta_2 a_2 \big\rangle_\E 
	& = a_1^* \big\langle \zeta_1, \zeta_2 \big\rangle_\E a_2, 
	&& \forall a_1, a_2 \in \A,\, \forall \zeta_1, \zeta_2 \in \E; \nonumber \\
	\big\langle \zeta_1, \zeta_2 \big\rangle_\E^* & = \big\langle \zeta_2, \zeta_1 \big\rangle_\E, 
	&& \forall \zeta_1, \zeta_2 \in \E; \\
	\big\langle \zeta, \zeta \big\rangle_\E & \ge 0, 
	&& \forall \zeta \in \E, \quad (\text{equal iff } \zeta = 0). \nonumber
\end{align}
and it is obtained by restricting the hermitian structure 
$\big\langle\cdot,\cdot\big\rangle_{\A^n}$ on $\A^n$ \eqref{eq:AN}, given by 
\begin{equation}
\label{eq:hermstruct1}
	\big\langle \zeta , \eta \big\rangle_{\A^n} 
	= \textstyle\sum_{j=1}^n \zeta_j^*\eta_j, \qquad \forall \zeta, \eta \in \A^n,
\end{equation}
to $\E_\A \simeq \wp\A^n$ \eqref{eq:E=pAN}, for some self-adjoint idempotent 
matrix $\wp \in M_n(\A)$.
\end{define}

A finite projective right $\A$-module $\E_\A$ is {\bf self-dual} with respect to 
the hermitian structure $\big\langle\cdot,\cdot\big\rangle_\E$ on it \cite[Prop.~7.3]{Ri}, 
that is
\begin{equation}
	\forall \varphi \in \Hom_\A(\E,\A) \quad \exists ! \;\xi_\varphi \in \E : 
	\quad \varphi(\zeta) = (\xi_\varphi,\zeta)_\E, \qquad \forall \zeta \in \E.
\end{equation}

Similarly, there exists a hermitian structure 
$_\E\big\langle\cdot,\cdot\big\rangle:\E\times\E\to\A$ on a finite projective left 
$\A$-module $_\A\E$, which is linear in the first entry, antilinear in the second, 
and obtained by restriction to $\A^n$, since ${_\A\E} \simeq \A^n\wp$ 
\eqref{eq:E=pAN}. 

\begin{define}
For $*$-algebras $\A$ and $\B$, a {\bf Hilbert bimodule} $\E$ is an 
$\A$-$\B$-bimodule $_\A\E_\B$ with a $\B$-valued inner product $\big\langle \cdot, 
\cdot \big\rangle_\E : \E \times \E \to \B$ on $\E$ satisfying
\begin{align}
	\big\langle\zeta_1, a\zeta_2\big\rangle_\E & = \big\langle a^*\zeta_1, \zeta_2\big\rangle_\E,
	&& \forall a \in \A, \; \forall \zeta_1, \zeta_2 \in \E; \nonumber \\
	\big\langle\zeta_1, \zeta_2b\big\rangle_\E & = \big\langle\zeta_1, \zeta_2\big\rangle_\E b,
	&& \forall b \in \B, \; \forall \zeta_1, \zeta_2 \in \E; \nonumber \\
	\big\langle\zeta_1, \zeta_2\big\rangle_\E^* & = \big\langle\zeta_2, \zeta_1\big\rangle_\E,
	&& \forall \zeta_1, \zeta_2 \in \E; \\
	\big\langle\zeta, \zeta\big\rangle_\E & \ge 0,
	&& \forall \zeta \in \E, \quad (\text{equal iff } \zeta = 0). \nonumber
\end{align}
\end{define}

In particular, a $*$-algebra $\A$ is Morita equivalent to itself, where the 
finite projective module $\E$ implementing Morita equivalence is taken to be the 
algebra $\A$ itself, carrying the hermitian structure:
\begin{equation}
\label{eq:Ainnprod}
	\big\langle a, b \big\rangle_\A = a^*b \quad \text{or} \quad
	{_\A\big\langle a, b \big\rangle} = ab^*, \qquad \forall a,b\in \A.
\end{equation} 

Morita equivalent algebras have equivalent representation 
theories: given Morita equivalent $*$-algebras $\A$ and $\B$, 
a right $\A$-module $\F_\A$ is converted to a right $\B$-module via 
$\F\x_\A\E \simeq \B$, for some $\A$-$\B$-bimodule $_\A\E_\B$. 

\smallskip

With a representation \eqref{eq:piAtoLH} of $\A$, the Hilbert space $\HH$ is, in 
fact, a Hilbert bimodule $_\A\HH_\mathbb{C}$ with the $\mathbb{C}$-valued
inner product given by $\big\langle\cdot,\cdot\big\rangle_\HH$.

\pagebreak

\subsection{Connections: from modules to Hilbert spaces}

\begin{define}
	A derivation $\delta$ of an algebra $\A$ taking values in an $\A$-bimodule 
	$\Omega$ is a map $\delta : \A \to \Omega$ satisfying 
\begin{equation}
\label{eq:deltAb}
	\delta(ab) = \delta(a)\cdot b + a\cdot\delta(b), \qquad \forall a,b\in \A,
\end{equation}
where $\cdot$ indicates both the left and the right $\A$-module structure of 
$\Omega$.
\end{define}

\begin{define}
\label{def:conn}
	An $\Omega$-valued connection $\nabla$ on a right $\A$-module $\E_\A$ is a 
	linear map $\nabla : \E \to \E\x_\A\Omega$ satisfying the Leibniz rule, i.e.
\begin{equation}
\label{eq:Leibniz}
 	\nabla(\zeta a) = \nabla(\zeta)\cdot a + \zeta\x\delta(a), 
	\qquad \forall a\in \A,\, \forall \zeta \in \E,
\end{equation}
	where the right action of $\A$ on $\E\x_\A\Omega$ is inherited from the right 
	$\A$-module structure of $\Omega$ as follows
\begin{equation}
	(\zeta \x \omega)\cdot a := \zeta \x (\omega\cdot a), \qquad 
	\forall \omega \in \Omega.
\end{equation}
Similarly, an $\Omega$-valued connection $\nabla$ on a left $\A$-module $_\A\E$ is
a linear map $\nabla : \E \to \Omega\x_\A\E$ such that
\begin{equation}
\label{eq:Leibniz1}
 	\nabla(a\zeta) = a\cdot\nabla(\zeta) + \delta(a)\x\zeta, 
	\qquad \forall a\in \A,\, \forall \zeta \in \E,
\end{equation}
where the left action of $\A$ on $\Omega\x_\A\E$ is being inherited from the left 
module structure of $\Omega$ as follows
\begin{equation}
	a\cdot(\omega \x \zeta) := (a\cdot \omega) \x \zeta, \qquad 
	\forall \omega \in \Omega.
\end{equation}
\end{define}

When both $\A$ and $\Omega$ are acting on a Hilbert space $\HH$ (on the left),
the connection $\nabla$ can then be moved from the right module $\E_\A$ to $\HH$, 
by virtue of the action of $\Omega$ on $\HH$:
\begin{equation}
	\E \x_\mathbb{C} \Omega \times \HH \to \E \x_\mathbb{C} \HH, \qquad
	(\zeta \x \omega) \psi = \zeta \x (\omega\psi),
\end{equation}
via the following map
\begin{equation}
\label{eq:nablaL}
	\nabla : \E \x_\mathbb{C} \HH \to \E \x_\mathbb{C} \HH, \qquad
	\nabla(\zeta\x\psi) := \nabla(\zeta)\psi,
\end{equation}
defined with a slight abuse of notation. 
Similar to \eqref{eq:nablaL}, for the right action of $\A$ and left action of 
$\Omega$ on $\HH$, the action
\begin{equation}
	\HH \times \Omega \x_\mathbb{C} \E \to \HH \x_\mathbb{C} \E, \qquad
	\psi (\omega \x \zeta) = (\psi\omega)\x\zeta,
\end{equation}
defines
\begin{equation}
\label{eq:nablaR}
	\nabla : \HH \x_\mathbb{C} \E \to \HH \x_\mathbb{C} \E, \qquad
	\nabla(\psi \x \zeta) := \psi\nabla(\zeta).
\end{equation}

\pagebreak
However, these maps (\ref{eq:nablaL} or \ref{eq:nablaR}) do not automatically 
extend over to the tensor products $\E\x_\A\HH$ or $\HH\x_\A\E$ (respectively) --
on account of failing to be $\A$-linear, which is captured by the derivation 
$\delta$ \eqref{eq:deltAb} generating $\Omega$, provided that the action of $\A$ 
and $\Omega$ be compatible \cite[Prop.~3.1,~3.2]{LM2}:
\begin{enumerate}
	\item If the left action of both $\A$ and $\Omega$ on $\HH$ are such that 
	$(\omega\cdot a)\psi = \omega(a\psi)$, then $\nabla$ in \eqref{eq:nablaL} 
	satisfies (Leibniz rule)
\begin{equation}
\label{eq:LeibL}
	\nabla(\zeta a)\psi = \nabla(\zeta)a\psi + \zeta\x\delta(a)\psi.
\end{equation}
	\item If the right action of $\A$ and the left action of $\Omega$ on $\HH$ are
	such that $(a\cdot\omega)\psi = \omega(\psi a)$, then $\nabla$ in 
	\eqref{eq:nablaR} satisfies (Leibniz rule)
\begin{equation}
\label{eq:LeibR}
	\psi\nabla(a\zeta) = \psi a\nabla(\zeta) + \delta(a)\psi\x\zeta.
\end{equation}
\end{enumerate}

\section{Morita equivalence of spectral triples}
\label{sec:MorEqSpecTr}

We are given a spectral triple $(\A,\HH,\D)$ and an algebra $\B\simeq\End_\A(\E)$ 
Morita equivalent to $\A$ via a finite projective module $\E$, as in 
\eqref{eq:Morita}. The task at hand is the construction of a spectral triple 
$(\B,\HH',\D')$, that is, to export the geometry $(\A,\HH,\D)$ on to an algebra 
Morita equivalent to $\A$.

\subsection{Morita equivalence via right $\A$-module}
\label{subsec:MoritaL}

Let us say that $\A$ and $\B$ are Morita equivalent via a Hilbert 
$\B$-$\A$-bimodule $\E_r$. Since $_\A\HH_\mathbb{C}$ carries a left $\A$-module 
structure induced by the representation of the algebra $\A$, the product
\begin{equation}
	\E_r \x_\A \HH =: \HH_r, 
\end{equation}
is a Hilbert $\B$-$\mathbb{C}$-bimodule with a left action of $\B$ inherited from $\E_r$, 
by extension:
\begin{equation}
	b(\eta\x\psi) := b\eta \x \psi, 
	\qquad \forall b\in\B,\,\forall(\eta\x\psi)\in\HH_r;
\end{equation}
and the $\mathbb{C}$-valued inner product given by
\begin{equation}
	\big\langle \eta_1\x\psi_1, \eta_2\x\psi_2 \big\rangle_{\HH_r} = 
	\big\langle \psi_1, \big\langle \eta_1,\eta_2 \big\rangle_{\E_r} \psi_2 \big\rangle_\HH, \qquad
	\forall \eta_1,\eta_2 \in\E_r,\, \forall \psi_1,\psi_2\in \HH.
\end{equation}
The following na\"{\i}ve attempt at furnishing an adaption $\D_r$ of the action of
$\D$ over to $\HH_r$:
\begin{equation}
	\D_r(\eta \x \psi) := \eta \x \D\psi,
\end{equation}
fails due to its incompatibility with $\A$-linearity \eqref{eq:Alinrty} of the 
balanced tensor product over $\A$, since its action on the elementary tensors 
generating the ideal, that is,
\begin{equation}
\label{eq:DrAct}
	\D_r(\eta a\x\psi - \eta\x a\psi) = \eta a\x\D\psi - \eta\x\D a\psi
	= -\eta\x[\D,a]\psi
\end{equation}
is not necessarily zero. However, \eqref{eq:DrAct} suggests for $\D_r$ to be a 
well-defined operator on $\HH_r$ that it must rather act as follows:
\begin{equation}
\label{eq:DrAct1}
	\D_r(\eta a\x\psi - \eta\x a\psi) =
	\eta a\x\D\psi - \eta\x\D a\psi + \eta\x[\D,a]\psi.
\end{equation}
Recalling Def.~\ref{def:conn} of a connection $\nabla$ on a right $\A$-module 
$\E_r$ taking values in the 
$\A$-bimodule $\Omega_\D^1(\A)$ generated by the derivation $\delta(a) = [\D,a]$, 
one has \eqref{eq:LeibL}
\begin{equation}
	\eta\x[\D,a]\psi = \nabla(\eta a)\psi -\nabla(\eta)a\psi.
\end{equation}
In that light, \eqref{eq:DrAct1} becomes
\begin{equation}
\begin{split}
	\D_r(\eta a\x\psi) -\D_r(\eta\x a\psi) 
	& = \eta a\x\D\psi -\eta\x\D a\psi +\nabla(\eta a)\psi -\nabla(\eta)a\psi \\
	& = \eta a\x\D\psi +\nabla(\eta a)\psi 
	-\big(\eta\x\D a\psi +\nabla(\eta)a\psi\big),
\end{split}
\end{equation}
implying that the correct action of $\D_r$ on $\HH_r$ must be as follows
\begin{equation}
\label{eq:D_r}
	\D_r(\eta\x\psi) := \eta\x\D\psi +\nabla(\eta)\psi,
\end{equation} 
which is $\A$-linear by \eqref{eq:LeibL}. 

\medskip

Now, if $\E_r$ is finite projective \eqref{eq:E=pAN}, any $\Omega_\D^1(\A)$-valued 
connection $\nabla$ is of the form $\nabla_0 + \bm\omega$, where 
$\nabla_0 := \wp \circ \delta$ is the
Gra{\ss}mann connection, i.e.
\begin{equation}
	\forall \eta = \wp\! 
	\begin{pmatrix}
		\eta_1 \\ \vdots \\ \eta_n
	\end{pmatrix}\! \in \E_r, \quad \text{with } 
	\eta_{j=1,\ldots,n} \in \A; \qquad
	\nabla_0\eta = \wp\!
	\begin{pmatrix}
		\delta(\eta_1) \\ \vdots \\ \delta(\eta_n)
	\end{pmatrix}\!, 
\end{equation}
and $\bm\omega$ is an $\A$-linear map $\E_r\to\E_r\x\Omega_\D^1(\A)$ such that
\begin{equation}
	\bm\omega(\eta a) = \bm\omega(\eta)\cdot a, \qquad 
	\forall a \in \A, \; \forall \eta \in \E_r.
\end{equation}

\begin{ftheorem}[Theorem 6.15, \cite{Su}]
Given a spectral triple $(\A,\HH,\D)$ and a connection $\nabla$ on finite 
projective $\E_r$, then $(\B,\HH_r,\D_r)$ is a spectral triple provided $\nabla$ is 
hermitian, that is, it satisfies
\begin{equation}
	\big\langle \eta_1, \nabla\eta_2 \big\rangle_{\E_r} - 
	\big\langle \nabla\eta_1, \eta_2 \big\rangle_{\E_r} =
	\delta \big\langle \eta_1, \eta_2 \big\rangle_{\E_r}, \qquad
	\forall \eta_1, \eta_2 \in \E_r. \medskip
\end{equation}
\end{ftheorem}

\pagebreak

\subsubsection{Morita self-equivalence via $\E_r = \A$}

The algebra $\A$ is Morita equivalent to itself via $\E_r = \A$ and 
any connection $\nabla$ on this right $\A$-module $\A$ is of the form $\delta + 
\omega$, for some $\omega \in \Omega^1_\D(\A)$. \eqref{eq:D_r} becomes
\begin{equation}
	\D_r(a\x\psi) = a\x\D\psi + (\delta + \omega)(a)\psi,
\end{equation}
which, identifying $a\x\psi\in\HH_r$ with $a\psi\in\HH$ and recalling that 
$\delta(a) = [\D,a]$, becomes
\begin{equation}
\begin{split}
	\D_r(a\psi) & = a\D\psi + [\D,a]\psi + \omega a\psi \\ 
	& = (\D+\omega)(a\psi),
\end{split}
\end{equation}
that is, on $\HH$, one has that
\begin{equation}
	\D_r = \D + \omega,
\end{equation}
which has a compact resolvent and bounded commutator with $\A$, since $\omega$ is
bounded. Thus, $(\A,\HH,\D_r)$ is a spectral triple, given $\omega$ is 
self-adjoint \cite{BMS}, and said to be {\bf Morita equivalent} to the
spectral triple $(\A,\HH,\D)$ \cite[\S3.2]{LM2}. 

\smallskip

If the initial spectral triple $(\A,\HH,\D)$ comes with a real structure $J$, the
latter does not necessarily get inherited by the Morita equivalent spectral triple
$(\A,\HH,\D_r)$ for $J\D_r = \epsilon'\D_r J$ holds iff $\omega = 
\epsilon'J\omega J^{-1}$, which, in general, may not be the case. In fact, it 
follows from \eqref{eq:RealStruct} and \eqref{eq:OppAlg} that
\begin{equation}
	J\omega J^{-1} = \epsilon'\sum_j(a_j^*)^\circ\left[\D,(b_j^*)^\circ\right],
\end{equation}
for some $\omega = \sum_j a_j[\D,b_j]\in\Omega_\D^1(\A)$.

\subsection{Morita equivalence via left $\A$-module}
\label{subsec:MoritaR}

Let us say that $\A$ and $\B$ are Morita equivalent via a Hilbert 
$\A$-$\B$-bimodule $\E_l$. Since $_\mathbb{C}\HH_\A$ is endowed with a right 
$\A$-module structure induced by the representation of the opposite algebra 
$\A^\circ$ (Def.~\ref{def:OppAlg}), the product
\begin{equation}
	\HH\x_\A\E_l =: {\HH_l}, 
\end{equation}
is a Hilbert $\mathbb{C}$-$\B$-bimodule with a right action of $\B$ inherited from $\E_l$, 
by extension:
\begin{equation}
	(\psi\x\eta)b := \psi\x\eta b, 
	\qquad \forall b\in\B,\,\forall(\psi\x\eta)\in\HH_l;
\end{equation}
and $\mathbb{C}$-valued inner product
\begin{equation}
	\big\langle \psi_1\x\eta_1, \psi_2\x\eta_2 \big\rangle_{\HH_l} = 
	\big\langle \psi_1 \big\langle \eta_1,\eta_2 \big\rangle_{\E_l}, \psi_2 \big\rangle_\HH, \qquad
	\forall \eta_1,\eta_2 \in\E_l,\, \forall \psi_1,\psi_2\in \HH.
\end{equation}
As before, the na\"{\i}ve attempt at furnishing an adaption $\D_l$ of the action of
$\D$ over to $\HH_l$:
\begin{equation}
	\D_l(\psi\x\eta) := \D\psi\x\eta 
\end{equation}
does not work for it is incompatible with the $\A$-linearity \eqref{eq:Alinrty} of 
the balanced tensor product over $\A$ as its action on the elementary tensors 
generating the ideal, that is,
\begin{equation}
\begin{split}
\label{eq:lDAct}
	\D_l(\psi\x a\eta - \psi a\x\eta)
	& = \D\psi\x a\eta - \D(\psi a)\x\eta, \\ 
	& = \big((\D\psi)a -\D(a^\circ\psi) \big)\x\eta, \\
	& = \big(a^\circ(\D\psi) -\D(a^\circ\psi) \big)\x\eta, \\ 
	& = -\big[\D,a^\circ\big]\psi\x\eta,
\end{split}
\end{equation}
is not necessarily vanishing. However, \eqref{eq:lDAct} suggests for $\D_l$ to be a 
well-defined operator on $\HH_l$ that it must rather act as follows:
\begin{equation}
\label{eq:lDAct1}
	\D_l(\psi\x a\eta - \psi a\x\eta) =
	\D\psi\x a\eta - \D(\psi a)\x\eta + [\D,a^\circ]\psi\x\eta.
\end{equation}
Recalling Def.~\ref{def:conn} of a connection $\nabla^\circ$ on a left $\A$-module 
$\E_l$ taking values in the 
$\A$-bimodule $\Omega_\D^1(\A^\circ)$ generated by the derivation $\delta^\circ(a) 
= [\D,a^\circ]$, one has \eqref{eq:LeibR}
\begin{equation}
	[\D,a^\circ]\psi\x\eta = \psi\nabla^\circ(a\eta) -\psi a\nabla^\circ(\eta).
\end{equation}
In that light, \eqref{eq:lDAct1} becomes
\begin{equation}
\begin{split}
	& \D_l(\psi\x a\eta) -\D_l(\psi a\x\eta) \\ 
	& \quad = \D\psi\x a\eta -\D(\psi a)\x\eta 
		+\psi\nabla^\circ(a\eta) -\psi a\nabla^\circ(\eta) \\
	& \quad = \D\psi\x a\eta +\psi\nabla^\circ(a\eta) 
		-\big(\D(\psi a)\x\eta +\psi a\nabla^\circ(\eta) \big),
\end{split}
\end{equation}
implying that the correct action of $\D_l$ on $\HH_l$ must be
\begin{equation}
\label{eq:_lD}
	\D_l(\psi\x\eta) := \D\psi\x\eta +\psi\nabla^\circ(\eta),
\end{equation} 
which is $\A$-linear by \eqref{eq:LeibR}. 

\medskip

Further, if $\E_l$ is finite projective \eqref{eq:E=pAN}, any 
$\Omega_\D^1(\A^\circ)$-valued connection $\nabla^\circ$ is of the form 
$\nabla_0^\circ + \bm\omega^\circ$, where $\nabla_0^\circ = \delta^\circ\circ\wp$ 
is the Gra{\ss}mann connection, i.e.
\begin{equation}
\begin{split}
	\forall \eta & = \big(\eta_1,\ldots,\eta_n\big)\wp \in \E_l, 
	\qquad \text{with} \quad \eta_{j=1,\ldots,n} \in \A, \\[3pt] 
	\nabla_0^\circ\eta & = 
	\big(\delta^\circ(\eta_1), \ldots, \delta^\circ(\eta_n)\big)\wp, 
\end{split}
\end{equation}
and $\bm\omega^\circ$ is an $\A$-linear map 
$\E_l\to\Omega^1_\D(\A^\circ)\x_\A\E_l$ such that
\begin{equation}
	\bm\omega^\circ(a\eta) = a\cdot\bm\omega^\circ(\eta), \qquad 
	\forall a \in \A, \; \forall \eta \in \E_l.
\end{equation}

\pagebreak

\subsubsection{Morita self-equivalence via $\E_l = \A$}

The algebra $\A$ is Morita equivalent to itself via $\E_l = \A$ and 
any connection $\nabla^\circ$ on this left $\A$-module $\A$ is of the form 
$\delta^\circ + \omega^\circ$, for some $\omega^\circ \in 
\Omega^1_\D(\A^\circ)$. Then, \eqref{eq:_lD} becomes
\begin{equation}
\begin{split}
	\D_l(\psi\x a) 
	& = \D\psi\x a + \big(\delta^\circ(a) + \omega^\circ(a)\big)\psi\x1, \\
	& = \big((\D\psi)a + [\D,a^\circ]\psi + a\cdot\omega^\circ \psi \big)\x1, \\
	& = \big(a^\circ\D\psi +(\D a^\circ -a^\circ\D)\psi + 
		\omega^\circ a^\circ\psi\big)\x1, \\
	& = \big(\D a^\circ\psi +\omega^\circ a^\circ\psi\big)\x1 
	  = (\D +\omega^\circ)(\psi\x a).
\end{split}
\end{equation}
Thus, identifying $\psi\x a \in {\HH_l}$ with $\psi a\in\HH$, one obtains the 
following action on $\HH$:
\begin{equation}
	\D_l = \D+\omega^\circ.
\end{equation}
Following from \eqref{eq:RealStruct} and \eqref{eq:OppAlg}, any $\omega^\circ = 
\sum_ja^\circ_j[\D,b_j^\circ] \in \Omega_\D^1(\A^\circ)$ has a left action on $\HH$
given by the bounded operator 
\begin{equation}
	\omega^\circ = \epsilon'J\omega J^{-1}
\end{equation}
for $\omega = \sum_ja_j^*[\D,b_j^*] \in \Omega_\D^1(\A)$. Therefore, 
$(\A,\HH,\D_l)$ is a spectral triple, given $\omega$ is self-adjoint \cite{BMS},
said to be Morita equivalent to $(\A,\HH,\D)$ \cite[\S3.2]{LM2}.

\smallskip

Yet again, the real structure $J$ of the initial spectral triple does not adapt to
$(\A,\HH,\D_l)$ for $J\D_l = \epsilon'\D_lJ$ holds iff $\omega = 
\epsilon'J\omega J^{-1}$, which is not necessarily true.

\subsection{Inner fluctuations by Morita self-equivalence}
\label{subsec:innerfluc}

In order to overcome the incompatibility (mentioned at the end of 
\S\ref{subsec:MoritaL} and \S\ref{subsec:MoritaR}) of the real structure with 
the construction of Morita equivalent spectral triples or, in other words, to 
construct Morita self-equivalent real spectral triples -- one combines the above 
two constructions of \S\ref{subsec:MoritaL} and \S\ref{subsec:MoritaR} together, 
cf.~\cite[\S A.1.3]{LM2}.

\smallskip

One begins with a real spectral triple $(\A,\HH,\D;J)$. Implementing Morita 
self-equivalence of $\A$ via a right $\A$-module $\E_r = \A$, one obtains 
(\S\ref{subsec:MoritaL})
\begin{equation}
	\big(\A,\,\HH,\,\D\big) \enskip 
	\xrightarrow[\text{via } \E_r = \A]{\text{ Morita self-equivalance }} \enskip 
	\big(\A,\,\HH,\,\D + \omega_r\big).
\end{equation}
Then, applying Morita self-equivalence of $\A$ via a left $\A$-module $\E_l = \A$
again gives (\S\ref{subsec:MoritaR})
\begin{equation}
	\big(\A,\,\HH,\,\D+\omega_r\big) \enskip 
	\xrightarrow[\text{via } \E_l = \A]{\text{ Morita self-equivalance }} 
	\enskip \big(\A,\,\HH,\,\D' := \D + \omega_r + \omega_l^\circ\big),
\end{equation}
where $\omega_l^\circ = \epsilon'J\omega_lJ^{-1}$. Both $\omega_r$ and $\omega_l$
are self-adjoint one-forms in $\Omega_\D^1(\A)$, but not necessarily related.

\pagebreak

The real structure $J$ of $(\A,\HH,\D)$ also becomes the real structure for
$(\A,\HH,\D')$ iff there exists $\omega \in \Omega_\D^1(\A)$ such 
that \cite[Prop.~A.5]{LM2}
\begin{equation}
\label{eq:InnerFluc}
	\D' = \D_\omega := \D + \omega + \epsilon'J\omega J^{-1}.
\end{equation}

Thus, Morita self-equivalence of real spectral triples induces {\it inner 
fluctuations} \eqref{eq:InnerFluc} of the Dirac operator (metric). The operator 
$\D_\omega$ is then referred to as the {\it gauged} or {\it covariant} or {\it 
fluctuated} Dirac operator and the self-adjoint one-forms $\omega \in 
\Omega_\D^1(\A)$ are identified as {\it generalized gauge fields}.

\section{Gauge transformations}
\label{sec:GT}

A \emph{gauge transformation} on a module is a change of
connection \eqref{eq:nablaU} on that module, induced by an adjoint action of a 
unitary endomorphism on it. 

\smallskip

When the module is taken to be the algebra of a spectral triple itself 
(implementing Morita self-equivalence), a gauge transformation is tantamount to 
transforming the fluctuated Dirac operator under an adjoint action of a unitary of
the algebra. This can equivalently be encoded in a law of transformation of the
generalized gauge fields and subsequently gives the transformation rules for the 
gauge potentials in physical theories.

\subsubsection{On hermitian modules}

Gauge transformations on a hermitian $\A$-module $\E$ are induced by the group of
its unitary endomorphisms \cite[\S A.2.1]{LM2}
\begin{equation}
\label{eq:UofE}
	\U(\E) := \big\{ u\in\End_\A(\E) \; \vert \; uu^* = u^*u = \id_\E \big\},
\end{equation}
which acts on an $\Omega$-valued connection $\nabla$ on $\E$ as
\begin{equation}
\label{eq:nablaU}
	\nabla \to \nabla^u := u\nabla u^*, \qquad \forall u \in \U(\E),
\end{equation} 
where the action on  $\E\x_\A\Omega$ or $\Omega\x_\A\E$, for a right $\A$-module
$\E_\A$ or a left $\A$-module $_\A\E$, respectively, is implemented as
\begin{equation}
	u\x_\A\id_\Omega \qquad \text{or} \qquad \id_\Omega\x_\A u.
\end{equation}
It follows that $\nabla^u$ in \eqref{eq:nablaU} is also an $\Omega$-valued 
connection on $\E$ \cite[Prop.~A.7]{LM2}. 

\bigskip

If $\E$ is finite projective \eqref{eq:E=pAN}, then for any connection
\begin{equation}
\label{eq:wtowU}
	\nabla = \nabla_0 + \omega \; \longrightarrow \; 
	\nabla^u = \nabla_0 + \omega^u 
	\qquad \Rightarrow \qquad \omega \to \omega^u,
\end{equation}
that is, the transformation law for a gauge potential $\omega$ solely 
captures a gauge transformation. Explicitly, the Gra{\ss}mann connection 
$\nabla_0 = \wp \circ \delta$, for a derivation $\delta$ of $\A$ in $\Omega$. 
Further, the group $\U(\E)$ \eqref{eq:UofE} consists of unitary matrices commuting 
with $\wp$, that is,
\begin{equation}
	\U(\E) := \big\{ u\in M_n(\A)\;\vert\;u\wp=\wp u,\,uu^*=u^*u=\id_\E \big\},
\end{equation}
acting by the usual matrix multiplication,
\begin{equation}
	u\zeta :=
\begin{cases}
	\wp u \zeta, 	&\qquad \forall \zeta \in \E_\A \\
	\zeta u^* \wp, 	&\qquad \forall \zeta \in {_\A}\E		
\end{cases},
\end{equation}
which implements the gauge transformation \eqref{eq:wtowU} as follows 
\cite[Prop.~A.8]{LM2}:
\begin{equation}
\label{eq:omegaU'}
	\omega^u \zeta :=
\begin{cases}
	\wp u\delta(u^*)\zeta + u\omega u^*\zeta,	&\qquad \forall \zeta \in \E_\A \\
	\zeta\delta(u)u^*\wp + u\omega u^*\zeta, 	&\qquad \forall \zeta \in {_\A}\E		
\end{cases},
\end{equation}
where $\delta(u), \delta(u^*) \in M_n(\Omega)$ for unitaries $u, u^* \in \A$.

\subsubsection{On real spectral triples}

Consider a real spectral triple $(\A,\HH,\D;J)$ with a right $\A$-module $\E_\A$ 
taken to be the algebra $\A$ itself (implementing Morita self-equivalence) and a 
derivation $\delta(\cdot)=[\D,\cdot]$ of $\A$ in $\Omega^1_\D(\A)$. Then, the 
first case of \eqref{eq:omegaU'} gives
\begin{equation}
\label{eq:omegaU}
	\omega^u = u[\D, u^*] + u\omega u^*,
\end{equation}
which induces \cite[Rem.~A.9]{LM2}
\begin{equation}
\label{eq:DomegaU}
	\D_\omega \mapsto \D_{\omega^u} = \D + \omega^u + \epsilon'J\omega^u J^{-1}.
\end{equation}
The same effect as above, i.e.\ the gauge transformation $\D_\omega \mapsto 
\D_{\omega^u}$ \eqref{eq:DomegaU}, can be achieved by the adjoint action of the 
group $\U(\A)$ of unitaries of $\A$
\begin{equation}
\label{def:U(A)}
	\U(\A) := \big\{ u\in\A \;|\; u^*u = uu^* = 1 \big\},
\end{equation}
on the Hilbert space $\HH \ni \psi$, defined as
\begin{equation}
\label{eq:gt1}
	\big(\Ad~u\big)\psi := u\psi u^* = uJuJ^{-1}\psi,
\end{equation}
recalling that $\A$ has both a left and a right representation on $\HH$ thanks to
the real structure $J$. On the 
Dirac operator $\D$, \eqref{eq:gt1} induces, following Axioms \ref{axm2'} and 
\ref{axm7'}, the transformation \cite[Prop.~1.141]{CMa}
\begin{equation}
	\D \mapsto (\Ad~u)\D(\Ad~u)^{-1} = \D + u[\D,u^*] + \epsilon'Ju[\D,u^*]J^{-1},
\end{equation}
which is basically the relation \eqref{eq:DomegaU} for $\omega = 0$ so that
$\omega^u = u[\D,u^*]$ as per \eqref{eq:omegaU}. On the gauged Dirac operator 
$\D_\omega$ \eqref{eq:InnerFluc}, one has \cite[Prop.~1.141]{CMa}
\begin{equation}
\label{eq:DomegaU1}
	\D_\omega \mapsto (\Ad~u)\D_\omega(\Ad~u)^{-1} = \D_{\omega^u},
\end{equation}
with $\D_{\omega^u}$ as in \eqref{eq:DomegaU} and $\omega^u$ as \eqref{eq:omegaU}.

\subsection{Gauge-invariants}
\label{subsec:gauge_inv}

Now that we have discussed, in the previous section, the generalized gauge fields 
carrying the action of the group of unitaries of the algebra; we can define their
gauge-invariants functionals on a spectral triple, viz.\ the (bosonic) spectral 
action and the fermionic action.

\subsubsection{Spectral action}

A general formalism for spectral triples is the \emph{spectral action principle} 
\cite{CC96,CC97,CC06a,CC06b}, which proposes a universal action functional on 
spectral triples that depends only on the spectrum of the Dirac operator $\D$ or 
-- more generally, if the inner fluctuations are turned on -- that of its 
fluctuation $\D_\omega$.

\smallskip

A straight forward way to construct such an action is to count the eigenvalues that 
are smaller than a fixed energy scale $\Lambda$. Thus, one defines the 
\emph{spectral action} as the fucntional
\begin{mdframed}
\begin{equation}
\label{eq:SB}
	S^b[\D_\omega] := \trc\;f\left(\frac{\D_\omega^2}{\Lambda^2}\right),
\end{equation}
\end{mdframed}
where $f$ is a positive and even real cutoff function taken to be the smooth approximation 
of the characteristic function on the interval $[0,1]$ such that the action 
$S^b[\D_\omega]$ vanishes sufficiently rapidly as the real cutoff parameter $\Lambda$ approaches
infinity.

\smallskip

It can be expanded asymptotically (in power series of $\Lambda$) -- using heat 
kernel expansion techniques given that $\D_\omega^2$ is a generalized Laplacian up
to an endomorphism term (generalized Lichnerowicz formula) -- as follows
\begin{equation}
\label{eq:SBexp}
	\trc\;f\left(\frac{\D_\omega^2}{\Lambda^2}\right) = \sum_{n\ge0} f_{4-n} 
	\Lambda^{4-n} a_n \left(\frac{\D_\omega^2}{\Lambda^2}\right),
\end{equation}
where $f_n$ are the momenta of $f$ (given by $f_k:=\int_0^\infty f(v)v^{k-1}dv$,
for $k>0$ and $f_0 = f(0)$), and $a_n$ are the Seeley-de Witt coefficients 
(non-zero only for even $n$) \cite{Gi,Va}, which yield the gauge theoretic 
lagrangians of the model. 

\smallskip

$S^b[\D]$ is the fundamental action functional that can be used both at the 
classical level to compare different geometric spaces and at the quantum level in
the functional integral formulation (after Wick rotation to euclidean signature)
\cite[\S11]{CMa}.

\smallskip

However, when applied to the inner fluctuations $\D_\omega$, the action 
$S^b[\D_\omega]$ only yields the bosonic content on the theory (hence, the 
superscript $b$). For instance, on 
classical riemannian manifolds where inner fluctuations vanish, $S^b[\eth]$ gives
the Einstein-Hilbert action of pure gravity. Thus, noncommutative geometries 
naturally contain gravity, while the other gauge bosons appear due to the
noncommutativity of the algebra of the spectral triples.  
The coupling with the fermions is accounted for by adding to the 
spectral action an extra term called the fermionic action.

\subsubsection{Fermionic action}

The {\it fermionic action} associated to a real graded 
spectral triple $(\A,\HH,\D;J,\g)$, defined as 
\begin{mdframed}
\begin{equation}
\label{Sf}
	S^f[\D_{\omega}] := \mathfrak{A}_{\D_\omega}(\tilde\psi, \tilde\psi),
	\smallskip
\end{equation}
\end{mdframed}
is a gauge-invariant quantity of Gra{\ss}mann nature \cite{CCM,Ba}, constructed 
from the following bilinear form
\begin{equation}
\label{eq:4}
	\mathfrak{A}_{\D_\omega}(\phi,\psi)
	:= \big\langle J\phi, \D_{\omega}\psi \big\rangle, \qquad \forall \phi,\psi \in \HH,
\end{equation}
defined by the covariant Dirac operator $\D_{\omega} := \D +\omega 
+\epsilon'J\omega J^{-1}$, where $\omega$ is a self-adjoint element of the set of 
generalized one-forms \cite{C96}
\begin{equation}
\label{eq:3}
	\Omega^1_\D({\cal A}) := \left\{ 
	\sum\nolimits_i a_i\big[\D, b_i\big], \quad a_i, b_i \in \cal A \right\}\!.
\end{equation}
Here, $\tilde{\psi}$ is a Gra{\ss}mann vector in the Fock space $\widetilde\HH_+$ 
of classical fermions, corresponding to the positive eigenspace 
${\cal H}_+ \subset {\cal H}$ of the grading operator $\g$, i.e.\
\begin{equation}
\label{eq:H_+}
	\widetilde{\cal H}_+ := \big\{ \tilde{\psi}, \,\, \psi \in {\cal H}_+ \big\}, 
	\quad \text{ where} \quad{\cal H}_+ := \big\{ \psi\in \HH, \,\, \g\psi = \psi \big\}.
\end{equation}

\smallskip

Both actions $S^b$ and $S^f$ are invariant under a gauge transformation, i.e.\ 
the simultaneous adjoint action of the unitary group $\U(\A)$ \eqref{def:U(A)}
both on $\HH$ as \eqref{eq:gt1} and on the fluctuated Dirac operator 
$\D_\omega$ as \eqref{eq:DomegaU1}.

\smallskip

\begin{rem}
\label{rem:bilingrassman}
Since the bilinear form \eqref{eq:4} is anti-symmetric for $KO$-dim.\ $2$ and $4$ 
(cf.\ Lem.~\ref{lemma:antisymm} below), $\frak A_{\D_\omega}(\psi,\psi)$ vanishes 
when evaluated on vectors, but it is non-zero when evaluated on Gra{\ss}mann 
vectors, see \cite[\S~I.16.2]{CMa}.

\smallskip

In particular, the fermionic action associated to the spectral triple of the
Standard Model (which  has $KO$-dim.\ $2$) contains the coupling of the fermionic 
matter with the fields (scalar, gauge, and gravitational).
\end{rem}

\pagebreak

\section{Almost-commutative geometries}
\label{sec:ACG}

Almost-commutative geometries are a special class of noncommutative geometries
arising by taking the product\footnote{The product of two graded real spectral 
triples is defined in the sense of the direct product of two manifolds, 
cf.~\cite[\S4]{Su}.}of the canonical triple \eqref{eq:CT} of an oriented closed 
spin manifold $\M$ with a \emph{finite geometry} $\F$ defined by a 
finite-dimensional unital spectral triple $(\A_\F,\HH_\F,\D_\F)$. The resulting 
\emph{product geometry}, denoted by $\M\times\F$, is then given by the spectral 
triple \cite{C96}:
\begin{mdframed}
\begin{equation}
	\big( \A := C^\infty(\M) \x \A_\F, \quad
	\HH := L^2(\M,{\cal S}) \x \HH_\F, \quad
	\D := \eth \x \Id_\F + \g_\M \x \D_\F \big), \label{eq:MxF}
\smallskip
\end{equation}
\end{mdframed}
where $\Id_\F$ is the identity in $\HH_\F$, $\g_\M$ is the grading on $\M$, 
and the representation $\pi_0$ of $\A$ on $\HH$ is the tensor product
\begin{equation}
	\pi_0 := \pi_\M \x \pi_\F
\end{equation}
of the multiplicative representation $\pi_\M$ \eqref{eq:pim} of $C^\infty(\M)$ on 
spinors with the representation $\pi_\F$ of $\A_\F$ on $\HH_\F$. If $\F$ is graded
and real with grading $\g_\F$ and real structure $J_\F$, then $(\A,\HH,\D)$ is
also graded and real, respectively, with
\begin{equation}
	\g = \g_\M \x \g_\F, \qquad J = \J \x J_\F.
\end{equation}

Here, the finite-dimensional Hilbert space $\HH_\F$ accounts for the fermionic 
content of the theory and its dimension reflects the number of elementary fermions 
in the model. An orthonormal basis can be chosen for $\HH_\F$ where the basis 
vectors represent these fermions. The finite Dirac operator $\D_\F$ is a square 
matrix acting on $\HH_\F$ and its entries encode the fermionic masses.
$\g_\F$ and $J_\F$ being the parity and the charge conjugation operator of the 
finite space $\F$, respectively, switch between right/left-handed particles and 
particles/antiparticles.

\smallskip

In the following subsections, we give some examples of almost-commutative 
geometries and the corresponding gauge theories they describe. Primarily, we 
briefly recall the spectral triples of a $U(1)$ gauge theory and that of 
electrodynamics, which we will make explicit use of later. We also mention the
spectral triples of Yang-Mills theory and the Standard Model of particle physics.

\subsection{$U(1)$ gauge theory}
\label{subsec:2.4.1}

One of the simplest finite noncommutative spaces is that consisting of two points
only -- the two-point space. The graded and real finite spectral triple $\F_2$ 
associated to a two-point space, given by the data:
\begin{equation}
\label{eq:35}
	\A_\F = \mathbb{C}^2,\; \HH_\F = \mathbb{C}^2,\; 
	\D_\F = 0\,; \quad \g_\F = \begin{pmatrix} 1 & 0 \\ 0 & -1 \end{pmatrix}\!,\;
	J_\F = \begin{pmatrix} 0 & cc \\ cc & 0 \end{pmatrix},
\end{equation}
when considered for the almost-commutative geometry $\M\times\F_2$, defined by
\eqref{eq:MxF}, describes a $U(1)$ gauge theory \cite[\S3]{DS}. Here, the grading 
$\g_\F$ and the real structure $J_\F$ of \eqref{eq:35} are in the orthonormal basis
$\left\{ e, \overline e \right\}$ of $\HH_\F = \mathbb{C}^2$ with $e$ being the 
basis element of $\HH_\F^+$ (representing an electron) and $\overline e$ of 
$\HH_\F^-$ (representing an anti-electron, i.e.\ a positron), where $\HH_\F^\pm$
denote the $\pm1$-eigenspace of the grading operator $\g_\F$. So, we have
\begin{equation}
\begin{split}
	\g_\F e = e, \qquad & \g_\F \overline e = \overline e, \\[3pt]
	J_\F e = \overline e, \qquad & J_\F \overline e = e.
\end{split}
\end{equation}

For the product geometry $\M\times\F_2$, the algebra $\A  = C^\infty(\M)
\x \mathbb{C}^2 \ni a := (f,g)$ acts on the Hilbert space $\HH = L^2(\M,{\cal S}) 
\x \mathbb{C}^2$ via the representation $\pi_0 : \A \to \B(\HH)$, defined by
\begin{equation}
\label{eq:p0_X}
	\pi_0(a) := \begin{pmatrix}
		\pi_\M(f) & 0 \\ 0 & \pi_\M(g)
	\end{pmatrix}\!, \qquad \forall f, g \in C^\infty(\M),
\end{equation}
with $\pi_\M$ as in \eqref{eq:pim}. The $KO$-dim.\ of \eqref{eq:35} is $6$, then
for a $4$-dim.\ manifold $\M$, the almost-commutative geometry $\M\times\F_2$ has
$KO$-dim.\ $6+4\text{ mod }8 = 2$.

\smallskip

The inner fluctuations of $\M\times\F_2$ are parametrized by a $U(1)$ gauge field 
$Y_\mu \in C^\infty(\M,\mathbb{R}) \simeq C^\infty\big(\M,i\,{\frak u}(1)\big)$ as
\cite[Prop.~3.3]{DS}
\begin{equation}
	\D \mapsto \D_\omega := \D + \gm Y_\mu\x\g_\F, 
\end{equation}
where $\D = \eth\x\Id_\F$ (here, setting $\D_\F = 0$ is the only choice 
for $\F_2$ to have a real structure \cite[Prop.~3.1]{DS}). Thus, this gauge field 
$Y_\mu$ implements the action of a unitary $u := e^{i\theta} \in 
C^\infty\big(\M,U(1)\big)$ of $\A$ on $\D_\omega$, by conjugation:
\begin{equation}
\label{U1act}
	Y_\mu \mapsto Y_\mu - iu\dm u^* = Y_\mu -\dm\theta, \qquad \text{for }
	\theta \in C^\infty(\M,\mathbb{R}).
\end{equation}

\smallskip

\begin{rem}
\label{rem:MxF2}
Although the almost-commutative geometry $\M\times\F_2$ successfully describes a 
$U(1)$ gauge theory, it falls short of a complete description of classical 
electrodynamics, as discussed at the end of \cite[\S3]{DS}. This is due to the 
following two reasons:
\begin{enumerate}
	\item Since the finite Dirac operator $\cal D_F$ is zero, the electrons cannot 
	be massive.
	\item The finite-dimensional Hilbert space $\cal H_F$ does not possess enough
	room to capture the required spinor degrees of freedom. More precisely, the 
	fermionic action (\ref{Sf}) of $\M\times\F_2$ describes
 	only one arbitrary Dirac spinor, whereas two of those, independent of each 
 	other, are needed to describe a free Dirac field, cf.~\cite[pg.~311]{Col}.
\end{enumerate}
However, none of the above arises as an issue if one only wishes to obtain the Weyl 
action, since the Weyl fermions are massless anyway, and they only need half of 
the spinor degrees of freedom as compared to the Dirac fermions.
\end{rem}

\pagebreak

\subsection{Electrodynamics}
\label{subsec:2.4.2}

Electrodynamics is one of the simplest field theories in physics. A slight
modification of the example of $\M\times\F_2$ given in \S\ref{subsec:2.4.1} can 
overcome the two obstructions of Rem.~\ref{rem:MxF2} and provide a unified (at the 
classical level) description of gravity and electromagnetism. 

\smallskip

Such a modification entails doubling the finite-dimensional Hilbert space $\HH_\F$ from 
$\mathbb{C}^2$ to $\mathbb{C}^4$; which not only allows for a non-zero finite Dirac 
operator $\D_\F$, but also gives the correct spinor degrees of freedom in the 
fermionic action. We refer to \cite{DS,Su} for details.

\smallskip

The spectral triple of electrodynamics is given by the product geometry $\M \times
\F_{ED}$ of a $4$-dim.\ compact riemannian spin manifold $\M$ (with grading $\g_\M 
= \g^5$ and real structure $\J$) and the graded real finite spectral triple 
$\F_{ED}$ defined by the following data \cite[\S4.1]{DS}:
\begin{equation}
\label{ED}
	\F_{ED} := \big( \A_\F = \mathbb{C}^2, \; \HH_\F = \mathbb{C}^4, \;
	\D_\F\,; \; \g_\F,\; J_\F \big),
\end{equation}
where, in the orthonormal basis $\big\{e_L,e_R,\overline{e_L},\overline{e_R}\big\}$
of $\HH_\F  = \mathbb{C}^4$ (denoting both the left and right handed electrons and 
positrons) and for a constant parameter $d \in \mathbb{C}$, we 
have
\begin{equation}
\label{eq:stEDF}
	\D_\F = \begin{pmatrix}
0 & d & 0 & 0 \\ \overline d & 0 & 0 & 0 \\ 0 & 0 & 0 & \overline d \\ 0 & 0 & d & 0
	\end{pmatrix}\!, \quad
	\g_\F = \begin{pmatrix}
1 & 0 & 0 & 0 \\ 0 & -1 & 0 & 0 \\ 0 & 0 & -1 & 0 \\ 0 & 0 & 0 & 1
	\end{pmatrix}\!, \quad
	J_\F = \begin{pmatrix}
0 & 0 & cc & 0 \\ 0 & 0 & 0 & cc \\ cc & 0 & 0 & 0 \\ 0 & cc & 0 & 0
	\end{pmatrix}\!,
\end{equation}
so that
\begin{align*}
	\g_\F e_L = e_L, \quad && \g_\F e_R = -e_R, \quad &&
	\g_\F \overline{e_L} = -\overline{e_L}, \quad &&
	\g_\F \overline{e_R} = \overline{e_R}, \\[3pt]
	J_\F e_L = \overline{e_L}, \quad && J_\F e_R = \overline{e_R}, \quad &&
	J_\F \overline{e_L} = e_L, \quad && J_\F \overline{e_R} = e_R.
\end{align*}
The algebra $\A = C^\infty(\M)\x\mathbb{C}^2 \ni a := (f,g)$ of $\M\times\F_{ED}$
acts on its Hilbert space $\HH = L^2({\cal M,S})\x\mathbb{C}^4$ via the representation
\begin{equation}
\label{p0}
	\pi_0(a) := \begin{pmatrix}
		f\mathbb{I}_4 & 0 & 0 & 0 \\ 0 & f\mathbb{I}_4 & 0 & 0 \\ 
		0 & 0 & g\mathbb{I}_4 & 0 \\ 0 & 0 & 0 & g\mathbb{I}_4
	\end{pmatrix}, \qquad \forall f, g \in C^\infty({\cal M}).
\end{equation}
The $KO$-dim.\ of $\F_{ED}$ \eqref{ED} is same as that of $\F_2$ \eqref{eq:35}, 
i.e.\ $6$. Therefore, the $KO$-dim.\ of $\M\times\F_{ED}$ is $2~(\text{mod}~8)$. 
The inner fluctuations of $\D = \eth\x\Id_4 + \g^5\x\D_\F$
\begin{equation}
\begin{split}
\label{Ymu}
	\D \to \D_\omega & = \D + \gm \x B_\mu, \\[3pt] \text{where } \qquad
	B_\mu & := {\sf diag}\big(Y_\mu,Y_\mu,-Y_\mu,-Y_\mu\big);
\end{split}
\end{equation}
are parametrized by a single $U(1)$ gauge field $Y_\mu$ carrying an adjoint action 
of the group $C^\infty\big(\M,U(1)\big)$ of unitaries of $\A$ on $\D_\omega$ 
\eqref{Ymu}, implemented by \eqref{U1act} \cite[\S4.2]{DS}. 

The full action (spectral plus fermionic) of the almost-commutative geometry $\M 
\times \F_{ED}$ yields the lagrangian for electrodynamics (on a curved background 
manifold $\M$) -- identifying $Y_\mu$ \eqref{Ymu} as the $U(1)$ gauge potential of 
electrodynamics -- along with a purely gravitational
lagrangian \cite[\S4.3]{DS}.

\subsection{Other physical models}
\label{subsec:2.4.3}

The great potential and flexibility of noncommutative geometry -- in particular, 
its applicability to theoretical high energy physics -- can be realized by looking
at the plethora of physically relevant models that can be described within its 
framework.

\subsubsection{Yang-Mills theory}

Electrodynamics, as discussed in the previous section, is an abelian $U(1)$ gauge
theory, which can be further generalized to the non-abelian cases. For instance, a 
non-abelian $SU(n)$ gauge theory -- also known as Yang-Mills theory among 
physicists, is described within the almost-commutative geometry 
$\M\times\F_{YM}$ defined by the data:
\begin{equation}
\label{eq:YangMills}
	\big( C^\infty(\M)\x M_n(\mathbb{C}),\;
		L^2(\M,{\cal S})\x M_n(\mathbb{C}),\;
		\eth\x\Id_n; \quad \J\x(\cdot)^*,\; \g_\M\x\Id_n \big).
\end{equation}
The spectral triple \eqref{eq:YangMills} describes the 
Einstein-Yang-Mills theory \cite{CC97}, which not only adapts to ($N=2$ and $N=4$)
supersymmetry \cite{Ch94}, but can also be extended to accommodate 
topologically non-trivial gauge configurations \cite{BS}.

\subsubsection{Standard Model}

Another very important non-abelian gauge theory is the Standard Model of particle 
physics, with structure (gauge) group $U(1)\times SU(2)\times SU(3)$, whose full 
lagrangian can be derived, together with the Higgs potential and the 
Einstein-Hilbert action of gravity with a minimal coupling; from the following 
spectral triple \cite{CCM}:
\begin{equation}
\label{eq:SM}
	\big( \A = C^\infty(\M) \x \A_{SM}, \quad
	\HH = L^2(\M,{\cal S}) \x \HH_\F, \quad 
	\D = \eth \x \Id_\F + \g_\M \x \D_\F \big),
\end{equation}
where the Standard Model algebra \cite{CC08}
\begin{equation}
	\A_{SM} := \mathbb{C}\+\mathbb{H}\+ M_3(\mathbb{C})
\end{equation}
acts on the space 
\begin{equation}
\label{eq:H_F}
	\HH_\F
	= \HH_L \+ \HH_R \+ \HH_{\bar L} \+ \HH_{\bar R} 
	= \mathbb{C}^{96}
\end{equation}
of elementary fermions: $8$ fermions (up and down quarks with $3$ colors each plus 
electron and neutrino) for $3$ generations and $2$ chiralities left/right $(L,R)$ 
plus their antiparticles (denoted by bar). So, each of the four subspaces in
\eqref{eq:H_F} is of dim.\ $24$ and, thus, isomorphic to $\mathbb{C}^{24}$. 
  
$\D_\F$ is a $96\times96$ matrix acting on $\HH_\F = \mathbb{C}^{96}$ whose 
entries correspond to the 31 real parameters\footnote{Yukawa couplings of the 
fermions, the Dirac and Majorana masses of the neutrinos, the quark mixing angles 
of the CKM matrix, and the neutrino mixing angles of the PMNS matrix.} 
of the Standard Model \cite[\S13.5]{CMa}. The grading is $\g = \g^5 \x \g_\F$ 
and the real structure is $J = \J \x J_\F$, where
\begin{equation}
\begin{split}
	\g_\F &:= {\sf diag}\big(\Id_{24},\,-\Id_{24},\,-\Id_{24},\,\Id_{24}\big), \\[3pt]
	J_\F &:= \begin{pmatrix} 0 & \Id_{48} \\ \Id_{48} & 0 \end{pmatrix} cc.
\end{split}
\end{equation}

The $KO$-dim.\ of $\M\times\F_{SM}$ \eqref{eq:SM} is $4+6~\text{mod}~8 = 2$.
The representation $\pi_0$ of $\A$ on $\HH$ for one generation\footnote{In 
\eqref{eq:H_F}, $\HH_\F = \mathbb{C}^{32}$ and all four of its subspaces are each 
isomorphic to $\mathbb{C}^8$. Then, the representation for all the three 
generations is just a direct sum of similar representations \eqref{eq:pi_SM} for 
each generation.} is written as, cf.~\cite{CCM}:
\begin{equation}
\label{eq:pi_SM}
	\pi_0 (f\x a) = \pi_\M(f) \x \pi_{SM}(a), \qquad 
	\forall f\in C^\infty(\M), \; a \in \A_{SM},
\end{equation}
where $\pi_\M$ is as in \eqref{eq:pim} and $\pi_{SM}$ for $a := (\lambda, q, m) 
\in \A_{SM}$ is given by
\begin{equation}
\label{eq:pi_SM(a)}
	\pi_{SM}(a) := \pi_L(q) \+ \pi_R(\lambda) \+ 
	\pi_{\bar L}(\lambda,m) \+ \pi_{\bar R}(\lambda,m),
\end{equation}
with $\lambda\in\mathbb{C}$ acting on $\HH_R\+\HH_{\bar L}\+\HH_{\bar R}$, 
quaternions $q\in\mathbb{H}$ on $\HH_L$, and matrices $m\in M_3(\mathbb{C})$ on
$\HH_{\bar L}\+\HH_{\bar R}$. The individual representations in 
\eqref{eq:pi_SM(a)}, identifying $\mathbb{H}$ with its usual representation as
$M_2(\mathbb{C})$, explicitly are:
\begin{equation}
\begin{split}
	\pi_L(q) & := q \x \Id_4, \\[3pt] 
	\pi_R(\lambda) & := {\sf diag}(\lambda,\bar\lambda) \x \Id_4, \\[3pt]
	\pi_{\bar L}(\lambda,m) = \pi_{\bar R}(\lambda,m) 
	& := \Id_2 \x {\sf diag}(\lambda,m),
\end{split}
\end{equation}
where $\Id_4$ in the first two eqs.\ indicates that $\mathbb{C}$ and $\mathbb{H}$
preserve color and do not mix the leptons (electrons and neutrinos) with the 
quarks (up and down), and $\Id_2$ in the last eq.\ indicates that $\mathbb{C}$ and
$M_3(\mathbb{C})$ preserve the flavor: $\lambda$ acts on the antileptons whereas 
$m$ mixes the color of the antiquarks. 

\subsubsection{Beyond Standard Model}

It is also possible to construct Grand Unified Theories in the framework of 
noncommutative geometry, such as the $SO(10)$ model \cite{CF1,CF2,W}. Among other
extensions, there is the Pati-Salam model \cite{AMST,CCS2,CCS3}, whose symmetry 
spontaneously breaks down to the Standard Model. 

\smallskip

Further, there are ways to modify and/or relax some of the axioms of 
noncommutative geometry to produce more flexible geometries that are 
capable of serving as platforms to explore the physics beyond that of the Standard
Model, see e.g.\ \cite{CCS1} and \cite{DLM1,DM}.
One such method of twisting will be the subject-matter of this thesis, which
we will be exploring in the next chapters.

\clearpage
\newpage	
	\chapter{Minimally Twisted Spectral Triples}
\label{chap:MTST}

Twisted spectral triples were first introduced -- from a purely 
mathematical motivation -- by Connes and Moscovici \cite{CMo} in the context of 
operator algebras to extend the local index formula for algebras of type III.\footnote{in 
the sense of the type classification of the von Neumann algebras, cf. \cite{C94,CMo}} 
Such algebras characteristically exhibit no nontrivial trace and, hence, are 
incompatible with the requirement of the boundedness of the commutator $[\D,a]$ in 
the Def.~\ref{def:1.1} of a spectral triple $(\A,\HH,\D)$. So, the `twist' 
basically comprises of trading off this requirement for the condition that there 
exists of an automorphism $\rho$ of $\A$ such that the `twisted' commutator, 
defined as
\begin{equation}
	[\D,a]_\rho := \D a-\rho(a)\D,
\end{equation}
is bounded for any $a\in\A$. Twisted commutators are well-defined on the domain of 
$\D$ and extend to bounded operators on $\HH$.

\smallskip

Later, noncommutative geometries twisted in this manner found applications to high 
energy physics in describing extensions of Standard Model, such as the Grand 
Symmetry Model~\cite{DLM1,DM}.

\smallskip

In \S \ref{subsec:twistedsp}, we define twisting real spectral 
triples using algebra automorphisms \cite{LM1} and, in \S \ref{sec:twstgaugetrans},
state the laws of gauge transformations for them \cite{LM2}. We recall how the
twist $\rho$ naturally induces a $\rho$-inner product $\langle \cdot, \cdot 
\rangle_\rho$ on the Hilbert space $\HH$ (\S\ref{subsec:rho}), which allows to 
define a fermionic action suitable for real twisted spectral triples (\S 
\ref{subsec:fermionaction}) \cite{DFLM}. 

\smallskip

The key difference from the Def.~\eqref{Sf} of the fermionic action
in the usual (i.e.\ non-twisted) case is that one no longer restricts to the 
positive eigenspace $\HH_+$ of the grading $\g$, but rather to that of the unitary $\R$ 
implementing the twist $\rho$. 

\smallskip

In \S \ref{sec:2.2}, we highlight the `twist by 
grading' procedure -- which canonically associates a twisted partner to any graded spectral triple whose 
representation is sufficiently faithful -- and the notion of twisting a spectral 
triple minimally \cite{LM1}.

\pagebreak

\section{Twisting by algebra automorphisms}
\label{subsec:twistedsp}

\begin{fdefine}[from~\cite{CMo}]
A {\bf twisted spectral triple} $(\A, \HH, \D)_{\rho}$ consists of a unital 
$*$-algebra ${\cal A}$ acting faithfully on a Hilbert space ${\cal H}$ as 
bounded operators, a self-adjoint operator $\D$ with compact 
resolvent on $\HH$ referred to as the {\bf Dirac operator}, and an automorphism 
$\rho$ of ${\cal A}$ such that the {\bf twisted commutator}, defined as
\begin{equation}
\label{twistcom}
	[{\cal D},a]_{\rho} := {\cal D}a - \rho(a){\cal D}, 
\end{equation}
is bounded for any $a\in\A$. 
\smallskip
\end{fdefine}

As for usual spectral triples, a {\bf graded} or {\bf even} twisted spectral triple
is one endowed with a $\mathbb{Z}_2$-grading $\g$ on ${\cal H}$, that is, a 
self-adjoint operator $\g : {\cal H} \rightarrow {\cal H}$, satisfying 
\eqref{eq:grad}.

\smallskip

The {\bf real structure} (Axiom \ref{ax7}) easily adapts to the twisted case 
\cite{LM1}. One considers an antilinear isometry $J:\HH\to\HH$ 
\eqref{eq:RealStruct} satisfying \eqref{eq:Reality}, where -- as in the non-twisted 
case -- the signs $\epsilon, \epsilon', \epsilon''$ determine the {\bf 
$KO$-dimension} of the twisted spectral triple. Additionally, $J$ is required to 
implement an isomorphism \eqref{eq:OppAlg} between $\A$ and its opposite algebra 
$\A^\circ$ such that Axiom \ref{axm7'} is satisfied. However, in the twisted case, 
Axiom \ref{axm2'} is modified to be compatible with the twist as follows 
\cite{DM,LM1}:
\begin{equation}
\label{eq:order1}
	\big[[{\cal D},a]_{\rho}, b^{\circ}\big]_{\rho^{\circ}} := 
	[{\cal D},a]_{\rho}b^{\circ} - \rho^{\circ}(b^{\circ})[{\cal
          D},a]_{\rho} = 0, \qquad \forall a, b \in\A,
\end{equation}
where $\rho^\circ$ is the automorphism induced by $\rho$ on $\A^\circ$ via
\begin{equation}
\rho^{\circ}(b^{\circ}) = \rho^{\circ}(Jb^*J^{-1})
:= J\rho(b^*)J^{-1}.
\label{eq:6}
\end{equation}

\begin{fdefine}[from~\cite{LM1}]
A {\bf real twisted spectral triple} is a graded twisted spectral triple with a 
real structure $J$ \eqref{eq:RealStruct} satisfying \eqref{eq:Reality}, 
Axiom \ref{axm7'}, and the `twisted' first order condition given by \eqref{eq:order1}.
\smallskip
\end{fdefine}

In case the automorphism $\rho$ coincides with an inner automorphism of 
$\B(\HH)$, that is
\begin{equation}
\label{eq:14}
	\pi(\rho(a))={\cal R} \pi(a)  {\cal R}^\dag, \qquad \forall a\in\A,
\end{equation}
where $\R\in\B(\HH)$ is unitary, then $\rho$ is said to be {\bf 
compatible with the real structure} $J$, as soon as \cite[Def.~3.2]{DFLM}
\begin{equation} 
\label{eq:2.11}
	J{\cal R} = \epsilon'''\,{\cal R}J, \qquad \text{ for } \quad
        \epsilon'''=\pm1.
\end{equation}
The inner automorphism $\rho$ and, hence, the unitary $\cal R$ are not 
necessarily unique. In that case, $\rho$ is compatible with the real structure $\J$
if there exists at least one $\R$ satisfying the conditions mentioned above.

\begin{rem}
\label{rem:autmodul}
In the original definition of the twist \cite[(3.4)]{CMo}, the automorphism $\rho$ is 
not required to be a $*$-automorphism, but rather to satisfy the regularity 
condition $\rho(a^*)=\rho^{-1}(a)^*$. If, however, one requires $\rho$ to be a 
$*$-automorphism, then the regularity condition renders
\begin{equation}
	\rho^2= {\sf Id}.
\label{eq:17}
\end{equation}
\end{rem}

Other modifications of spectral triples by twisting the real structure have 
been proposed in \cite{BCDS} and some interesting relations with the twisted 
spectral triples mentioned above have been worked out in \cite{BDS}.

\subsection{Twisted fluctuation of the metric}
\label{subsec:twist-inner-fluc}

The fluctuations of the Dirac operator in the non-twisted case -- as discussed 
in \S\ref{subsec:innerfluc} -- can be extended for twisted spectral triples. 
Initially done by analogy in \cite{DM}, the twisted fluctuations have successfully 
been put on the same footing in \cite{LM2}, as Connes' original ``fluctuations of 
the metric'' \cite{C96}. This essentially entails transporting a real twisted spectral triple over to a
Morita equivalent algebra. Particularly, for Morita self-equivalence, one has the 
following.

\begin{define}[from \cite{LM1}]
	Given a real twisted spectral triple $(\A,\HH,\D;J)_\rho$, a twisted 
	fluctuation of $\D$ by $\A$ is a self-adjoint operator of the form 
\begin{equation}
\label{eqw:twistfluct}
	\D_{\omega_\rho} := \D + \omega_\rho + \epsilon'J\omega_\rho J^{-1},
\end{equation}
for some twisted one-form
\begin{equation}
\label{eq:8}
	\omega_\rho \in \Omega^1_\D(\A,\rho) := \left\{ 
	\textstyle\sum_ja_j\big[\D,b_j\big]_\rho, \enskip a_j,b_j\in\A \right\}.
\end{equation}
The operator $\D_{\omega_\rho}$ is then referred to as the twisted-covariant Dirac 
operator.
\end{define}

The set $\Omega^1_\D$ \eqref{eq:8} of twisted one-forms is an $\A$-bimodule with a 
twisted action on the left:
\begin{equation}
	a\cdot\omega_\rho\cdot b = \rho(a)\omega_\rho b, \qquad 
	\forall a,b\in\A,
\end{equation}
so that the twisted commutator $[\D, \cdot]_\rho =: \delta_\rho(\cdot)$ is a 
derivation of $\A$ in $\Omega^1_\D$ \cite[Prop.~3.4]{CMo}:
\begin{equation}
	\delta_\rho(ab) = \rho(a)\cdot\delta_\rho(b) + \delta_\rho(a)\cdot b.
\end{equation}
Thus, $\Omega^1_\D(\A,\rho)$ is the $\A$-bimodule generated by $\delta_\rho$ and it
acts as bounded operator on $\HH$, since so do both $\A$ and $[\D,\A]_\rho$.

\smallskip

Here, it is also worth noticing that the self-adjointness is being imposed on the 
twist-fluctuated operator $\D_{\omega_\rho}$ (\ref{eqw:twistfluct}), 
which has the same domain as that of $\D$, and not 
necessarily on the twisted one-form $\omega_\rho$ \eqref{eq:8}. We shall further 
emphasize and explore this point in detail right after Lem.~\ref{lem:3.1} below.

\pagebreak

\subsection{Twisted gauge transformations}
\label{sec:twstgaugetrans}

For twisted spectral triples, the gauge transformations have been worked out in
\cite{LM2}, which -- as in the non-twisted case (\S\ref{sec:GT}) -- involve a change 
of the connection in the bimodule implementing the Morita equivalence. These are 
induced by the simultaneous action of the unitary group $\U(\A)$ \eqref{def:U(A)}  
on both the Hilbert space $\HH$ and the space $L(\HH)$ of linear operators in 
$\HH$.  

\smallskip

On $\psi\in\HH$, a unitary $u\in\U(\A)$ acts via the usual adjoint action 
\eqref{eq:gt1} of $\A$. However, 
on $T\in L(\HH)$, the action is twisted and implemented by the map
\begin{equation}
	T \mapsto \big(\Ad~\rho(u)\big)T\big(\Ad~u^*\big), \quad \text{with} \quad
	\Ad~\rho(u) := \rho(u)J\rho(u)J^{-1},
\end{equation}
and this evaluated for the twisted-covariant Dirac operator $\D_{\omega_\rho}$ 
gives \cite[\S4]{LM2}
\begin{equation}
\label{eq:gt2}
	\D_{\omega_\rho} \mapsto \big(\Ad~\rho(u)\big)\D_{\omega_\rho}\big(\Ad~u^*\big) 
	=: \D_{\omega_\rho^u},
\end{equation}
where, one has
\begin{equation}
	\omega_\rho\mapsto\omega_\rho^u := \rho(u)\omega_\rho u^* +\rho(u)[\D,u^*]_\rho,
\end{equation}
which is the twisted analogue of how the one-forms of the usual spectral 
triples transform as \eqref{eq:omegaU} in noncommutative geometry \cite{C96}.

\smallskip

Although gauge transformations leave the fluctuated Dirac operator 
$\D_\omega$ self-adjoint, the twist-fluctuated Dirac operator $\D_{\omega_\rho}$
is not self-adjoint under twisted gauge transformations. However, there does 
exist a more natural property than self-adjointness that is preserved under the 
twisted gauge transformations: {\bf$\rho$-adjointness} defined with respect to the 
{\bf$\rho$-inner product} induced by the twist $\rho$ on the Hilbert space $\HH$,
cf. \cite[Rem.~2.1]{DFLM}. 

\subsection{$\rho$-inner product}
\label{subsec:rho}

Given a Hilbert space $\HH$ with the inner product $\langle\cdot,\cdot\rangle$ and 
an automorphism $\rho$ of $\B(\HH)$, a {\it $\rho$-inner product} $\langle\cdot, 
\cdot\rangle_{\rho}$ is an inner product satisfying \cite[Def.~3.1]{DFLM}
\begin{equation} 
\label{3.2}
	\langle\psi,  \mathcal{O}\phi \rangle_{\rho}
	= \langle {\rho(\mathcal{O})}^\dag\psi, \phi \rangle_{\rho},
        \qquad \forall \mathcal{O} \in \mathcal{B}({\cal H}), \; \forall
        \psi, \phi \in {\cal H},
\end{equation}
where $^\dag$ denotes the hermitian adjoint with respect to $\langle\cdot,
\cdot\rangle$. One denotes
\begin{equation}
	\mathcal{O}^{+} := {\rho(\mathcal{O})}^\dag
\label{eq:13}
\end{equation}
to be the {\it $\rho$-adjoint} of the operator $\mathcal{O}$.

\smallskip

If $\rho$ is an inner automorphism and implemented by a unitary operator $\R \in 
\B(\HH)$, that is, $\rho(\mathcal{O}) = {\cal R}\mathcal{O}{\cal R}^\dag$ for any
$\mathcal{O} \in \mathcal{B}({\cal H})$. Then, a canonical $\rho$-inner product is
given by
\begin{equation} 
\label{rho-p}
	\langle \psi, \phi \rangle_{\rho} = \langle \psi, {\cal R}\phi \rangle, \qquad
	\forall \psi,\phi\in\HH.
\end{equation}

The $\rho$-adjointness is not necessarily an involution. If $\rho$ is a
$*$-automorphism (e.g.\ when $\rho$ is an inner automorphism), then $^+$ is an 
involution iff \eqref{eq:17} holds, since
\begin{equation}
  \label{eq:18}
  ({\cal O}^+)^+ = \rho({\cal O}^+)^\dag = \rho\big(({\cal O}^+)^\dag\big) 
  = \rho\big(\rho({\cal O})\big).
\end{equation}
The same condition arises for a twisted spectral
triple, when one defines the $\rho$-adjointness solely at the algebraic level, 
i.e.\ $a^+:=\rho(a)^*$, without assuming that $\rho\in\Aut(\A)$ extends to an
automorphism  of ${\cal B}(\HH)$. 

\smallskip

Indeed, assuming the regularity condition
in Rem.~\ref{rem:autmodul} (written as
$\rho(b)^*=\rho^{-1}(b^*)$ for any $b=a^*\in \A$), one then gets
\begin{equation}
  \label{eq:16}
  (a^+)^+ = (\rho(a)^*)^+ = (\rho^{-1}(a^*))^+ = \rho(\rho^{-1}(a^*))^* 
  = \rho(\rho(a)^*)^* = \rho^2(a).
\end{equation}

\subsection{Fermionic action}
\label{subsec:fermionaction}

In twisted spectral geometry, the fermionic action is defined by replacing 
$\D_{\omega}$ in \eqref{Sf} with the twist-fluctuated Dirac operator 
$\D_{\omega_{\rho}}$ \eqref{eqw:twistfluct} and by replacing the inner 
product with the $\rho$-inner product \eqref{3.2} or, in particular, with 
\eqref{rho-p} when the compatibility condition \eqref{eq:2.11} holds 
\cite[\S4.1]{DFLM}. Thus, instead of \eqref{eq:4}, one defines 
\begin{equation}
\label{Sfrho}
	{\mathfrak A}^\rho_{\D_{\omega_\rho}}\!(\phi, \psi)
	:= \langle J\phi, \D_{\omega_{\rho}}\psi \rangle_{\rho}
	= \langle J\phi, {\cal R}\D_{\omega_{\rho}}\psi \rangle, \qquad
	\forall\phi, \psi \in \mathsf{Dom}(\D_{\omega_{\rho}}).
\end{equation} 

In the case when the twist $\rho$ is compatible with the real structure $J$, in the
sense of \eqref{eq:2.11}, the bilinear form \eqref{Sfrho} is invariant under the
`twisted' gauge transformation, given by the simultaneous actions of \eqref{eq:gt1} 
and \eqref{eq:gt2} \cite[Prop.~4.1]{DFLM}. 

\smallskip

However, the antisymmetry  of the form ${\mathfrak A}^\rho_{\D_{\omega_\rho}}$ is
not guaranteed, unless one restricts to the positive eigenspace $\HH_\R$ of $\R$
\cite[Prop.~4.2]{DFLM}
\begin{equation}
\label{eq:2.16}
	\mathcal{H_R} := \big\{\chi\in\Dom(\D_{\omega_{\rho}}), \quad {\cal R}\chi = \chi \big\},
\end{equation}
which led to the following:

\begin{fdefine}
For a real twisted spectral triple $(\A, \HH, \D;J)_\rho$, the fermionic action is
\begin{equation}
	S^f_\rho(\D_{\omega_\rho}) := 
	\frak A^\rho_{\D_{\omega_\rho}}\!(\tilde \psi, \tilde \psi),
\label{eq:10}
\end{equation}
where $\tilde\psi$ is the Gra{\ss}mann vector associated to $\psi\in{\cal H_R}$.
\smallskip
\end{fdefine}

In the spectral triple of the Standard Model, the restriction to ${\cal H}_+$ is 
there to solve the fermion doubling problem \cite{LMMS}. It also selects out the 
physically meaningful elements of the space $\HH = L^2(\M,{\cal S}) \x \HH_\F$, 
i.e.\ those spinors whose chirality in $L^2(\M,{\cal S})$ coincides with their 
chirality as elements of the finite-dimensional Hilbert space $\HH_{\cal F}$.

In the twisted case, the restriction to ${\cal H_R}$ is there to guarantee the 
antisymmetry of the bilinear form $\frak A^\rho_{\D_{\omega_\rho}}$ \eqref{Sfrho}. 
However, the eigenvectors of $\cal R$ may not have a well-defined chirality. If 
fact, they cannot have it when the twist comes from the grading (see \S 
\ref{sec:2.2}), since the unitary ${\cal R}$ implementing the twist, given by 
\eqref{eq:12}, anticommutes with the chirality $\g = {\sf diag} \left( 
{\mathbb I}_{\HH_+}, -{\mathbb I}_{\HH_-} \right)$, so that we have
\begin{equation}
	\HH_+\cap \HH_{\cal R} = \big\{0\big\}.
 \label{eq:11}
\end{equation}

From a physical point-of-view, by restricting to $\HH_{\cal R}$ rather than 
$\HH_+$, one loses a clear interpretation of the elements of the Hilbert space: a
priori, an element of $\HH_\R$ is not physically meaningful since its chirality is
not well-defined. However, we shall demonstrate in what follows that -- at least in two 
examples: the almost-commutative geometry of a $U(1)$ gauge theory and that of 
electrodynamics -- the restriction to ${\cal H_R}$ is actually meaningful, for it 
allows us to obtain the Weyl and Dirac equations in the lorentzian signature, even 
though one starts with a riemannian manifold.

\smallskip

We conclude this subsection with two easy but useful lemmas. The first one recalls 
how the symmetry property of the bilinear form ${\frak A}_\D = \langle J\cdot, 
\D\cdot\rangle$ does not depend on the explicit form of the Dirac operator $\D$, 
but solely on the signs $\epsilon, \epsilon'$ in \eqref{eq:Reality}. And, the second one
emphasizes that once restricted to $\HH_\R$, the bilinear forms \eqref{eq:4}
and \eqref{Sfrho} differ only by a sign.

\begin{lem}
\label{lemma:antisymm}
Let $J$ be an antilinear isometry on the Hilbert space $(\HH, \langle\cdot, 
\cdot\rangle)$ such that $J^2=\epsilon\mathbb I$, and $\D$ be a self-adjoint 
operator on $\HH$ such that $J\D = \epsilon'\D J$. Then, we have
\begin{equation}
	\langle J\phi, \D\psi \rangle = \epsilon\epsilon'\langle J\psi, \D\phi\rangle,
	\qquad \forall \phi,\psi\in\HH.
\label{eq:15}
\end{equation}
\end{lem}

\begin{proof}
By definition, an antilinear isometry satisfies 
$\langle J\phi, J\psi\rangle = \overline{\langle\phi, \psi\rangle} 
= \langle\psi, \phi\rangle$. Thus,
\begin{equation*}
\begin{split}
	\langle J\phi, \D\psi\rangle &= \epsilon\langle J\phi, J^2\D\psi\rangle 
	= \epsilon\langle J\D\psi, \phi\rangle \\[3pt]
	&= \epsilon\epsilon'\langle \D J\psi, \phi\rangle 
	= \epsilon\epsilon'\langle J\psi, \D\phi\rangle. \qed
\end{split}
\end{equation*}
\end{proof}

In particular, for $KO$-dim.\ $2$ and $4$ one has $\epsilon=-1, \epsilon'=1$, so 
$\frak A_{\D}$ is antisymmetric. The same is true for $\frak A_{\D_\omega}$ in 
\eqref{eq:4}, because the covariant operator $\D_\omega$ also satisfies the same 
rules of sign \eqref{eq:Reality} as $\D$. On the other hand, for $KO$-dim.\ $0$ and
$6$ one has $\epsilon=\epsilon'=1$, and so $\frak A_{\D}$ is symmetric.

\begin{lem}
\label{lem:bilinearform}
Given $\D$, and $\cal R$ compatible with $J$ in the sense of \eqref{eq:2.11}, 
one has 
\begin{equation}
	{\mathfrak A}_\D^\rho(\phi, \psi)
	= \epsilon'''\,{\mathfrak A}_\D(\phi, \psi),
	\qquad \forall \phi, \psi \in {\cal H_R}.
\end{equation}
\end{lem}

\begin{proof}
	For any $\phi, \psi \in {\cal H_R}$, we have	
\begin{equation}
\begin{split}
	{\mathfrak A}_\D^\rho(\phi, \psi)
	&= \langle J\phi, {\cal R}\D\psi \rangle
	= \langle {\cal R}^\dag J\phi, \D\psi \rangle \\[3pt]
	&= \epsilon'''\langle J{\cal R}^\dag\phi, \D\psi \rangle
	= \epsilon'''\langle J\phi, \D\psi \rangle,
\end{split}
\end{equation}
where we used \eqref{eq:2.11} as
${\cal R}^\dag J = \epsilon'''J{\cal R}^\dag$ and 
\eqref{eq:2.16} as ${\cal R}^\dag\phi = \phi. \qed$
\end{proof}

\section{Minimal twist by grading}
\label{sec:2.2}

The twisted spectral triples that are recently employed in physics have been built 
by minimally twisting a usual spectral triple $(\A, \HH, \D)$. The idea of minimal 
twisting is to substitute the commutator $[{\cal D},\cdot]$ with the twisted 
commutator $[{\cal D},\cdot]_{\rho}$, while keeping the Hilbert space $\HH$ and the
Dirac operator $\D$ intact, since they encode the fermionic content of the theory
and there has, so far, been no experimental indications of the existence of extra 
fermions beyond those of the Standard Model. 

\smallskip

Such a substitution yields new fields \cite{DLM1, DM} that not only make the 
theoretical mass of the Higgs boson compatible with its experimental value 
\cite{CC12}, but also offer a way out of the problem of the instability 
(or, meta-stability) of the electroweak vacuum at intermediate energies,
as mentioned in the Introduction. However, for physically
relevant spectral triples, both $[\D,\cdot]$ and $[\D,\cdot]_{\rho}$ cannot be
bounded simultaneously and so one needs to enlarge the algebra \cite{LM1}.

\begin{fdefine}[from~\cite{LM1}]
\label{defi:minimaltwist}
	A \emph{minimal twist} of a spectral triple $(\cal A, H, D)$ by
	a unital $*$-algebra $\mathcal{B}$ is a twisted spectral triple 
	$(\A\x\B, \HH, \D)_\rho$ where the initial 
	representation $\pi_0$ of $\cal A$ on $\cal H$ is related to the 
	representation $\pi$ of $\cal A \otimes \cal B$ on $\cal H$ 
	by
\begin{equation}
	\pi(a \otimes \mathbb{I}_{\mathcal{B}}) = \pi_0(a), \qquad 
	\forall a \in {\cal A},
\end{equation}
where $\mathbb{I}_{\mathcal{B}}$ is the identity of the algebra $\cal B$. 
\smallskip
\end{fdefine}

If the initial spectral triple is graded, then a canonical minimal twist can be 
obtained naturally as follows. The grading $\g$ commutes with the representation of
the algebra $\A$, so the latter is a direct sum of two representations on 
the positive and negative eigenspaces, respectively, $\HH_+$ and $\HH_-$, of the
grading $\g$, see \eqref{eq:H_+}. 

\smallskip

Therefore, one has enough room on $\HH = \HH_+ 
\+ \HH_-$ to represent the algebra $\A$ twice. It is tantamount to taking $\B = 
{\mathbb C}^2$ in the Def.~\ref{defi:minimaltwist} above, with $\A\x\mathbb{C}^2 
\simeq \A \oplus \A \ni (a,a')$ represented on ${\cal H}$ as  
\begin{equation} 
\label{repaa'}
	\pi(a,a') := {\frak p}_+\pi_0(a) + {\frak p}_-\pi_0(a') =
	\left( \begin{array}{cc}
		\pi_+(a) & 0 \\ 0 & \pi_-(a')
	\end{array} \right)\!,
\end{equation}
where
\begin{equation}
	{\frak p}_{\pm} := \frac{1}{2} \big(\mathbb{I}_\HH \pm \g \big) \qquad 
	\text{and} \qquad 
	\pi_\pm(a) := \pi_0(a)_{|\HH_\pm}
\end{equation}
are, respectively, the projections on ${\cal H}_{\pm}$ and the restrictions on 
${\cal H}_{\pm}$ of $\pi_0$. 

\medskip

If $\pi_{\pm}$ are faithful,\footnote{The requirement
    that $\pi_\pm$ are faithful was not explicit in \cite{LM1}. If it
    does not hold, then $(\A\otimes\mathbb C^2, \HH, D)_\rho$ still
    satisfies all the properties of a twisted spectral triple, except
    that $\pi$ in (\ref{repaa'}) might not be faithful.}
then $({\cal
  A}\otimes\mathbb{C}^2, {\cal H, D})_{\rho}$ with the {\bf flip automorphism} $\rho$
  given by
\begin{equation}
\label{eq:flip}
	\rho(a,a') := (a',a), \qquad 
	\forall (a,a') \in {\cal A}\otimes\mathbb{C}^2,
\end{equation}
is indeed a twisted spectral triple, with grading $\g$. Furthermore,
if the initial spectral triple is real, then so is this minimal twist, with the 
same real structure \cite{LM1}.

\medskip

The flip $\rho$ \eqref{eq:flip} is a $*$-automorphism satisfying \eqref{eq:17} 
and coinciding on $\pi(\A\otimes\mathbb C^2)$ with the inner automorphism of ${\cal B}(\HH)$
implemented by the unitary operator 
\begin{equation}
\label{eq:12}
	{\cal R} = \left( \begin{array}{cc}
					0 & {\mathbb I}_{\HH_+} \\ {\mathbb I}_{\HH_-} & 0
				\end{array} \right)\!,
\end{equation}
where $\mathbb{I}_{\HH_{\pm}}$ is the identity operator in $\HH_{\pm}$.

\clearpage
\newpage 
	\chapter{Lorentzian Fermionic Actions from Euclidean Spectral Triples}
\label{chap:examples}

In this chapter, we study three examples of minimally 
twisted spectral triples: a manifold, $U(1)$ gauge theory, and electrodynamics -- 
with their corresponding fermionic actions. This yields the Weyl and the Dirac
action in lorentzian signature, respectively, in the last two cases.

\smallskip

We recall that the canonical $\rho$-inner product \eqref{rho-p} associated to the 
minimal twist of a $4$-dimensional closed {\bf riemannian} spin manifold turns out 
to coincide with the {\bf lorentzian} Kre\u{\i}n product on the Hilbert space of 
lorentzian spinors \cite[\S3.2]{DFLM}. One of the main results of this thesis is to 
demonstrate that a similar transition of metric signature -- from the euclidean to 
the lorentzian -- also occurs at the level of the fermionic action \cite{MS}. 

\smallskip

In \S \ref{sec:Weyl}, we will first investigate how this idea comes about, by 
looking at the simplest example of the minimal twist of a closed riemannian spin
manifold $\M$ and computing the associated fermionic action. 

\smallskip

Then, in \S\ref{sec:doublem}, 
we demonstrate how we obtain a lorentzian Weyl action from the 
minimally twisted $U(1)$ gauge theory. Similarly, in \S\ref{sec:electrody},
we also derive the lorentzian Dirac action from the spectral triple of 
electrodynamics \cite{DS} -- by minimally twisting it.

\smallskip

Since we intend to arrive at the physically relevant Weyl and Dirac actions, 
we chose to work in dimension $4$, assuming gravity is negligible (hence the flat 
metric). This is tantamount to choosing in \eqref{eq:Reality} the following signs:
    \begin{equation}
      \label{eq:22}
      \epsilon=-1,\qquad \epsilon'=1,\qquad \epsilon''=1.
    \end{equation}

\pagebreak

\section{Minimal twist of a manifold}
\label{sec:Weyl}

The minimal twist of the canonical triple \eqref{eq:CT} of a closed riemannian 
spin manifold $\cal M$ is the following real twisted spectral triple:
\begin{equation}
\label{mnfld}
	\big(\A = C^{\infty}(\M)\x\mathbb{C}^2, \quad
		\HH = L^2(\mathcal{M,S}), \quad \D = \eth \big)_{\rho}
\end{equation}
with the inner product on the Hilbert space $L^2(\M,\cal S)$ given by
\begin{equation}
\label{eq:12bis}
	\langle \psi, \phi \rangle = \int_\M\dmu \; \psi^\dag \phi,
\end{equation}
where $\dmu$ is the volume form. With the self-adjoint euclidean Dirac matrices 
$\g$'s (see App.~\ref{GammaMatrices}), the real structure is given by \cite{DM}
\begin{equation}
\label{eqn:3.9}
	\J = i\g^0\g^2 cc
	= i\begin{pmatrix} 
		\tilde\sigma^2 & 0 \\ 0 & \sigma^2 
	\end{pmatrix}cc,
\end{equation}
where $cc$ denotes complex conjugation and the grading is
\begin{equation}
	\g^5 = \g^1\g^2\g^3\g^0
= \begin{pmatrix}
		\mathbb{I}_2 & 0 \\ 0 & -\mathbb{I}_2
	\end{pmatrix}.
\end{equation}
The representation \eqref{repaa'} for the algebra $C^\infty(\M)\x\mathbb{C}^2$ on
the Hilbert space, decomposed as 
$L^2(\M,{\cal S}) = L^2(\M,{\cal S})_+ \+ L^2(\M,{\cal S})_-$, is given by
\begin{equation}
\label{eq:piM}
	\pi_{\M}(f,f')
=  \begin{pmatrix}
		f\,\mathbb{I}_2 & 0 \\ 0 & f'\,\mathbb{I}_2
	\end{pmatrix},
\end{equation}
where each of the two copies of $C^{\infty}(\M)$ acts independently and faithfully
by the pointwise multiplication on the eigenspaces $L^2(\M,{\cal S})_{\pm}$ of the
grading $\g^5$.

\smallskip

The automorphism $\rho$ of $C^{\infty}(\M)\x\mathbb{C}^2$ is the flip
\begin{equation}
	\rho(f,f') = (f',f), \qquad \forall f,f' \in C^{\infty}(\M),
\end{equation}
which coincides with the inner automorphism of $\B(\HH)$ implemented by the 
following unitary matrix
\begin{equation}
\label{eq:definR}
	\R = \begin{pmatrix}
		0 & \mathbb{I}_2 \\ \mathbb{I}_2 & 0
	\end{pmatrix}.
\end{equation}
This $\R$ is nothing but the first Dirac matrix $\g^0$ and it is compatible 
with the real structure $\J$, in the sense of \eqref{eq:2.11}, with
\begin{equation}
\label{eq:31} 
	\epsilon'''=-1.
\end{equation}

\pagebreak

\begin{lem}
\label{lemma:commgamma}
	For any $a=(f,f') \in C^{\infty}(\M)\x{\mathbb C}^2$ and $\mu=0,1,2,3$, one has  
 \begin{equation}
 \label{eq:commgamma}
   \gm a = \rho(a) \gm, \qquad \gm\rho(a) =
   a\gm, \qquad
  \gm \J = -\epsilon'\J\gm.
 \end{equation}
\end{lem}

\begin{proof}
	The first equation is checked by direct calculation, using the explicit form 
	of $\gm$, along with \eqref{eq:piM}. Omitting the symbol of 
	representation, by $\pi_\M(\rho(a))$, we mean
\begin{equation}
\label{eq:twistmanifold}    
	\rho(a) = \begin{pmatrix}
		f'\,\mathbb{I}_2 & 0 \\ 0 & f\,\mathbb{I}_2
	\end{pmatrix}.
\end{equation}
The second equation then follows from \eqref{eq:17}. 
And, finally, the third one is obtained from \eqref{eq:Reality}, 
noticing that $\J$ commutes with $\dm$, 
having constant components:
\begin{equation*}
  0 = \J\eth - \epsilon' \eth\J =i \left(\J \gm +
    \epsilon'\gm \J\right)\dm.  \qed
\end{equation*}
\end{proof}

As an immediate corollary, one checks the boundedness of the twisted commutator, 
thus
\begin{align}
\label{eq:test1}
	\big[\eth, a\big]_\rho 
& = \big(\gm\dm a -\rho(a)\gm\dm \big) \nonumber \\
& = -i\gm \big[ \dm, a \big] \nonumber \\
& = -i\gm(\dm a), \qquad\qquad\qquad 
	\forall a\in C^\infty(\M)\x\mathbb{C}^2.
\end{align}

\subsection{The $X_\mu$ field}
\label{subsec:Xmu}

Following the standard terminology of the non-twisted case,
given a twisted spectral triple $(\A, \HH,\D)_\rho$, 
the substitution of $\D$ with $\D_{\omega_\rho}$ is referred to as the
{\bf twisted fluctuation}. 

\smallskip

The Dirac operator $\eth$ of a four-dimensional manifold $\M$  
has non-vanishing self-adjoint twisted fluctuation \eqref{eqw:twistfluct} of the 
form \cite[Prop.~5.3]{LM1}:
\begin{equation}
\label{eq:dflucM}
	\eth \to \eth_X := \eth + {\bf X},
\end{equation}
where
\begin{equation}
\label{Xmu}
	{\bf X} := -i\gm X_{\mu}, \qquad \text{ with } \quad X_{\mu} :=
f_\mu \g^5,
\end{equation}
for some $f_\mu \in C^\infty(\M,\mathbb{R})$.

\smallskip

In contrast, the self-adjoint (non-twisted) fluctuations of the Dirac operator 
$\eth$ are always vanishing, irrespective of the dimension of the manifold $\M$ 
\cite{C96}. 

\smallskip

However, in \cite{LM1} one imposes the self-adjointness of $\eth_X$, without 
necessarily requiring $\omega_\rho$ to be self-adjoint. One might wonder 
that the non-vanishing of $\bf X$ is then an artifact of such a choice
and that $\bf X$ might vanish as soon as one also imposes
$\omega_\rho = \omega_\rho^\dag$. The following lemma clarifies this issue.

\pagebreak

\begin{lem} 
\label{lem:3.1} 
The twisted one-forms $\omega_\rho$ \eqref{eq:8} and the twisted fluctuations 
$\omega_\rho + \J\omega_\rho \J^{-1}$ of the minimally twisted canonical triple
\eqref{mnfld} are of the form
\begin{align}
\label{eq:omrgo}	
	& \omega_\rho = {\bf W} := -i\gm W_\mu, 
	& \text{ with } & \quad W_\mu 
	= {\sf diag} \big( h_\mu\mathbb{I}_2, \, h'_\mu\mathbb{I}_2 \big), \\
\label{eq:omrgo2}
	& \omega_\rho + \J\omega_\rho \J^{-1} = {\bf X} 
	:= -i\gm X_\mu, & \text{ with } & \quad X_\mu 
	= {\sf diag} \big( f_\mu\mathbb{I}_2, \, f'_\mu\mathbb{I}_2 \big),
\end{align}
for some $h_\mu,\,h'_\mu\in C^\infty(\M)$ with $f_\mu := 2\Re{h_\mu}$ 
and $f'_\mu := 2\Re{h'_\mu}$. They are self-adjoint, respectively, iff
\begin{align}
\label{eq:29}
	h'_\mu =-\overline{h}_\mu, \qquad \text{ and } \qquad f'_\mu = - f_\mu.
\end{align}
\end{lem}

\begin{proof}
For some $a_i:=(f_i,f'_i)\in C^{\infty}(\M)\x{\mathbb C}^2$, using 
\eqref{eq:commgamma} along with $[\nabla_\mu^{\cal S},f_i] = 
(\dm f_i)$, one gets
\begin{align}
	\big[\eth, a_i\big]_\rho 
& = -i\big( \gm\nabla_\mu^{\cal S}a_i 
	-\rho(a_i)\gm\nabla_\mu^{\cal S} \big) \nonumber \\[3pt]
& = -i\big( \gm\nabla_\mu^{\cal S}a_i 
	-\gm a_i\nabla_\mu^{\cal S} \big) \nonumber \\[3pt]
& = -i\gm\big[\nabla_\mu^{\cal S}, a_i\big] \\[3pt]
& = -i\gm(\dm a_i) \nonumber \\[3pt]
& = -i\gm\begin{pmatrix}
		(\dm f_i)\mathbb{I}_2 & 0 \\ 0 & (\dm f_i')\mathbb{I}_2 
	\end{pmatrix}. 
\label{eq:test1}
\end{align}
Then, with some $b_i:=(g_i,g'_i)\in C^{\infty}(\M)\x{\mathbb C}^2$, one has
\begin{equation}
\label{eq:omrgobis}	
	\omega_\rho = \sum_ib_i[\eth,a_i]_\rho
	= -i \gm\sum_i\rho(b_i)(\dm a_i) = -i\gm W_\mu,
\end{equation}
where $W_\mu$ is defined in \eqref{eq:omrgo}, with
\begin{equation}
	h_\mu := \sum_i g_i'(\dm f_i) \qquad \text{ and } 
	\qquad h'_\mu := \sum_i g_i(\dm f_i'). 
\end{equation}
The adjoint is
\begin{equation}
	\omega^\dag_\rho = (-i\gm  W_\mu)^\dag = iW_\mu^\dag\gm  
	= i\gm \rho(W_\mu^\dag),
\end{equation}
where the last equality follows from \eqref{eq:commgamma} applied to $W_\mu$ 
viewed as an element of the algebra $C^{\infty}(\M)\x{\mathbb C}^2$. Thus,
$\omega_\rho$ is self-adjoint iff $\gm\rho(W_\mu^\dag) = -\gm W_\mu$,
that is, going back to the explicit form of $\gm$, equivalent to
\begin{equation}
\label{eq:gsym}
	\sigma^\mu \overline{h}_\mu = -\sigma^\mu h'_\mu, \qquad \text{ and } \qquad
	\tilde\sigma^\mu \overline{h'}_\mu = -\tilde \sigma^\mu h_\mu.
\end{equation}
Multiplying the first equation by $\sigma^\lambda$ and using 
$\trc(\sigma^\lambda\sigma^\mu) = 2\delta^{\mu\lambda}$, \eqref{eq:gsym} then 
implies that $\overline{h}_\mu = -h'_\mu$, from where \eqref{eq:29} follows. 
Hence, $\omega_\rho = \omega_\rho^\dag$ is equivalent to the first equation of 
\eqref{eq:29}. Further, we have	
\begin{equation*}
\begin{split}
	\J \omega_\rho \J^{-1} 
& = \J (-i\gm W_\mu) \J^{-1}
  = i\J(\gm W_\mu) \J^{-1} \\[3pt] 
& = -i\gm \J W_\mu \J^{-1}
  = -i\gm  W_\mu^\dag,
\end{split}
\end{equation*}
using $\J\gm = -\gm\J$ -- from \eqref{eq:22} and \eqref{eq:commgamma} -- 
along with $\J W^\mu= W_\mu^\dag\J$ -- from \eqref{eqn:3.9} and the explicit form 
\eqref{eq:omrgo} of $W_\mu$. Therefore, 
\begin{equation}
	\omega_\rho + J\omega_\rho J^{-1} = -i\gm (W_\mu + W_\mu^\dag),
\end{equation}
which is nothing but \eqref{eq:omrgo2}, identifying
\begin{equation}
\begin{split}
	X_\mu & := W_\mu + W_\mu^\dag \\ 
	& = {\sf diag}
	\big( (h_\mu + \overline{h}_\mu)\mathbb{I}_2,
		(h'_\mu + \overline{h'}_\mu)\mathbb{I}_2 \big).
\end{split}
\end{equation}
One checks, in a similar way as above, that $\omega_\rho + \J\omega_\rho\J^{-1}$ 
is self-adjoint iff the second eq.\ of \eqref{eq:29} holds. $\qed$
\end{proof}

\smallskip

Imposing that $\omega_\rho\neq 0$ be self-adjoint, that is imposing \eqref{eq:29} 
with $h_\mu\neq 0$, does not imply that $X_\mu$ vanishes. It does vanish, if 
$h_\mu$ is purely imaginary, for then $h_\mu+ \overline{h}_\mu=0$ and 
\eqref{eq:29} imposes that $h'_\mu$ is also purely imaginary, consequently, the 
sum $(h'_\mu + \overline{h'}_\mu)$ also vanishes, hence $X_\mu = 0$. However, 
$h_\mu$ is not necessarily purely imaginary, in which case the self-adjointness of
$\omega_\rho$ does not forbid a non-zero twisted fluctuation.

\subsection{Gauge transformation}
\label{sec:3.1.2}

For a minimally twisted manifold, not only is the fermionic action \eqref{eq:10}
invariant under a gauge transformation (\ref{eq:gt1},~\ref{eq:gt2}), but also the 
twist-fluctuated Dirac operator $\D_{\omega_\rho}$ (in dim.\ $0$ and $4$) 
\cite[Prop.~5.4]{LM2}. It is interesting to check this explicitly by studying how 
the field $h_\mu$ parametrizing the twisted one-form $\omega_\rho$ in 
\eqref{eq:omrgo} transforms. This will also be useful later in the example of 
electrodynamics.

\smallskip

A unitary $u$ of the algebra $C^\infty(\M) \x\mathbb C^2$ is of the form
$(e^{i\theta},e^{i\theta'})$ with $\theta, \theta'\in C^\infty(\M,\mathbb R)$. 
It (and its twist) acts on $\HH$ according to \eqref{eq:piM} as (omitting the 
symbol of representation):
\begin{equation}
\begin{split}
\label{eq:pi_M(u)}
	u &= 
	\begin{pmatrix}
		e^{i\theta}\mathbb{I}_2 & 0 \\ 0 & e^{i\theta'}\mathbb{I}_2
	\end{pmatrix}, \\[3pt]
	\rho(u) &=
	\begin{pmatrix}
		e^{i\theta'}\mathbb{I}_2 & 0 \\ 0 & e^{i\theta}\mathbb{I}_2
	\end{pmatrix}.
\end{split}
\end{equation}

\pagebreak

\begin{prop}
	Under a gauge transformation with unitary $u \in
        C^{\infty}(\mathcal{M})\x\mathbb C^2$, the fields 
	$h_\mu$ and $h'_\mu$ parametrizing the twisted one-form $\omega_\rho$ in 
	\eqref{eq:omrgo} transform as
	\begin{equation}
		h_\mu \to h_\mu -i\dm\theta, \qquad
		h'_\mu \to h_\mu' +i\dm\theta'\!.
	\end{equation}
\end{prop}

\begin{proof}
Under a gauge transformation, a twisted one-form $\omega_\rho \in 
\Omega_\D^1(\A,\rho)$ transforms as \cite[Prop.~4.2]{LM2}
\begin{equation}
\label{LM2Prop4.2}
	\omega_\rho \to \omega^u_\rho
	:= \rho(u) \big( [\D,u^*]_\rho + \omega_\rho u^* \big).
\end{equation}
For $\D = \eth = -i\gm \dm$ and $\omega_\rho = -i\gm W_\mu$, 
we have
\begin{equation*}
\begin{split}
	\omega^u_\rho 
& = -i\rho(u) \big( [\gm\dm,u^*]_\rho + \gm W_\mu u^* \big) \\
& = -i\rho(u)\gm (\dm + W_\mu)u^* \\
& = -i\gm (u\dm u^* + W_\mu),
\end{split}
\end{equation*}
where we have used \eqref{eq:test1} for $a_i=u^*$, namely
\begin{equation}
\label{Id1}
	\big[ \gm \dm, u^* \big]_\rho 
	= \gm (\dm u^*),
\end{equation}
as well as \eqref{eq:commgamma} for $a=u$, together with $uW_\mu u^*=W_\mu$, since
$u$ commutes with $W_\mu$. Therefore, $W_\mu \to W_\mu + u\dm u^*$,
which with the explicit representation \eqref{eq:omrgo} of $W_{\mu}$ and 
\eqref{eq:pi_M(u)} of $u$, respectively, reads
\begin{equation*}
	\begin{pmatrix}
		h_\mu\mathbb{I}_2 & 0 \\ 0 & h_\mu'\mathbb{I}_2
	\end{pmatrix} \longrightarrow
	\begin{pmatrix}
		(h_\mu-i\dm\theta)\mathbb{I}_2 & 0 \\
		0 &  (h_\mu'+i\dm\theta')\mathbb{I}_2
	\end{pmatrix}, \qed
\end{equation*}
\end{proof}

Although $h_\mu$ and $h'_\mu$ transform in a nontrivial manner, their real parts 
remain invariant, as we have
\begin{equation*}
	h_\mu + \overline{h}_\mu \quad \longrightarrow \quad
	h_\mu -i\dm\theta + \overline{h}_\mu + i\dm\theta \; = \;
	h_\mu + \overline{h}_\mu,
\end{equation*}
and similarly for $h'_\mu$. Since it is the real parts that enter in the 
Def.\ \eqref{eq:omrgo} of $\bf X$, this explains why the latter is invariant
under a gauge transformation. 

\smallskip

Notice that this is true whether $\bf X$ is self-adjoint or not. In case 
$\omega_\rho$ is not self-adjoint, the imaginary part $$g_\mu := \Im{h_\mu} = 
\frac 1{2i}\big(h_\mu -\overline{h}_\mu\big)$$ of $h_\mu$ is not invariant under 
a gauge transformation, but transforms as
\begin{equation*}
	g_\mu \to g_\mu -\dm\theta.
\end{equation*}

We return to this point while discussing the gauge transformations for the example
of electrodynamics, where a similar phenomenon occurs in 
\eqref{eq:4.51}--\eqref{eq:4.52}.

\pagebreak

\subsection{Fermionic action with no spinor freedom}
\label{subsec:fermioncactionmanif}

First we work out how the the positive eigenspace $\HH_\R$ \eqref{eq:2.16} of the 
unitary matrix $\R = \g^0$, as in \eqref{eq:definR}, looks like.

\begin{lem}
\label{lem3.1}
	An eigenvector $\phi \in {\cal H_R}$ is of the form $\phi := 
	\begin{pmatrix} \varphi \\ \varphi \end{pmatrix}$, where $\varphi$ is a Weyl 
	spinor.
\end{lem}

\begin{proof}
The $+1$-eigenspace of $\R=\g^0$ is spanned by
\begin{equation*}
	\upsilon_1 = \begin{pmatrix} 1 \\ 0 \end{pmatrix} \x
	\begin{pmatrix} 1 \\ 1 \end{pmatrix}, \qquad
	\upsilon_2 = \begin{pmatrix} 0 \\ 1 \end{pmatrix} \x
	\begin{pmatrix} 1 \\ 1 \end{pmatrix}. 
\end{equation*}
Therefore, a generic vector $\phi \in\HH_\R$ is given by
\begin{equation*}
	\phi = \phi_1\upsilon_1 + \phi_2\upsilon_2 =:
	\begin{pmatrix} \varphi \\ \varphi \end{pmatrix}, \qquad \text{ with } \quad
	\varphi := \begin{pmatrix} \phi_1 \\ \phi_2 \end{pmatrix}. \qed
\end{equation*}
\end{proof}

Now, we compute the fermionic action \eqref{eq:10} of the minimally twisted 
manifold \eqref{eq:CT}.

\begin{prop}
\label{prop:actspecmanif}
	Let $\eth_X$ be the twist-fluctuated Dirac operator \eqref{eq:dflucM}.  
	The symmetric form \eqref{Sfrho} is
\begin{equation}
\label{3.1}
	{\frak A}^\rho_{\eth_X}(\phi,\xi) 
	= 2 \int_\M \dmu \left[ \bar\varphi^\dag\sigma_2
		\left( if_0\mathbb{I}_2 - \sum_{j=1}^3\sigma_j\partial_j \right)
	\zeta \right],
\end{equation}
where $\varphi$, $\zeta$ are, respectively, the Weyl components of the Dirac 
spinors $\phi$, $\xi\in{\cal H_R}$, and $f_0$ is the zeroth component of $f_\mu $ in 
\eqref{Xmu}.
\end{prop}

\begin{proof}
	One has the following relations:
\begin{align}
	\J\phi & = i\g^0\g^2 cc
	\begin{pmatrix} \varphi \\ \varphi \end{pmatrix} 
\label{eq:JPhi}
 = i\begin{pmatrix} 
		\tilde\sigma^2 & 0 \\ 0 & \sigma^2
	\end{pmatrix} \!\!
	\begin{pmatrix} \bar \varphi \\ \bar \varphi \end{pmatrix}
= i\begin{pmatrix}
		\tilde\sigma^2\,\bar \varphi \\ \sigma^2\,\bar\varphi
	\end{pmatrix}, \\[5pt]
	\eth \xi 
& = -i\gm \dm
	\begin{pmatrix} \zeta \\ \zeta \end{pmatrix} 
\label{eq:DPhi}
 = -i\begin{pmatrix}
		0 & \sigma^\mu \\ \tilde\sigma^\mu & 0
	\end{pmatrix} \!\!
	\begin{pmatrix}	
		\dm\zeta \\ \dm\zeta
	\end{pmatrix}
= -i\begin{pmatrix}
		\sigma^\mu\dm\zeta \\ \tilde\sigma^\mu\dm\zeta
	\end{pmatrix}, \\[5pt]
	{\bf X} \xi 
& = -i\gm X_{\mu}
	\begin{pmatrix} \zeta \\ \zeta \end{pmatrix} 
\label{eq:Xxi}
 = -i\begin{pmatrix}
		0 & \sigma^\mu \\ \tilde\sigma^\mu & 0
	\end{pmatrix} \!\!
	\begin{pmatrix}
		f_\mu\mathbb{I}_2 & 0 \\ 0 & -f_\mu\mathbb{I}_2
	\end{pmatrix} \!\!
	\begin{pmatrix} \zeta \\ \zeta \end{pmatrix}
= -i\begin{pmatrix}
		-f_\mu\sigma^\mu\zeta \\ f_\mu\tilde\sigma^\mu\zeta
	\end{pmatrix}.
\end{align}
Hence, noticing that $(\tilde\sigma^2)^\dag = \sigma^2$ and 
${\sigma^2}^\dag = -\sigma^2$ (see App.\ \ref{GammaMatrices}), and using 
\begin{equation}
\begin{split}
\label{eq:sumsigma}
	\sigma^\mu + \tilde\sigma^\mu & = 2{\mathbb I}_2 \delta^{\mu 0}, \\
	\sigma^\mu - \tilde\sigma^\mu & = -2i\delta^{\mu j} \sigma^j,
\end{split}
\end{equation}
one gets
\begin{align}
	{\frak A}_\eth(\phi,\xi) = \langle \J\phi, \eth\xi \rangle
& = -\begin{pmatrix}
		\bar\varphi^\dag \tilde\sigma^{2\dag} & \bar\varphi^\dag \sigma^{2\dag}
	\end{pmatrix} \!
	\begin{pmatrix}
		\sigma^\mu\dm\zeta \\ \tilde\sigma^\mu\dm\zeta
	\end{pmatrix} \nonumber \\
& = -\int_\M \dmu \left( \bar\varphi^\dag \sigma^2\sigma^\mu \dm\zeta 
	-\bar\varphi^\dag\sigma^2\tilde\sigma^\mu\dm\zeta \right) \nonumber \\
& = -\int_\M\dmu \left( \bar\varphi^\dag \sigma^2(\sigma^\mu 
		-\tilde\sigma^\mu)\dm \zeta \right) \nonumber \\ 
& = 2 \int_\M \dmu \left( \bar\varphi^\dag \sigma_2
	 	\textstyle\sum_{j=1}^3 \sigma_j\partial_j \zeta \right); 
\label{derniere1} \\[5pt]
	{\frak A}_{\bf X}(\phi,\xi) = \langle \J\phi, {\bf X}\xi \rangle
& = -\begin{pmatrix}
		\bar\varphi^\dag \tilde\sigma^{2\dag} & \bar\varphi^\dag \sigma^{2\dag}
	\end{pmatrix} \!
	\begin{pmatrix}
		-f_\mu\sigma^\mu \zeta \\ f_\mu\tilde\sigma^\mu\zeta
	\end{pmatrix} \nonumber \\
& = \int_\M \dmu \left( \bar\varphi^\dag\sigma^2 f_\mu\sigma^\mu\zeta 
	+ \bar\varphi^\dag\sigma^2 f_\mu\tilde\sigma^\mu\zeta \right) \nonumber \\
& = \int_\M \dmu \left( 
	\bar\varphi^\dag\sigma^2 f_\mu(\sigma^\mu +\tilde\sigma^\mu)\dm\zeta
	\right) \nonumber \\ 
\label{derniere2}
& = -2i \int_\M \dmu \left( f_0 \bar\varphi^\dag \sigma_2 \zeta \right);
\end{align} 
where in \eqref{derniere1} and \eqref{derniere2} we have used $\sigma^2 = 
-i\sigma_2$. The result then follows from Lem.\ \ref{lem:bilinearform} and 
\eqref{eq:31}, as
\begin{equation*}
\begin{split}
\label{eq:arho}
	{\frak A}^\rho_{\eth_X}(\phi, \xi)
&	= -	{\frak A}_{\eth_X}(\phi, \xi) \\
&	= -{\frak A}_{\eth}(\phi, \xi) -{\frak A}_{X}(\phi,\xi). \qed
\end{split}
\end{equation*}
\end{proof}

The fermionic action is then obtained by substituting $\phi=\xi$ in \eqref{3.1} 
and replacing the components $\zeta$ of $\xi$ by the associated Gra{\ss}mann 
variable $\tilde \zeta$, as follows

\begin{mdframed}
\begin{equation}
\label{eq:56}
	S^f_\rho(\D_{\omega_\rho}) 
= 2 \int_\M \dmu \left[ \tilde{\bar\zeta}^\dag\sigma_2
		\left( if_0\mathbb{I}_2 -\textstyle\sum_{j=1}^3\sigma_j\partial_j 
			\right) \tilde\zeta \right].
\end{equation}
\end{mdframed}

The most interesting observation regarding this action \eqref{eq:56} is the disappearance of the 
derivative in the $x_0$ direction, and the appearance of the zeroth-component of 
the real field $f_\mu$ parametrizing the twisted fluctuation ${\bf X}$, instead. 
This derivative, however, can be restored by interpreting $-if_0\zeta$ as 
$\partial_0\zeta$, that is, assuming
\begin{equation}
  \label{eq:33}
  \zeta(x_0, x_i) = \exp(- if_0 x_0) \, \zeta(x_i).
\end{equation}

Denoting by $\sigma^\mu_M = \{\mathbb I_2, \sigma_j\}$ the upper-right 
components of the minkowskian Dirac matrices (see \eqref{MDirac}), the integrand 
in the fermionic action $S^f_\rho$ \eqref{eq:56} then reads (with summation on the
index $\mu$)
\begin{equation}
	-\tilde{\bar\zeta}^\dag \sigma^2_M
	\left( \sigma^\mu_M\dm\right)\tilde\zeta,
\label{eq:034}
\end{equation}
which reminds us of the Weyl lagrangian densities \eqref{eq:Weyl}:
\begin{equation}
	S^F = i\Psi_r^\dag \left({\sigma}_M^\mu\dm \right)\Psi_r, 
\end{equation}
but with an extra $\sigma_M^2$ matrix factor that prevents us from simultaneously 
identifying $\tilde\zeta$ with $\Psi_r$ and $-\tilde{\bar\zeta}^\dag\sigma^2_M$ 
with~$i\Psi_r^\dag$. 

\bigskip

To make such an 
identification 
possible, one needs more spinorial degrees of freedom, which can be obtained by 
considering a tensor product of the manifold $\M$ with a two-point space $\F_2$.

\bigskip

\section{Minimal twist of a $U(1)$ gauge theory}
\label{sec:doublem}

\medskip

Following the minimal twist by grading procedure of \S\ref{sec:2.2}, the 
minimal twist of the spectral triple of a $U(1)$ gauge theory (\S\ref{subsec:2.4.1}) is 
given by the algebra ${\cal A} \x \mathbb{C}^2$, where 
$\A := C^{\infty}(\M)\x\mathbb{C}^2$, represented on the Hilbert space 
$\HH := L^2(\M,{\cal S}) \x \mathbb{C}^2$ as
\begin{equation}
\label{eq:mintwst_X}
	\pi(a,a') =
	\begin{pmatrix}
		f\mathbb{I}_2 & 0 & 0 & 0 \\ 0 & f'\mathbb{I}_2 & 0 & 0 \\ 
		0 & 0 & g'\mathbb{I}_2 & 0 \\ 0 & 0 & 0 & g\mathbb{I}_2
	\end{pmatrix} =:
	\begin{pmatrix}
		F & 0 \\ 0 & G' 
	\end{pmatrix},
\end{equation} 
for $a:=(f,g),\; a':=(f',g') \in {\cal A}$; along with its twist represented as
\begin{equation}
\label{eq:mintwst_Xbis}
	\pi(\rho(a,a')) = \pi(a',a) 
= 	\begin{pmatrix}
		f'\mathbb{I}_2 & 0 & 0 & 0 \\ 0 & f\mathbb{I}_2 & 0 & 0 \\ 
		0 & 0 & g\mathbb{I}_2 & 0 \\ 0 & 0 & 0 & g'\mathbb{I}_2
	\end{pmatrix}
=:	\begin{pmatrix}
		F' & 0 \\ 0 & G 
	\end{pmatrix}.
\end{equation}
In both of the equations above, we have denoted
\begin{equation}
\label{eq:4.10}
\begin{split}
	F := \pi_{\mathcal{M}}(f,f'), \qquad \qquad
& 	F' := 
 \pi_{\mathcal{M}}(f',f), \\[3pt]
	G := \pi_{\mathcal{M}}(g,g'), \qquad \qquad
& 	G' := 
 \pi_{\mathcal{M}}(g',g),
\end{split}
\end{equation}
where $\pi_\M$ is the representation \eqref{eq:piM} of
$C^{\infty}(\M)\x\mathbb{C}^2$ on $L^2(\M,{\cal S})$.

\clearpage

\subsection{Twisted fluctuation}
\label{subsec:twistfluc2man}

Following the notation of \eqref{Xmu}, given $Z_\mu = \pi_\M(f_\mu,f_\mu')$ and 
$Z_\mu' = \pi_\M(f'_\mu,f_\mu)$ for some $f_\mu, f'_\mu \in C^\infty(\M)$, we 
denote the following
\begin{equation}
\label{eq:notaZ}
	\mathbf{Z} := -i\gm  Z_\mu, \qquad 	\mathbf{Z}' :=
        -i\gm  Z_\mu', \qquad \overline{\mathbf{Z}}
        :=	-i\gm  \overline{Z}_{\mu}.
\end{equation} 
Notice that $\overline{\bf Z}$ \emph{is not} the complex conjugate of ${\bf Z}$, 
since in \eqref{eq:notaZ} the complex conjugation does not act on the Dirac 
matrices. This guarantees that $\,\bar{}\,$ and $\,'\,$ commute not only for 
$Z_\mu$ but also for $\bf Z$, that is,
\begin{equation}
  \label{eq:45}
  {\overline{Z'_\mu}}=(\overline Z_\mu)', \qquad \text{ and } \qquad
  {\bf (\overline Z)'}={\bf \overline{Z'}}.
\end{equation}
Thus, the notation ${\bf \overline Z'}$ is then unambiguous and denotes 
indistinctly both the members of the second eq.\ in \eqref{eq:45}.

\begin{lem} 
For any $F$, $G$, $Z_\mu$, as in (\ref{eq:4.10},~\ref{eq:notaZ}), one has
\label{lem:notatio}
\begin{equation}
\label{eq:4.11}
	F[\eth,G]_{\rho} = -i\gm F'\partial_{\mu}G, \qquad 
	\J\mathbf{Z}\J^{-1} = \overline{\mathbf{Z}}, \qquad 
	{\bf Z}^\dag = -{\bf \overline Z'}.
\end{equation}
\end{lem}

\begin{proof}
Eq.\ \eqref{eq:commgamma} for $a=F'$ gives $F\gm  = \gm F'$, whereas eq.\
\eqref{eq:test1} for $a=G$ yields $[\eth,G]_{\rho} 
  -i\gm\dm G$.
Thus, we have the first of \eqref{eq:4.11}:
\begin{equation*}
	F[\eth, G]_\rho = -i F\gm\dm G = -i\gm F'\dm G.
\end{equation*}
The second relation in \eqref{eq:4.11} follows from
\begin{equation}
\begin{split}
\label{eq:41}
	\J {\bf Z}\J^{-1} = i\J\gm Z_\mu\J^{-1} = - i\gm\J Z_\mu J^{-1} 
	= -i\gm \overline Z_\mu \quad = \quad {\bf \overline Z},
\end{split}
\end{equation}
where we used (\ref{eq:commgamma}), as well as (recalling that in
$KO$-dime.\ $4$, one has $\J^{-1} = -\J$)
\begin{align*}
	\J Z_\mu\J^{-1} 
& = -i\begin{pmatrix} 
		\tilde\sigma^2 & 0 \\ 0 & \sigma^2
	\end{pmatrix} cc 
	\begin{pmatrix}
    	f_\mu\,\mathbb I_2 & 0 \\ 0 & f'_\mu\,\mathbb I_2
	\end{pmatrix} 
	i \begin{pmatrix} 
		\tilde\sigma^2 & 0 \\ 0 & \sigma^2
	\end{pmatrix} cc, \\[3pt]
& = -\begin{pmatrix} 
		\tilde\sigma^2 & 0 \\ 0 & \sigma^2
	\end{pmatrix} \!\!
	\begin{pmatrix}
		\overline f_\mu \,\mathbb I_2 & 0 \\
		0 & \overline{f'}_\mu\,\mathbb I_2
	\end{pmatrix} \!\!
	\begin{pmatrix} 
		\bar{\tilde\sigma}^2 & 0 \\
		0 & \bar\sigma^2
	\end{pmatrix} \\[3pt]
& = \begin{pmatrix}
		\overline{f}_\mu\, \mathbb I_2 & 0 \\
		0 & \overline{f'}_\mu\,\mathbb I_2
	\end{pmatrix} \quad = \quad \overline Z_\mu,
\end{align*}
noticing that $\bar{\tilde\sigma}^2 = \tilde\sigma^2$ and $\bar\sigma^2 = 
\sigma^2$, so $\tilde\sigma^2\bar{\tilde\sigma}^2 = \sigma^2\bar\sigma^2 = 
-\mathbb I_2$. Finally, the third eq.\ of \eqref{eq:4.11} follows from 
\begin{equation}
	{\bf Z}^\dag = iZ_\mu^\dag\gm = i \overline Z_\mu\gm =
	i\gm (\overline Z_\mu)' =i\gm \overline{Z_\mu'} = -{\bf \overline Z'},
\label{eq:37}
\end{equation}
where we notice that $Z_\mu^\dag = \overline Z_\mu$, from the explicit form 
\eqref{eq:piM} of $\pi_\M$. $\qed$
\end{proof}

\begin{prop}
\label{prop:twistfluctweyl}
	For $a=(f,g)$,  $a'=(f',g')$, $b=(v, w)$, $b'=(v', w')$ in $\A$, let
\begin{equation*}
	\omega_\rho := \pi(a,a') \big[ \eth\x\mathbb{I}_2,\;\pi(b,b')\big]_{\rho}
\end{equation*}
be a twisted one-form. Then,
\begin{equation}
\label{eq:51bis}
    \omega_\rho + \J \omega_\rho \J^{-1} = {\bf X}\x {\mathbb I}_2
    +i{\bf Y}\x \g_\F  ,
\end{equation}
where ${\bf X}=-i\gm X_\mu$ and ${\bf Y} =-i\gm Y_\mu$ with
\begin{equation}
  \label{eq:40}
  X_\mu=\pi_\M(f_\mu, f'_\mu), \qquad Y_\mu=\pi_\M(g_\mu, g'_\mu),
\end{equation}
where $f_\mu, f'_\mu$ and $g_\mu, g'_\mu$ denote, respectively, the real and the
imaginary parts of
\begin{equation}
 \label{eq:47}
 z_\mu := f'\dm v + \bar g\dm\bar w', 
 \qquad \text{ and } \qquad z'_\mu =
 f\dm v' + \bar g'\dm\bar w'.
\end{equation}
\end{prop}

\begin{proof}
  We first set the following notation:
  \begin{equation}
  \begin{split}
    \label{eq:38}
    V:=\pi_\M(v, v'), \qquad & V':=\pi_\M(v', v), \\
    W:=\pi_\M(w,w'), \qquad & W':=\pi_\M(w', w).
\end{split}
  \end{equation}
From (\ref{eq:mintwst_X},~\ref{eq:mintwst_Xbis}), we have
\begin{equation} 
\label{5.11}
	\big[ \eth \x \mathbb{I}_2, \; \pi(b,b') \big]_{\rho} 
= \begin{pmatrix}
		[\eth,V]_{\rho} & 0 \\ 0 & [\eth,W']_{\rho} 
	\end{pmatrix},
\end{equation}
so that, for $(a, a')$ as in \eqref{eq:mintwst_X}, and using
(\ref{eq:4.11}) we get
\begin{equation} 
\begin{split}
\label{eq:1form2point}
\omega_\rho 
& := \begin{pmatrix} F & 0 \\  0 & G' \end{pmatrix} \!\!
	\begin{pmatrix} 
		[\eth,V]_{\rho} & 0 \\ 0 & [\eth,W']_{\rho}
	\end{pmatrix} \\[5pt]
& = \begin{pmatrix}
		-i\gm P_{\mu} & 0 \\ 0 & -i\gm Q_{\mu}'  
	\end{pmatrix} 
= \begin{pmatrix}
	{\bf P}& 0 \\ 0 & {\bf Q'} 
	\end{pmatrix},
\end{split}
\end{equation}
with
\begin{equation}
  \label{eq:39}
  P_\mu := F'\dm V, \qquad \text{ and } \qquad Q'_\mu:= G\dm W'.
\end{equation}
The explicit form of the real structure $J$ and its inverse $J^{-1}$, that is,
\begin{equation} 
\label{J}
	J =	\J \x J_F
= \begin{pmatrix}
		0 & \J \\ \J & 0 
	\end{pmatrix}, \qquad J^{-1} = 
	\begin{pmatrix}
		0 & \J^{-1} \\ \J^{-1} & 0 
	\end{pmatrix},
\end{equation}
along with the second relation of \eqref{eq:4.11}, yield
\begin{equation} 
\label{eq;3-5.18}
\begin{split}
	J\omega_\rho J^{-1}
 = \begin{pmatrix}
		\J {\bf Q'} \J^{-1} & 0 \\
		 0 & \J{\bf P'} \J^{-1}
	\end{pmatrix} = 
	\begin{pmatrix}
		\overline{\mathbf{Q}}' & 0 \\ 
		 0 & \overline{\mathbf{P}} 
	\end{pmatrix}.
\end{split}
\end{equation}
Summing up \eqref{eq:1form2point} and \eqref{eq;3-5.18}, one obtains \eqref{eq:51}
\begin{equation}
  \label{eq:49}
  \omega_\rho + \J\omega_\rho \J^{-1} =
    \begin{pmatrix}
      {\bf Z} & 0 \\ 0 & {\bf \overline Z}
    \end{pmatrix},
\end{equation}
with
${\bf Z} := {\bf P} + {\bf \overline Q'} = -i\gm Z_\mu$
and
\begin{equation}
\begin{split}
\label{sac1bis}
	Z_\mu & = P_{\mu} + \overline Q_{\mu}' 
	= {F}'\partial_{\mu}{V} + \overline G\partial_{\mu}\overline W' \\[3pt]
& = \begin{pmatrix}
		( f'\dm v + \bar g\dm \bar w')\mathbb{I}_2 & 0 \\
		0 & (f\dm v' + \bar g'\dm \bar w)\mathbb{I}_2
	\end{pmatrix},
\end{split}
\end{equation}
where the last equality follows from the explicit form \eqref{eq:38} of $V,W'$ and
\eqref{eq:4.10} of $F', G$, respectively. Then, with \eqref{eq:47}, this gives
\begin{equation}
\begin{split}
	Z_\mu & = \pi_\M(z_\mu, z'_\mu) \\
	& = \pi_\M(f_\mu, f'_\mu) + i\pi_\M(g_\mu, g'_\mu) = X_\mu + iY_\mu.
\label{eq:42}
\end{split}
\end{equation}
Similarly, ${\bf \overline Z} = -i\gm\overline Z_\mu$ with $\overline Z_\mu =
X_\mu - iY_\mu$. Hence, \eqref{eq:49} becomes
 \begin{equation}
    \label{eq:51}
    \omega_\rho + \J \omega_\rho \J^{-1} =
    \begin{pmatrix}
    -i\gm (X_\mu + i Y_\mu) &0\\ 
       0 & -i\gm(X_\mu - iY_\mu)
       \end{pmatrix}\!,
        \end{equation}
which is nothing but \eqref{eq:51bis}. $\qed$
\end{proof}

\begin{prop}
\label{prop:twistfluctweylsa}
	The self-adjoint twisted fluctuations of the Dirac operator for the $U(1)$
	gauge theory (\S\ref{subsec:2.4.1}) are parametrized by two real fields $f_\mu, 
	g_\mu \in C^\infty(\M,\mathbb R)$, and are of the form:
\begin{equation}
\label{eq:twstflct}
	\eth_X \x \mathbb{I}_2 \; + \; g_\mu\gm\x \g_\F,
\end{equation}
where $\eth_X$ is the twisted-covariant Dirac operator \eqref{eq:dflucM} of the 
manifold $\M$.
\end{prop}

\begin{proof} 
A generic twisted fluctuation 
\eqref{eq:49}\footnote{Technically, one should add a summation index $i$ and 
redefine it as ${\bf Z} := \textstyle\sum_i{\bf Z}_i$.} is self-adjoint iff 
${\bf Z} = {\bf Z}^\dag$ and ${\bf \overline Z} = {\bf \overline Z}^\dag$.
By \eqref{eq:45}, and the third eq.\ of \eqref{eq:4.11}, both
conditions are equivalent to $\bf Z = -\bf \overline Z'$, that is, 
$$-i\gm \big(Z_{\mu} + \overline Z'_{\mu}\big) =0.$$ 
As discussed in the argument following eq.\ \eqref{eq:gsym}, this is equivalent to 
$Z_\mu = -\overline Z'_\mu$, which using the explicit form \eqref{sac1bis} boils 
down to $z_\mu = -\overline z_\mu'$, that is,
\begin{equation}
\label{eq:44}
	f_\mu = -f'_\mu, \qquad \text{ and } \qquad g_\mu = g'_\mu.
\end{equation}
Then, in \eqref{eq:40} one has the following explicit forms:
\begin{equation}
\begin{split}
\label{eq:34}
	X_\mu & = \pi_\M(f_\mu, -f_\mu)= f_\mu\g^5, \\ 
	Y_\mu & = \pi_\M(g_\mu, g_\mu) = g_\mu {\mathbb I}_4, 
\end{split}
\end{equation}
so that (\ref{eq:51bis}) reads
\begin{equation}
  \label{eq:46}
  \omega_\rho + \J \omega_\rho \J^{-1} = -i\gm
  f_\mu\g^5\x{\mathbb I}_2
  +g_\mu\gm\x\g_\F .
\end{equation}
Hence, the result follows by adding $\eth\x {\mathbb I}_2$ to \eqref{eq:46}. 
$\qed$
\end{proof}

\medskip

Prop.~\ref{prop:twistfluctweylsa} shows that the self-adjointness can directly be
read into the bold notation. Meaning, \eqref{eq:44} shows that ${\bf Z} = {\bf X}
\x \mathbb I_2 + i{\bf Y} \x \g_\F $ is self-adjoint iff ${\bf X'} = -{\bf 
\overline X}$ and ${\bf Y'} = {\bf \overline Y}$, that is, from the third 
eq.\ in \eqref{eq:4.11}, iff ${\bf X}={\bf X}^\dag$ and ${\bf Y}=-{\bf Y}^\dag$.

\subsection{Weyl action in Lorentz signature}
\label{subsec:Weyl}

Here we show that the fermionic action associated to the twisted-covariant operator 
$\eth_X \x\mathbb{I}_2$ (assuming that $g_\mu=0$) yields the Weyl equations in the 
lorentzian signature.

\smallskip

For the $U(1)$ gauge theory, the unitary operator implementing the action of the
twist $\rho$ on the Hilbert space $\HH$ is given by the matrix $\R = \g^0 \x 
\mathbb{I}_2$, which has eigenvalues $\pm1$ and is compatible with the real 
structure $J$, in the sense of (\ref{eq:2.11}), with the sign $\epsilon'''=-1$. 

\smallskip

Similar to Lem.~\ref{lem3.1}, a generic element $\eta$ in the $+1$-eigenspace 
$\cal H_R$ of $\R$ is written as 
\begin{equation}
\label{eq:eta}
	\eta = \xi \x e + \phi \x \overline e, \qquad \text{ with } \quad
	\xi := \begin{pmatrix} \zeta \\ \zeta \end{pmatrix}, \quad
	\phi := \begin{pmatrix} \varphi \\ \varphi \end{pmatrix},
\end{equation} 
where $\xi,\phi \in L^2({\cal M,S})$ are Dirac spinors with corresponding Weyl 
components $\zeta,\varphi$.

\begin{prop}
\label{prop:3.6} 
	Let $\eta$, $\eta' \in \HH_\R$, with $\zeta', \varphi'$ being the Weyl 
	components of the Dirac spinors $\xi', \phi'$ -- as in the decomposition 
	\eqref{eq:eta} of $\eta'$. Then,
\begin{equation*}
	{\frak A}^\rho_{\eth_X\x\mathbb{I}_2}(\eta,\eta') 
	= 2\int_\M\dmu \left[ \bar\zeta^\dag\sigma_2 \left( if_0\mathbb I_2
- \sum_j \sigma_j\partial_j\right)\varphi' + \bar\varphi^\dag\sigma_2 \left(
if_0\mathbb I_2 -\sum_j\sigma_j\partial_j\right)\zeta' \right]\!.
\end{equation*}
\end{prop}

\begin{proof}
	For $\eta \in {\cal H_R}$ \eqref{eq:eta}, recalling
        that $J_\F e = \overline e$ and $J_\F \overline e = e$, one has
\begin{equation*}
\begin{split}
	J\eta & = \J\xi\x\overline e + \J\phi\x e, \\ 
	(\eth_X\x\mathbb{I}_2)\eta' & = \eth_X\xi'\x e + \eth_X\phi'\x\overline e.
\end{split}
\end{equation*}
Then, using Lem.~\ref{lem:bilinearform} with $\epsilon'''=-1$ yields
\begin{equation*} 
\begin{split}
	{\frak A}^\rho_{\eth_X\x\mathbb{I}_2}(\eta,\eta') &
	= -\left\langle J\eta, (\eth_X\x\mathbb{I}_2)\eta' \right\rangle \\
&	= - \left\langle \J\xi, \eth_X\phi' \right\rangle
	- \left\langle \J\phi, \eth_X\xi' \right\rangle \\
& = -\frak A_{\eth_X}(\xi, \phi') - \frak A_{\eth_X}(\phi, \xi') \\
& = \frak A^\rho_{\eth_X}(\xi, \phi') + \frak A^\rho_{\eth_X}(\phi, \xi'),
\end{split}
\end{equation*}
where the inner product in the first line is on $\HH$, the ones in the second line
are on $L^2(\M,S)$, and the second equality is due to \eqref{eq:arho}. Thus, the 
result then follows from Prop.~\ref{prop:actspecmanif}. $\qed$
\end{proof}

\bigskip

The twisted fermionic action $S^f_\rho$ is then obtained substituting $\eta'=\eta$
in the Prop.~\ref{prop:3.6} and promoting $\zeta$ and $\varphi$ to their 
corresponding Gra{\ss}mann variables $\tilde\zeta$ and $\tilde\varphi$, 
respectively. The bilinear form $\frak A^\rho_{\eth_X}$ then becomes symmetric 
when evaluated on these Gra{\ss}mann variables -- as in the proof of 
\cite[Prop.~4.3]{DS}. Hence,

\begin{mdframed}
\begin{equation}
\begin{split}
  \label{eq:57}
  S^f_\rho(\eth_X\x \mathbb I_2) 
& = 2\;\frak A^\rho_{\eth_X}(\tilde\phi,\tilde \xi) 
 = 4\int_\M\dmu \left[ \bar{\tilde\varphi}^\dag\sigma_2 \left( if_0\mathbb I_2 -
\sum_{j=1}^3 \sigma_j\partial_j \right)\tilde\zeta \right].
\end{split}
\end{equation}
\end{mdframed}

\begin{prop}
\label{Prop:Weyl}
Identifying the physical Weyl spinors $\Psi_l, \Psi_r$ as       
\begin{equation}
\Psi_l := \tilde\zeta, \quad \Psi_l^\dag := 
	-i\bar{\tilde\varphi}^\dag\sigma_2\quad \text{ or } \quad
        \Psi_r := \tilde\zeta, \quad \Psi_r^\dag := 
i\bar{\tilde \varphi}^\dag\sigma_2,
\label{eq:48}        
      \end{equation} 
	the lagrangian
	$${\cal L}^f_\rho := \bar{\tilde\varphi}^\dag\sigma_2 
	\left(if_0 - \sum_{j=1}^3\sigma_j\partial_j \right) \tilde\zeta\;$$
  in the fermionic action \eqref{eq:57} describes, for a
non-zero constant function $f_0$, a
plane wave solution of the Weyl equation, with $x^0$ being the time coordinate.
\end{prop}

\begin{proof}
With the first identification in \eqref{eq:48}, ${\cal L}_\rho^f$ coincides with 
the Weyl lagrangian ${\cal L}_M^l$ \eqref{eq:Weyl}, as soon as one imposes
\begin{equation*}
	\partial_0\Psi_l = if_0\Psi_l, \quad
  \text{ i.e. } \quad
	\Psi_l(x_0, x_j) = \Psi_l(x_j)e^{if_0x_0},
\end{equation*}
which is the plane-wave solution \eqref{eq:50} with $f_0, x_0$ being identified
with $E, t$  and $\Psi_l(x_j) = \Psi_0e^{-ip_jx^j}$. 
The second identification in \eqref{eq:48} yields the other Weyl
lagrangian ${\cal L}_M^r$,  imposing 
\begin{equation*}
	\partial_0\Psi_r = -if_0\Psi_r, \quad
\text{ i.e. } \quad\Psi_r(x_0, x_i) = \Psi_r(x_i)e^{-if_0x_0},
\end{equation*}
which is again the plane wave solution \eqref{eq:50}, where one identifies $f_0$ 
with $-E$ and $\Psi_r(x_j) = \Psi_0e^{-ip_jx^j}$. $\qed$
\end{proof}

The above Prop.~\ref{Prop:Weyl} adds weight to the observation made at the end of 
the previous section, right after \eqref{eq:56}. That is, without fluctuation,
the fermionic  action $S^f_\rho(\eth\x\mathbb I_2)$ of a minimally twisted $U(1)$
gauge theory yields the spatial part of the Weyl lagrangian, which is nothing but 
the lagrangian in \eqref{eq:57} with $f_0=0$. For a non-zero constant $f_0$, the 
twisted fluctuation not only brings back a fourth component, but it also allows 
its interpretation as a time direction. This further provides a clear 
interpretation of the zeroth component $f_0$ (of the real field $f_\mu$ that 
parametrizes the twisted fluctuation) as an energy.

\bigskip

These above two examples discussed so far -- that is, the manifold and the $U(1)$
gauge theory -- indicate that the main difference between the non-twisted and the 
twisted fermionic actions does not lie so much in the twisting of the inner 
product to a $\rho$-inner product, but rather in the restriction to different
subspaces, viz.\ $\HH_\R$ instead of $\HH_+$. Indeed, by 
Lem.~\ref{lem:bilinearform}, the twisting of the inner product $\langle \cdot, 
\cdot \rangle$ on $\HH$ to $\langle \cdot, \cdot \rangle_\rho$ solely brings forth 
to a global sign. However, as highlighted in the following remark: it is the 
restriction to $\HH_\R$ instead of $\HH_+$ that explains the change of signature.

\medskip

\begin{rem}
The disappearance of a derivative has no analogue in the non-twisted case, 
i.e.\ for $\psi \in \HH_+$:
\begin{itemize}
\setlength\itemsep{-0.5em}
	\item the usual fermionic action \eqref{Sf} vanishes on a manifold,
    since $\eth \psi \in \HH_-$ while $\J \psi \in \HH_+$;
	\item in case of a $U(1)$ gauge theory, $\HH_+$ is spanned by 
	$\left\{\xi\x e, \phi\x \bar e\right\}$ with
    $\xi = \begin{pmatrix} \zeta \\ 0 \end{pmatrix}$, $\phi=
    \begin{pmatrix} 0 \\ \varphi \end{pmatrix}$. Then 
\begin{equation}
	S^f(\eth\x\mathbb I_2) = 2 \langle \J\tilde\phi, \eth\tilde\xi\rangle 
	= -2\int_\M\dmu \big( \tilde{\bar\varphi}^\dag \sigma^2 \tilde\sigma^\mu
	\dm \tilde\zeta \big).
\end{equation}
    Up to the identification \eqref{eq:48}, the integrand is the euclidean
    version $${\cal L}_E^l := i\Psi_l^\dag \tilde\sigma^\mu\dm \Psi_l$$ of the 
    Weyl lagrangian ${\cal L}_M^l$.
\end{itemize}
\end{rem}

\bigskip

According to the result of \S \ref{sec:3.1.2}, we anticipate the invariance of the
real field $f_\mu$ under a gauge transformation. We check this for the case of the
spectral triple of electrodynamics in section \S \ref{subsec:gaugetransformED}.
We will also discuss the meaning of the other field, viz.\ $g_\mu$, which 
parametrizes the twisted fluctuation in Prop.~\ref{prop:twistfluctweylsa}. As in 
the non-twisted case, this will be identified with the $U(1)$ gauge field of 
electrodynamics. 

\clearpage

\section{Minimal twist of electrodynamics}
\label{sec:electrody}
\label{subsec:mintwised}

In this subsection, we first write down the minimal twist of 
electrodynamics (\S\ref{subsec:2.4.2}) following the recipe prepared in 
\S \ref{sec:2.2}. Then, we compute the twisted fluctuation in
\S\ref{sec:4.3} and
investigate the gauge transformations in \S \ref{subsec:gaugetransformED}.
Finally, we compute the fermionic action in \S \ref{sec:Dirtaceq} and 
derive the Dirac equation in lorentzian signature.


\smallskip

A minimally twisted spectral triple of electrodynamics is obtained by doubling its
algebra ${\A}_{ED} := C^\infty(\M)\x\mathbb{C}^2$ to $\A={\cal A}_{ED}\x\mathbb{C}^2$ along with its flip 
automorphism $\rho$~\eqref{eq:flip}, with the representation $\pi_0$ of $\A$ 
defined by \eqref{repaa'}. Explicitly, the grading $\g$ is given by the tensor
product\footnote{The product has been taken in the following sense:
\begin{equation*}
	A \otimes B =
	\begin{pmatrix}
		b_{11}A & \dots & b_{1n}A \\ 
		\vdots & \ddots & \vdots \\ 
		b_{m1}A & \dots & b_{mn}A 
	\end{pmatrix},
\end{equation*}
where $B := (b_{ij})$ is an $m\times n$~matrix.}
\begin{equation}
\begin{split}
\g = \g^5 \x \g_\F & =
	\begin{pmatrix}
		\mathbb{I}_2 & 0 \\ 0 & -\mathbb{I}_2
	\end{pmatrix} \x 
	\begin{pmatrix}
		1 & 0 & 0 & 0 \\ 0 & -1 & 0 & 0 \\ 0 & 0 & -1 & 0 \\ 0 & 0 & 0 & 1
	\end{pmatrix} \\[5pt]
& =	\begin{pmatrix}
		\mathbb{I}_2 & 0 & 0 & 0 & 0 & 0 & 0 & 0 \\
		0 & -\mathbb{I}_2 & 0 & 0 & 0 & 0 & 0 & 0 \\
		0 & 0 & -\mathbb{I}_2 & 0 & 0 & 0 & 0 & 0 \\
		0 & 0 & 0 & \mathbb{I}_2 & 0 & 0 & 0 & 0 \\
		0 & 0 & 0 & 0 & -\mathbb{I}_2 & 0 & 0 & 0 \\
		0 & 0 & 0 & 0 & 0 & \mathbb{I}_2 & 0 & 0 \\
		0 & 0 & 0 & 0 & 0 & 0 & \mathbb{I}_2 & 0 \\
		0 & 0 & 0 & 0 & 0 & 0 & 0 & -\mathbb{I}_2
	\end{pmatrix},
\end{split}
\end{equation}
so that the projections 
${\frak p}_{\pm} = \frac{1}{2}\big(\mathbb{I}_{16}\pm\g\big)$ on the
eigenspaces $\HH _{\pm}$ of $\HH$ are
\begin{equation}
\begin{split}
	{\frak p}_+ & = \mathsf{diag}
	(\mathbb{I}_2, 0_2, 0_2, \mathbb{I}_2,
	0_2, \mathbb{I}_2, \mathbb{I}_2, 0_2), \\[5pt]
	{\frak p}_- & = \mathsf{diag}
	(0_2, \mathbb{I}_2, \mathbb{I}_2, 0_2,
	\mathbb{I}_2, 0_2, 0_2, \mathbb{I}_2).
\end{split}
\end{equation}
Therefore, for $(a, a') \in \A$, where $a := (f, g)$, $a' := (f', g')$ with 
$f, g, f' , g' \in C^{\infty}(\M)$, one has the representation,
$$\pi(a,a') = {\frak p}_+ \pi_0(a) + {\frak p}_- \pi_0(a'),$$
explicitly given by
\begin{equation}
\begin{split}
\label{5.4}
\pi(a,a') 
& =	\begin{pmatrix}
		f\mathbb{I}_2 & 0 & 0 & 0 & 0 & 0 & 0 & 0 \\
		0 & f'\mathbb{I}_2 & 0 & 0 & 0 & 0 & 0 & 0 \\
		0 & 0 & f'\mathbb{I}_2 & 0 & 0 & 0 & 0 & 0 \\
		0 & 0 & 0 & f\mathbb{I}_2 & 0 & 0 & 0 & 0 \\
		0 & 0 & 0 & 0 & g'\mathbb{I}_2 & 0 & 0 & 0 \\
		0 & 0 & 0 & 0 & 0 & g\mathbb{I}_2 & 0 & 0 \\
		0 & 0 & 0 & 0 & 0 & 0 & g\mathbb{I}_2 & 0 \\
		0 & 0 & 0 & 0 & 0 & 0 & 0 & g'\mathbb{I}_2
	\end{pmatrix} \\[5pt]
& =:\begin{pmatrix}
		F & 0 & 0 & 0 \\ 0 & F' & 0 & 0 \\ 0 & 0 & G' & 0 \\ 0 & 0 & 0 & G
	\end{pmatrix},
\end{split}
\end{equation}
where $F, F', G$ and $G'$ are given as in (\ref{eq:4.10}). The image of 
$(a, a')\in\A$ under the flip $\rho$ is represented by
\begin{equation}
\label{5.4b}
	\pi\big(\rho(a,a')\big) = \pi(a',a) =
	\begin{pmatrix}
		F' & 0 & 0 & 0 \\ 0 & F & 0 & 0 \\ 0 & 0 & G & 0 \\ 0 & 0 & 0 & G'
	\end{pmatrix}.
\end{equation}
The unitary $\R \in \B(\HH)$ implementing the action of $\rho$ on $\HH  = 
L^2(\mathcal{M,S}) \x \mathbb{C}^4$ is 
\begin{equation}
\label{Red}
	\R = \g^0 \x \mathbb{I}_4 =
	\begin{pmatrix}
		0 & \mathbb{I}_2 \\ \mathbb{I}_2 & 0
	\end{pmatrix} \x \mathbb{I}_4,
\end{equation}
which, as before, is compatible with the real structure, in the sense of 
\eqref{eq:2.11}, with $\epsilon''' = -1$.

\subsection{Twisted fluctuation}
\label{sec:4.3}

The twisted commutator $[\D,a]_\rho$ being linear in $\D$, we treat the free part 
$\eth\x\mathbb{I}_4$ and the finite part $\g^5\x\D_\F$ of the Dirac operator $\D$ 
separately.
The results are summarized in Prop.~\ref{prop:twistflucED}.

\subsubsection{The free part}
\label{subsubsec:freepart}

The self-adjoint twisted fluctuations of the free part $\eth\x\mathbb{I}_4$ of the
Dirac operator $\D$ are parametrized by two real fields (Prop.~\ref{prop:4.1}). 
One we relate with the anticipated $X_{\mu}$ field arising from the minimal twist 
of the manifold $\M$ \eqref{eq:dflucM} and the other one with the $U(1)$ gauge 
field $Y_{\mu}$ \eqref{Ymu} of electrodynamics. To arrive there, we need two 
lemmas that we discuss below.

\pagebreak

The following lemma gives a general expression for a twisted 
one-form $\omega_{\rho_\M}$ associated to the free Dirac operator $\eth\x\mathbb{I}_4$.

\begin{lem}
\label{lem:4.2}
	For $a  = (f,g)$, $b=(v,w)$ in $\A_{ED}$ with similar definitions for $a',b'$,
	one has
\begin{equation} 
\label{w_M}
\omega_{\rho_\M} := 
	\pi(a,a')\big[\eth\x\mathbb{I}_4,\;\pi(b,b')\big]_{\rho}
 =	\begin{pmatrix}
		\mathbf{P} & 0 & 0 & 0 \\ 0 & \mathbf{P}' & 0 & 0 \\ 
		0 & 0 & \mathbf{Q}' & 0 \\ 0 & 0 & 0 & \mathbf{Q}
	\end{pmatrix},
\end{equation}
where we use the notation \eqref{eq:notaZ} for
\begin{equation}
\begin{split}
\label{eq:59}
P_{\mu} := F'\partial_{\mu}V, \qquad & P_{\mu}' := F \partial_{\mu}V', \\[5pt]
Q_{\mu} : = G'\partial_{\mu}W, \qquad & Q_{\mu}': = G \partial_{\mu}W' ,
\end{split}
\end{equation}
with $F,F',G,G'$ as in \eqref{eq:4.10}, and $V,V',W,W'$ as in \eqref{eq:38}.
\end{lem}

\begin{proof}
	Using (\ref{5.4}, \ref{5.4b}) written for $(b,b')$, one computes the twisted
	commutator as follows
\begin{equation} 
\label{5.110}
	\big[ \eth \x \mathbb{I}_4, \; \pi(b,b') \big]_{\rho} =:
	\begin{pmatrix}
		[\eth,V]_{\rho} & 0 & 0 & 0 \\ 0 & [\eth,V']_{\rho} & 0 & 0 \\ 
		0 & 0 & [\eth,W']_{\rho} & 0 \\ 0 & 0 & 0 & [\eth,W]_{\rho}
	\end{pmatrix}.
\end{equation}
The result simply follows by multiplying \eqref{5.110} with \eqref{5.4b} and then
using \eqref{eq:4.11}. $\qed$
\end{proof}

\bigskip

The lemma below gives a general expression for the twisted 
fluctuation $\big( \omega_{\rho_\M} + J\omega_{\rho_\M}J^{-1} \big)$ associated to
the free part $\eth\x\mathbb{I}_4$ of the Dirac operator $\D$.

\begin{lem}
	With the same notations as in Lem.~\ref{lem:4.2}, one has 
\begin{equation} 
\label{5.16}
	{\cal Z}  := \omega_{\rho_\M} + J\omega_{\rho_\M}J^{-1}
 = \begin{pmatrix}
		\mathbf{Z} & 0 & 0 & 0 \\ 0 & \mathbf{Z'} & 0 & 0 \\[5pt]
		0 & 0 & \bar{\mathbf{Z}} & 0 \\ 0 & 0 & 0 & \bar{\mathbf{Z'}}
	\end{pmatrix}, 
\end{equation}
denoting
\begin{equation}
\begin{split}
  \label{eq:60}
	\mathbf{Z} := \mathbf{P} + \bar{\mathbf{Q}}', \qquad 
	&	\mathbf{Z'}  :=  \mathbf{P}' + \bar{\mathbf{Q}}, \\[3pt]
		\bar{\mathbf{Z}} : =  \bar{\mathbf{P}} + \mathbf{Q}', \qquad
	&	\bar{\mathbf{Z'}} : =  \bar{\mathbf{P}}' + \mathbf{Q}.
\end{split}
\end{equation}
\end{lem}
\pagebreak
\begin{proof}
Using the explicit form of $J=\J\x J_\F $ with $J_\F $ as in 
\eqref{eq:stEDF}, one gets, from \eqref{w_M} and Lem.~\ref{lem:notatio}, the 
following
\begin{equation*} 
\label{5.18}
\begin{split}
	J\omega_{\rho_\M}J^{-1}
& = \begin{pmatrix}
		0 & 0 & \J & 0 \\ 0 & 0 & 0 & \J \\ \J & 0 & 0 & 0 \\ 0 & \J & 0 & 0
	\end{pmatrix} \!\!\!
	\begin{pmatrix}
		\mathbf{P} & 0 & 0 & 0 \\ 0 & \mathbf{P}' & 0 & 0 \\
		0 & 0 & \mathbf{Q}' & 0 \\ 0 & 0 & 0 & \mathbf{Q}
	\end{pmatrix} \!\!\!
	\begin{pmatrix}
			0 & 0 & \J^{-1} & 0 \\ 0 & 0 & 0 & \J^{-1} \\ \J^{-1} & 0 & 0 & 0 \\ 0 & \J^{-1} & 0 & 0
	\end{pmatrix},\\[3pt]	
& = \begin{pmatrix}
		\J\mathbf{Q}'\J^{-1} & 0 & 0 & 0 \\ 0 & \J\mathbf{Q}\J^{-1} & 0 & 0 \\
		0 & 0 & \J\mathbf{P}\J^{-1} & 0 \\ 0 & 0 & 0 & \J\mathbf{P}'\J^{-1}
	\end{pmatrix} = 
	\begin{pmatrix}
		\bar{\mathbf{Q}}' & 0 & 0 & 0 \\ 0 & \bar{\mathbf{Q}} & 0 & 0 \\
		0 & 0 & \bar{\mathbf{P}} & 0 \\ 0 & 0 & 0 & \bar{\mathbf{P}}'
	\end{pmatrix}.
\end{split}
\end{equation*}
Adding this up with $\omega_{\rho_\M}$ \eqref{w_M}, the result follows. $\qed$
\end{proof}

\medskip

In the following proposition we constrain the form of the twisted
fluctuation $\cal Z$ discussed above, by imposing self-adjointness on it.

\begin{prop}
\label{prop:4.1}
	Any self-adjoint twisted fluctuation \eqref{5.16} of the 
	free Dirac operator ${\eth \x \mathbb{I}_4}$ is of the form
\begin{equation}
\label{5.20a}
	{\cal Z} = {\bf X} \x \mathbb{I}'
	+ i{\bf Y} \x \mathbb{I}'',
\end{equation}
where
\begin{equation*}
		X_{\mu} := f_\mu \g^5 \qquad \text{ and } \qquad 
		Y_{\mu} := g_{\mu}\mathbb{I}_4
\end{equation*}
are parametrized, respectively, by real fields $f_\mu, g_\mu \in C^{\infty}(\M,\mathbb{R})$ and 
$$\;\mathbb{I}' :=
 \mathsf{diag}(1,-1,1,-1), \qquad \mathbb{I}'' := \mathsf{diag}(1,1,-1,-1).$$
\end{prop}
\medskip
\begin{proof}
${\cal Z}$ as given by \eqref{5.16} is self-adjoint iff
	\begin{equation} \label{cnd}
\mathbf{Z} = \mathbf{Z}^\dag, \qquad 
		\mathbf{Z'} = \mathbf{Z'}^\dag, \qquad
		\bar{\mathbf{Z}} = \bar{\mathbf{Z}}^\dag, \qquad 
		\bar{\mathbf{Z'}} = \bar{\mathbf{Z'}}^\dag\!.
	\end{equation}	
From \eqref{eq:45} and the third eq. in \eqref{eq:4.11}, all of these four
conditions are equivalent to $\mathbf{Z} = \mathbf{Z}^\dag$, which is a condition 
that we have already encountered in the proof of Prop.~\ref{prop:twistfluctweyl}, 
and it yields, cf.\ \eqref{eq:34}:
\begin{equation}
\label{eq:zmued}
\begin{split}
	Z_\mu & = X_\mu +iY_\mu \\[3pt] & = 
	\begin{pmatrix} 
		(f_\mu + ig_\mu)\mathbb{I}_2 & 0 \\ 0 & -(f_\mu - ig_\mu)\mathbb{I}_2
	\end{pmatrix},
\end{split}
\end{equation}
where $X_\mu := f_\mu\g^5$ and $Y_\mu := g_\mu\mathbb{I}_4$ with $f_\mu$ and 
$g_\mu$ as defined in \eqref{eq:47}. Going back to \eqref{5.16}, one obtains 
\begin{equation*}
\label{eq:4.31}
\begin{split}
  {\cal Z} & = \begin{pmatrix}
                     \mathbf{Z} & 0 & 0 & 0 \\ 0 & -\bar{\mathbf{Z}} & 0 & 0 \\
                     0 & 0 & \bar{\mathbf{Z}} & 0 \\ 0 & 0 & 0 &
                                                                 -\mathbf{Z}
                   \end{pmatrix} \\[3pt] 
& =                 \begin{pmatrix}
                         -i\g^{\mu}Z_{\mu} & 0 & 0 & 0 \\
                         0 & i\g^{\mu}\overline{Z}_{\mu} & 0 & 0 \\
                         0 & 0 & -i\g^{\mu}\overline{Z}_{\mu} & 0 \\
                         0 & 0 & 0 & i\g^{\mu}Z_{\mu}
                       \end{pmatrix} \\[3pt]
& = \begin{pmatrix}
		-i\gm  (X_\mu +iY_\mu) & 0 & 0 & 0 \\
		0 & i\gm  (X_\mu -iY_\mu) & 0 & 0 \\
		0 & 0 & -i\gm  (X_\mu -iY_\mu) & 0 \\ 
		0 & 0 & 0 & i\gm (X_\mu +iY_\mu)
	\end{pmatrix} \\[3pt]
	& = -i\gm  X_\mu\x
	\begin{pmatrix}
		1 & 0 & 0 & 0 \\ 0 & -1 & 0 & 0 \\
		0 & 0 & 1 & 0 \\ 0 & 0 & 0 & -1
	\end{pmatrix} + i(-i\gm  Y_\mu)\x
	\begin{pmatrix}
		1 & 0 & 0 & 0 \\ 0 & 1 & 0 & 0 \\
		0 & 0 & -1 & 0 \\ 0 & 0 & 0 & -1
	\end{pmatrix}. \qed
\end{split}
\end{equation*}
\end{proof}

\bigskip

We conclude the discussion on the free part with the following 
remark on self-adjointness of the twisted one-form $\omega_{\rho\M}$ vs.\ that 
of the twisted fluctuation $\cal Z$.

\begin{rem}
 Imposing self-adjointness of the twisted one-form
  $\omega_{\rho_\M}$ amounts to the following constraints:
  \begin{equation}
    \label{eq:61}
    {\bf P}^\dag ={\bf P}, \qquad {\bf Q}^\dag ={\bf Q}.
  \end{equation}
 These conditions imply, but are not equivalent to, imposing the self-adjointness 
 of the twisted fluctuation $\cal Z$,
 that is,
  \begin{equation}
    \label{eq:62}
    {\bf Z}^\dag ={\bf Z}.
  \end{equation}
 As discussed right after Lem.~\ref{lem:3.1} for the minimal twist of a manifold, 
 the relevant point is that the stronger condition \eqref{eq:61} does not imply
  that the twisted fluctuation $\cal Z$ be zero. The final form of the 
  twist-fluctuated operator is the same, whether one requires \eqref{eq:61} or 
  \eqref{eq:62}. What differs is the relations among the real fields $f_\mu, 
  g_\mu$ and the components of $(a,a'), (b, b')$ appearing in the definition of 
  the twisted-one form $\omega_{\rho\M}$.
\end{rem}

\subsubsection{The finite part}
\label{subsubsec:finitepart}

In the spectral triple of electrodynamics, the finite part $\g^5\x\D_\F$ of the 
Dirac operator $\D$ \S\ref{subsec:2.4.2} does not fluctuate \cite{DS}, for it commutes with 
the representation $\pi_0$~(\ref{p0}) of the algebra $\A_{ED}$. The same is true 
in case of the minimal twist of electrodynamics also -- as shown in the following
proposition.

\begin{prop}
\label{prop:4.2}
	The finite Dirac operator $\g^5\x \D_\F $ has no twisted fluctuation.
\end{prop}

\begin{proof}
	With the representations (\ref{5.4}, \ref{5.4b}), one calculates that
\begin{equation*}
\begin{split}
& \big[ \g^5 \x \D_\F,\; \pi(a,a') \big]_\rho \\[3pt]
& = (\g^5 \x \D_\F)\,\pi(a,a') - \pi(a',a)\,(\g^5 \x \D_\F) \\[3pt]
& = \begin{pmatrix}
		0 & d\g^5 & 0 & 0 \\ \bar{d}\g^5 & 0 & 0 & 0 \\
		0 & 0 & 0 & \bar{d}\g^5 \\ 0 & 0 & d\g^5 & 0
	\end{pmatrix} \!\!
	\begin{pmatrix}
		F & 0 & 0 & 0 \\ 0 & F' & 0 & 0 \\ 0 & 0 & G' & 0 \\ 0 & 0 & 0 & G
	\end{pmatrix} -
	\begin{pmatrix}
		F' & 0 & 0 & 0 \\ 0 & F & 0 & 0 \\ 0 & 0 & G & 0 \\ 0 & 0 & 0 & G'
	\end{pmatrix} \!\!
	\begin{pmatrix}
		0 & d\g^5 & 0 & 0 \\ \bar{d}\g^5 & 0 & 0 & 0 \\
		0 & 0 & 0 & \bar{d}\g^5 \\ 0 & 0 & d\g^5 & 0
	\end{pmatrix} \\[3pt]
& = \begin{pmatrix}
		0 & d[\g^5,F'] & 0 & 0 \\ \bar{d}[\g^5,F] & 0 & 0 & 0 \\
		0 & 0 & 0 & \bar{d}[\g^5,G] \\ 0 & 0 & d[\g^5,G'] & 0
	\end{pmatrix} = 0,
\end{split}
\end{equation*}
where $F$, $F'$, $G$, $G'$ \eqref{eq:4.10} being diagonal, commute with 
$\g^5$. $\qed$
\end{proof}

\bigskip

The results of \S\ref{sec:4.3} summarize as follows:

\begin{mdframed}
\begin{prop}
\label{prop:twistflucED}
The Dirac
  operator
  ${\cal D} = {\eth\x\mathbb{I}_4} + {\g^5\x {\cal D}_\F }$ of
  electrodynamics (\S\ref{subsec:2.4.2}), under the minimal twist (\ref{5.4}--\ref{Red}), 
  twist-fluctuates to
  \begin{equation}
    \label{eq:4.34}
 {\cal D}_{\cal Z} := {\cal D} + {\cal Z}, \quad\text{ where }\quad  {\cal Z} :=
    {\bf X}\x\mathbb{I}' + i{\bf Y}\x\mathbb{I}''\!,
  \end{equation}
  as given in Prop.~\ref{prop:4.1}. \smallskip
\end{prop}
\end{mdframed}

 The explicit form of {\bf Y} is the same as that of the gauge potential $Y_\mu$ 
 \eqref{U1act} of electrodynamics in the non-twisted case. This is confirmed in 
 the next section, where we show that $\bf Y$ transforms exactly as the $U(1)$ 
 gauge potential of electrodynamics. 

\smallskip

 The $\bf X$ field is similar to that of the minimally twisted manifold. We show 
 below that this field is gauge-invariant and induces a transition of signature 
 from the euclidean to the lorentzian, in the same way as exhibited in 
 \S\ref{subsec:Weyl}. 

\begin{rem}
   Expectedly, substituting $\rho = \mathsf{Id}$, one returns to the
   non-twisted case of electrodynamics (\S\ref{subsec:2.4.2}): the triviality of $\rho$ is
   tantamount to equating~\eqref{5.4} with~\eqref{5.4b}, that is to
   identify the `primed' functions $(f',g',\cdots)$ with their
   `un-primed' partners $(f,g,\cdots)$. Hence, ${\bf Z'}={\bf Z}$.
   
   \smallskip
   
   Imposing the self-adjointness, the third eq.  \eqref{eq:4.11} gives
   ${\bf Z}=-{\bf \overline Z}$. Going back to \eqref{eq:zmued}, this
   yields $f_\mu=0$. Therefore, $X_\mu$ vanishes and only the
   $U(1)$ gauge field $\bf Y$ survives.
 \end{rem}

\subsection{Gauge transformation}
\label{subsec:gaugetransformED}

Along the lines of \S\ref{sec:3.1.2}, here we discuss the transformations of the 
fields parametrizing the twisted fluctuation $\cal Z$. Let $u:=(e^{i\alpha}, 
e^{i\beta})$ and $u':=(e^{i\alpha'}, e^{i\beta'})$ be two unitaries of the algebra
$\A_{ED}$, with $\alpha,\alpha',\beta,\beta'\in C^\infty(\M,\mathbb{R})$. A unitary
of $\A_{ED}\x\mathbb C^2$ is of the form $(u,u')$, with the
representation
\begin{equation}
\begin{split}
\label{eq:piuu'}
	\pi(u,u') & = 
	\begin{pmatrix}
		A & 0 & 0 & 0 \\ 0 & A' & 0 & 0 \\ 0 & 0 & B' & 0 \\ 0 & 0 & 0 & B
	\end{pmatrix}, \\[3pt] \text{and } \qquad \pi(\rho(u,u')) = \pi(u',u) & =
	\begin{pmatrix}
		A' & 0 & 0 & 0 \\ 0 & A & 0 & 0 \\ 0 & 0 & B & 0 \\ 0 & 0 & 0 & B'
	\end{pmatrix},
\end{split}
\end{equation}
where, similar to~\eqref{eq:4.10}, we have denoted
\begin{equation}
\label{rep:A,B}
\begin{split}
&	A := \pi_{\mathcal{M}}(e^{i\alpha},e^{i\alpha'}), \qquad 
	A' := \rho(A) = \pi_{\mathcal{M}}(e^{i\alpha'},e^{i\alpha}),\\
&	B := \pi_{\mathcal{M}}(e^{i\beta},e^{i\beta'}), \qquad \,
	B' := \rho(B) = \pi_{\mathcal{M}}(e^{i\beta'},e^{i\beta}).
\end{split}
\end{equation}

\begin{prop}
	Under a gauge transformation \eqref{eq:DomegaU1} with a
        unitary $(u,u')$~\eqref{eq:piuu'}, the fields $z_\mu, z'_\mu$
        parametrizing the twisted-covariant operator $\D_{\cal Z}$ of
        Prop.~\ref{prop:twistflucED} transform as
	\begin{equation}
		z_\mu \to z_\mu -i\dm\vartheta,
                \qquad z'_\mu \to z'_\mu
                -i\dm\vartheta'
	\end{equation}
for $\vartheta:=\alpha - \beta'$ and $\vartheta'=\alpha'-\beta$ in 
$C^\infty(\M,\mathbb{R})$.
\end{prop}

\begin{proof}
	Since the finite part $\g_\F \x \D_\F $ twist-commutes with
        the algebra, in the law of transformation of the gauge potential
        \eqref{LM2Prop4.2}, it is then enough to consider only the free part $\eth
        \x \mathbb{I}_4$. Thus, $\omega_{\rho\M}$ in \eqref{5.16}
        transforms to
	\begin{equation}
	\label{eq:4.45}
	\begin{split}
\omega_{\rho_\M}^{(u,u')} 
		 & = \rho(u,u') \big( \big[\eth\x\mathbb{I}_4, (u,u')^*\big]_\rho 
		+ \omega_{\rho_\M}(u,u')^* \big) \\
		& = (u',u) \big( \eth\x\mathbb{I}_4 
		+ \omega_{\rho_\M} \big)(u,u')^*,
	\end{split}
	\end{equation}
	where, as in \eqref{Id1}, we made use of 
	$\big[\eth\x\mathbb{I}_4, (u,u')^*\big]_\rho 
		= (\eth\x\mathbb{I}_4)(u,u')^*$.
	With the representation~\eqref{eq:piuu'} of $(u,u')$ and 
	$\omega_{\rho_\M}$ from Lem.~\ref{lem:4.2}, the above
        transformation \eqref{eq:4.45} reads as
	\begin{equation*}
		\begin{pmatrix}
			{\bf P} &0 &0 &0 \\0 & {\bf P'} & 0& 0\\ 
			0&0 & {\bf  Q'} &0 \\ 0& 0&0 & {\bf Q}	
		\end{pmatrix} \to
		\begin{pmatrix}
			A'(\eth + {\bf P})\bar A &0 &0 &0 \\
			0& A(\eth +{\bf  P'})\bar A' &0 &0 \\ 
			0& 0& B(\eth+ {\bf Q'})\bar B' &0 \\ 
			0&0 &0 & B'(\eth + {\bf Q})\bar B
		\end{pmatrix},
	\end{equation*}
where recalling that matrices $A'$ and $B'$ twist-commute with $\gm$ and $A$ 
commutes with $P_\mu$ (and, $B$ with $Q_\mu$), one obtains
\begin{equation}
P_\mu \to P_\mu + A\dm\bar A, \qquad \text{ and } \qquad 
	Q'_\mu \to Q_\mu + B'\dm\bar B'.\label{eq:63}
              \end{equation}
implying, for $Z_\mu =  P_\mu+Q'_\mu$, that
	\begin{equation}
\label{eq:transforz}
		Z_\mu \quad \to \quad
				Z_\mu + \left(A\dm\bar A + B'\dm\bar B'\right)\!,
	\end{equation}
With the representations \eqref{eq:zmued} of $Z_\mu$
(recalling that $z_\mu=f_\mu + ig_\mu$), and  \eqref{rep:A,B}
	of $A, B$, the transformation \eqref{eq:transforz} reads
	\begin{equation*}
		\begin{pmatrix}
			z_\mu\mathbb{I}_2 & 0 \\ 0 &  z'_\mu\mathbb{I}_2
		\end{pmatrix} \longrightarrow
		\begin{pmatrix}
			\left( z_\mu-i\dm\vartheta \right) \mathbb{I}_2 & 0 \\ 
			0 &  z'_\mu -i\dm\vartheta' \mathbb{I}_2
		\end{pmatrix}. \qed
	\end{equation*} 
\end{proof}

By imposing that both $\cal Z$ and its gauge transform are self-adjoint, that is, 
by Lem.~\ref{lem:notatio}: $z'_\mu = -\bar z_\mu$ and $z'_\mu -
        i\dm\vartheta' =- \overline{z_\mu
          -i\dm\vartheta}$, one is forced to identify
        $\vartheta'=\vartheta$. 
	Then, the law of transformation of $z_\mu$ in terms of
	its real and imaginary components reads:
	\begin{equation}
	\label{eq:4.51}
		f_\mu + ig_\mu \quad \longrightarrow \quad
		f_\mu + i(g_\mu - \dm\vartheta),
	\end{equation}
which implies for the fields $X_\mu = f_\mu\g^5$ and $Y_\mu = 
	g_\mu\mathbb{I}_4$ of Prop.~\ref{prop:4.1} that $X_\mu$ stays invariant, 
	while $Y_\mu$ undergoes a nontrivial transformation, induced by
	\begin{equation}
	\label{eq:4.52}
		g_\mu \to g_\mu - \dm\vartheta, \qquad 
		\vartheta \in C^\infty(\M,\mathbb{R}).
	\end{equation}
In the light of~\eqref{U1act}, this identifies $g_\mu$ as the $U(1)$ gauge field 
	of electrodynamics. 
	
\subsection{Dirac action in Lorentz signature}
\label{sec:Dirtaceq}

To calculate the twisted fermionic action, we first  identify
the eigenvectors of the unitary $\R$ implementing the twist.
 
\begin{lem}
	Any $\eta$ in the $+1$-eigenspace 
	$\mathcal{H_R}$~\eqref{eq:2.16} of the unitary $\R$~\eqref{Red}
	is of the form
\begin{equation}
\label{6.3}
	\eta = \Phi_1 \x e_l + \Phi_2 \x e_r 
	+ \Phi_3 \x \overline{e_l} + \Phi_4 \x \overline{e_r},
\end{equation}
with $\Phi_{k=1,\ldots,4} := \begin{pmatrix} \varphi_k \\ \varphi_k \end{pmatrix}$
	where $\Phi_k\in L^2(\M, {\cal S})$ are Dirac spinors with Weyl
        components $\varphi_k$, while  
	$\{e_l, e_r, \overline{e_l}, \overline{e_r}\}$ denotes the orthonormal basis 
	for the finite-dimensional Hilbert space $\mathcal{H_F} = \mathbb{C}^4$.
\end{lem}

\begin{proof}
	$\R$
	has eigenvalues $\pm1$ and its eigenvectors corresponding to the eigenvalue 
	$+1$ are:
\begin{equation*}
\begin{split}
	\varepsilon_1 = \upsilon_1 \x e_l, \qquad
	\varepsilon_2 = \upsilon_2 \x e_l, \qquad
&	\varepsilon_3 = \upsilon_1 \x e_r, \qquad
	\varepsilon_4 =	\upsilon_2 \x e_r,	\\[5pt]
	\varepsilon_5 = \upsilon_1 \x \overline{e_l}, \qquad
	\varepsilon_6 = \upsilon_2 \x \overline{e_l}, \qquad
&	\varepsilon_7 = \upsilon_1 \x \overline{e_r}, \qquad
	\varepsilon_8 = \upsilon_2 \x \overline{e_l},
\end{split}
\end{equation*}
	where $$\upsilon_1 := 
	\begin{pmatrix} 1 \\ 0 \end{pmatrix} \x 
	\begin{pmatrix} 1 \\ 1 \end{pmatrix} \qquad \text{ and } \qquad
	\upsilon_2 := 
	\begin{pmatrix} 0 \\ 1 \end{pmatrix} \x	
	\begin{pmatrix} 1 \\ 1 \end{pmatrix}$$ denote 
	the eigenvectors of $\g^0$, as in Lem.~\ref{lem3.1}. Therefore, we have
\begin{equation*}
\begin{split}
	\eta \enskip  & = \enskip \textstyle\sum_{j = 1}^8 \phi_j \varepsilon_j \\[3pt]
       & = \enskip
	(\phi_1 \upsilon_1 + \phi_2 \upsilon_2) \x e_l +
 	(\phi_3 \upsilon_1 + \phi_4 \upsilon_2) \x e_r +
	(\phi_5 \upsilon_1 + \phi_6 \upsilon_2) \x \overline{e_l} +
	(\phi_7 \upsilon_1 + \phi_8 \upsilon_2) \x \overline{e_r} \\[3pt] 
	& = \enskip \Phi_1 \x e_l + \Phi_2 \x e_r + 
	\Phi_3 \x \overline{e_l} + \Phi_4 \x \overline{e_r}, \end{split}
\end{equation*} 
denoting $$\Phi_k := 
	\begin{pmatrix} \varphi_k \\ \varphi_k \end{pmatrix} \text{ with }
	\varphi_1 := \begin{pmatrix} \phi_1 \\ \phi_2 \end{pmatrix},
	\varphi_2 := \begin{pmatrix} \phi_3 \\ \phi_4 \end{pmatrix},
	\varphi_3 := \begin{pmatrix} \phi_5 \\ \phi_6 \end{pmatrix},
	\varphi_4 := \begin{pmatrix} \phi_7 \\ \phi_8 \end{pmatrix}. \qed$$
\end{proof}

The twisted-covariant Dirac operator \eqref{eq:4.34} contains
two extra terms that were not present in the $U(1)$ gauge theory: 
$\g^5\x {\cal D}_\F$ and the term in $\bf Y$. The following lemma is
useful to compute the
contribution of these two terms to the fermionic action.

\begin{lem}
\label{lem:4.6}
	For Dirac spinors $\phi :=
	\begin{pmatrix} \varphi \\ \varphi \end{pmatrix}$ and $\xi :=
	\begin{pmatrix} \zeta \\ \zeta\end{pmatrix}$ in
        $L^2(\mathcal{M,S})$, one has
\begin{equation}
\begin{split}
\label{6.4.2}
	\langle \J \phi, i{\bf Y} \xi \rangle
& = 2i\int_\M \dmu\; \big[
	\bar\varphi^\dag \sigma_2 
	\big(\textstyle\sum_{j=1}^3\sigma_j g_j\big)\zeta \big], \\[3pt]
\langle \J \phi, \g^5 \xi \rangle
& = -2 \int_\M \dmu \; \big(
	\bar\varphi^\dag \sigma_2 \zeta\big).
\end{split}
\end{equation}
\end{lem}

\begin{proof}
Using \eqref{5.20a} for $Y_\mu$ and \eqref{EDirac} for
the Dirac matrices, one gets
\begin{equation*}
\begin{split}
	i {\bf Y} \phi
& = \gm Y_\mu \begin{pmatrix} \varphi \\ \varphi \end{pmatrix}
  = \begin{pmatrix} 
  		0 & \sigma^\mu \\ \tilde\sigma^\mu & 0 
	\end{pmatrix} \!\!
	\begin{pmatrix} 
		g_\mu\mathbb{I}_2 & 0 \\ 0 & g_\mu\mathbb{I}_2 
	\end{pmatrix} \!\!
  	\begin{pmatrix} \varphi \\ \varphi \end{pmatrix}
  = \begin{pmatrix}
  		g_\mu\sigma^\mu\varphi \\ g_\mu\tilde\sigma^\mu\varphi
	\end{pmatrix}.
\end{split}
\end{equation*}
Along with \eqref{eq:JPhi}, recalling that $\sigma^{2\dag} =
i\sigma_2$ and $\tilde\sigma^{2\dag} = -i\sigma_2$, yields
\begin{equation*}
 \begin{split}
 	(\J \phi)^\dag (i{\bf Y} \xi) \quad
& = \quad -i\begin{pmatrix}
 		\tilde\sigma^2 \bar \varphi \\ \sigma^2 \bar\varphi 
 	\end{pmatrix}^{\!\!\dag} \!\!
 	\begin{pmatrix} 
 		g_\mu \sigma^\mu \zeta\\ g_\mu \tilde\sigma^\mu \zeta 
 	\end{pmatrix} \quad
   = \quad	-i \bar\varphi^\dag \left( 
   \tilde\sigma^{2\dag} \sigma^\mu+\sigma^{2\dag} \tilde\sigma^\mu\right) g_\mu\zeta \\
 &    = \quad	\bar\varphi^\dag \sigma_2
     	\big(-\sigma^\mu + \tilde\sigma^\mu\big) g_\mu \zeta \quad
     = \quad 2i\bar\varphi^\dag \sigma_2 \left(\textstyle\sum_{j=1}^3\sigma_j g_j\right)\zeta,
 \end{split}
 \end{equation*}
where we used \eqref{eq:sumsigma} and obtained the first equation of
\eqref{6.4.2}. The second one follows from
\begin{equation*}
	(\J \phi)^\dag (\g^5 \xi)
 = -i\begin{pmatrix}
     \tilde\sigma^2 \bar\varphi \\ \sigma^2 \bar\varphi
   \end{pmatrix}^{\!\!\dag} \!\!
	\begin{pmatrix} \zeta \\ -\zeta \end{pmatrix}
 = - \bar\varphi^\dag \sigma_2 \zeta 
	- \bar\varphi^\dag \sigma_2 \zeta
  = -2 \bar\varphi^\dag \sigma_2 \zeta. \qed
\end{equation*}
\end{proof}

\begin{prop}
\label{prop:eq-Dirac}
The fermionic action of the minimal twist of $\M \times\F_{ED}$ is
\begin{equation}
\begin{split}
S^f_\rho(\D_{\cal Z}) \quad = \quad & {\frak A}^\rho_{\D_{\cal Z}}(\tilde\Phi,\tilde\Phi) 
\enskip = \enskip 4\int_\M \dmu \; {\cal L}_\rho^f,\\[5pt]
\label{eq:lagDir1}
\text{where } \qquad {\cal L}_\rho^f \quad := \quad 
&	{\bar{\tilde\varphi}}_1^\dag \sigma_2 \Big( if_0 - \sum_j\sigma_j\frak D_j \Big) \tilde\varphi_3 
 -	{\bar{\tilde{\varphi}}}_2^\dag \sigma_2 \Big( if_0 + \sum_j\sigma_j\frak D_j \Big) \tilde\varphi_4 \\
& +	\left( \bar d \bar{\tilde{\varphi}}_1^\dag \sigma_2 \tilde\varphi_4 + 
		d \bar{\tilde{\varphi}}_2^\dag \sigma_2 \tilde\varphi_3 \right),
\end{split}
\end{equation}
with
${\frak D}_\mu := \dm - ig_\mu$ being
the covariant derivative associated to the electromagnetic four-potential $g_\mu$~\eqref{eq:4.52}.
Further, identifying the physical spinors as 
\begin{equation}
\label{eq:spinors}	\Psi = \begin{pmatrix} \Psi_l \\ \Psi_r \end{pmatrix} 
	:= \begin{pmatrix} \tilde\varphi_3 \\ \tilde\varphi_4 \end{pmatrix}\!, \qquad \Psi^\dag 
	= \begin{pmatrix} \Psi_l^\dag, & \Psi_r^\dag \end{pmatrix} 
:= \begin{pmatrix} 
		-i\bar{\tilde{\varphi}}_1^\dag\sigma_2, & i\bar{\tilde\varphi}_2^\dag\sigma_2 
	\end{pmatrix},
\end{equation}
the lagrangian \eqref{eq:lagDir1} describes a plane-wave
solution of the Dirac equation, in lorentzian signature, and with the
temporal gauge. 
\end{prop}

\begin{proof}
	Let  ${\frak A}^\rho_{{\cal D}_{\cal Z}}$ be
	 the antisymmetric bilinear form~\eqref{Sfrho}
         defined by the twisted-covariant Dirac operator 
	\eqref{eq:4.34}: $$\D_{\cal Z} = \eth\x\mathbb{I}_4 + {\bf X}\x\mathbb{I}' 
	+i{\bf Y}\x\mathbb{I}'' + \g^5\x \D_\F,$$ which breaks down into the following
	four terms:
\begin{equation}
\label{6.8}
	{\frak A}^\rho_{\D_{\cal Z}} = 
	{\frak A}^\rho_{\eth \x \mathbb{I}_4} + 
	{\frak A}^\rho_{{\bf X} \x \mathbb{I}'} +
	{\frak A}^\rho_{i{\bf Y} \x \mathbb{I}''} + 
	{\frak A}^\rho_{\g^5 \x \D_\F }.
\end{equation} 
For $\Phi,\xi \in\cal H_R$ with $\Phi$ as in \eqref{6.3} and $\Xi$ with
        components $\xi_i=
          \begin{pmatrix}
            \zeta_i\\ \zeta_i
          \end{pmatrix} \in L^2(\M,\cal S)$, one has
\begin{equation}
\begin{split}
	J \Phi \quad
& = \quad \J \phi_1 \x \overline{e_l} + 
	\J \phi_2 \x \overline{e_r} + 
	\J \phi_3 \x e_l +
	\J \phi_4 \x e_r, \\[5pt]
	(\eth \x \mathbb{I}_4) \Xi \quad
& = \quad \eth \xi_1 \x e_l + \eth \xi_2 \x e_r
	+ \eth \xi_3 \x \overline{e_l} 
	+ \eth \xi_4 \x \overline{e_r}, \\[5pt]
	({\bf X} \x \mathbb{I}') \Xi \quad
& = \quad {\bf X} \xi_1 \x e_l - {\bf X} \xi_2 \x e_r
	+ {\bf X} \xi_3 \x \overline{e_l} 
	- {\bf X} \xi_4 \x \overline{e_r}, \\[5pt]
	(i{\bf Y} \x \mathbb{I}'') \Xi \quad
& = \quad i{\bf Y} \xi_1 \x e_l + i{\bf Y} \xi_2 \x e_r
	- i{\bf Y} \xi_3 \x \overline{e_l} 
	- i{\bf Y} \xi_4 \x \overline{e_r}, \\[5pt]
	(\g^5 \x \D_\F ) \Xi \quad
& = \quad \g^5 \zeta_1 \x \bar{d}e_r + \g^5 \zeta_2 \x de_l
	+ \g^5 \zeta_3 \x d\overline{e_r} 
	+ \g^5 \zeta_4 \x \bar{d}\overline{e_l}
\end{split}
\end{equation}
where the first and the last eqs.\ come from the explicit forms \eqref{eq:stEDF}
of $J_\F $ and $\D_\F$, respectively, while the third and the fourth ones follow
from the explicit form \eqref{5.20a} of $\bf X$ and $\bf Y$, respectively.
These eqs.\ allow to reduce each of the four terms in \eqref{6.8}
to a bilinear form on $L^2(\M, \cal S)$ rather than on the tensor product
$L^2(\M,{\cal S})\x \mathbb C^4$. More precisely, omitting the summation symbol on
the index $j$ and recalling Lem.~\ref{lem:bilinearform} with $\epsilon'''=-1$, 
one computes as below:
\begin{align} 
\nonumber
{\frak A}^\rho_{\eth \x \mathbb{I}_4}(\Phi,\Xi) \quad
& = \quad - {\frak A}_{\eth \x \mathbb{I}_4}(\Phi,\Xi) \quad
= \quad - \langle J\Phi, (\eth \x \mathbb{I}_4)\Xi \rangle \\[3pt]
\nonumber & = \quad
-	\langle \J \phi_1, \eth \xi_3 \rangle 
-	\langle \J \phi_2, \eth \xi_4 \rangle 
-	\langle \J \phi_3, \eth \xi_1 \rangle 
-	\langle \J \phi_4, \eth \xi_2 \rangle \\[3pt]
\label{eq:bilinear1}
& = \quad -\frak A_\eth(\phi_1, \xi_3 )  
-	\frak A_\eth(\phi_2, \xi_4) 
-	\frak A_\eth(\phi_3, \xi_1) 
-	\frak A_\eth(\phi_4, \xi_2 ); \\[5pt]
\nonumber	
{\frak A}^\rho_{{\bf X} \x \mathbb{I}'}(\Phi,\Xi) \quad 
& = \quad -{\frak A}_{{\bf X} \x \mathbb{I}'}(\Phi,\Xi) \quad
 = \quad -\langle J\Phi, ({\bf X} \x \mathbb{I}')\Xi \rangle \\[3pt] 
\nonumber
& = \quad -\langle \J \phi_1, {\bf X} \zeta_3 \rangle +
	\langle \J \phi_2, {\bf X} \zeta_4 \rangle -
	\langle \J \phi_3, {\bf X} \zeta_1 \rangle +
	\langle \J \phi_4, {\bf X} \zeta_2 \rangle \\[3pt]
\label{eq:bilinear2}
& = \quad -\frak A_{\bf X}(\phi_1, \zeta_3) 
	+ \frak A_{\bf X}(\phi_2, \zeta_4) -
	\frak A_{\bf X}(\phi_3, \zeta_1) +
	\frak A_{\bf X}(\phi_4,\zeta_2); \\[5pt]
\nonumber	
{\frak A}^\rho_{i{\bf Y} \x \mathbb{I}''}(\Phi,\Xi) \quad
& = \quad -{\frak A}_{i{\bf Y} \x \mathbb{I}''}(\Phi,\Xi) \quad
= \quad -\langle J\Phi, (i{\bf Y} \x \mathbb{I}'')\xi \rangle \\[3pt]
  \nonumber & = \quad
	\langle \J \phi_1, i{\bf Y} \xi_3 \rangle +
	\langle \J \phi_2, i{\bf Y} \xi_4 \rangle -
	\langle \J \phi_3, i{\bf Y} \xi_1 \rangle -
	\langle \J \phi_4, i{\bf Y} \xi_2 \rangle \\[3pt]
\label{eq:bilinear3}
& = \quad
	\frak A_{\bf Y} (\phi_1, \xi_3) +
	\frak A_{\bf Y}(\phi_2,\xi_4) -
	\frak A_{\bf Y} (\phi_3, \xi_1)- 
	\frak A_{\bf Y}(\phi_4,\xi_2);
  \\[5pt]
\nonumber
	{\frak A}^\rho_{\g^5 \x \D_\F }(\Phi,\Xi) \quad 
& = \quad -{\frak A}_{\g^5 \x \D_\F }(\Phi,\Xi) \quad 
= \quad -\langle J\Phi, (\g^5 \x \D_\F )\Xi \rangle \\[3pt] 
\nonumber
& = \quad -\bar d\langle \J \phi_1, \g^5 \xi_4 \rangle 
	- d\langle \J \phi_2, \g^5 \xi_3 \rangle -
	d\langle \J \phi_3, \g^5 \xi_2 \rangle -
	\bar d\langle \J \phi_4, \g^5 \xi_1 \rangle  \\[3pt]
\label{eq:bilinear4}
 & = \quad
	-\bar d\,\frak A_{\g^5}(\phi_1, \xi_4) -
	d\,\frak A_{\g^5}(\phi_2, \xi_3) -
	d\,\frak A_{\g^5}(\phi_3, \xi_2) -
	\bar d\,\frak A_{\g^5}(\phi_4, \xi_1 ).
\end{align}
Substituting $\Xi=\Phi$, and then promoting the spinor $\Phi$
to a Gra{\ss}mann spinor $\tilde\Phi$, the sum of eqs.\
\eqref{eq:bilinear1}, \eqref{eq:bilinear2}, and \eqref{eq:bilinear4} is
\begin{equation}
\begin{split}
-2\,\frak A_\eth(\tilde \phi_1, \tilde \phi_3 )  -2\,\frak A_{\eth}(\tilde \phi_2, \tilde \phi_4 ) 
-2\,\frak A_{\bf X}(\tilde \phi_1, \tilde \phi_3 )+2\,\frak A_{\bf X}(\tilde
\phi_2, \tilde \phi_4)\\[5pt]
 - 	2\bar d\,\frak A_{\g^5}(\phi_1, \phi_4) -
	2d\,\frak A_{\g^5}(\phi_2, \phi_3);
\label{eq:64}
\end{split}
\end{equation}
where we have used the fact that the bilinear forms $\frak A_\eth$, 
$\frak A_{\bf X}$, and $\frak A_{\g^5}$ are antisymmetric on vectors (by 
Lem.~\ref{lemma:antisymm}, since $\eth$, ${\bf X}$, and $\g^5$ are all commuting
with $\J$ in $KO$-dim.\ $4$), and so they are symmetric when evaluated on the 
corresponding Gra{\ss}mann variables. On the other hand, \eqref{eq:bilinear3} is 
symmetric on vectors (since in $KO$-dim.\ $4$: $i{\bf Y}\J = g_\mu \gm\J = -\J g_\mu\gm 
= -\J i{\bf Y}$), while antisymmetric in Gra{\ss}mann variables. Thus,
it yields
\begin{equation}
  \label{eq:19}
  	2\frak A_{\bf Y} (\phi_1, \phi_3) +
	2\,\frak A_{\bf Y}(\phi_2,\phi_4).
\end{equation}
The lagrangian \eqref{eq:lagDir1} follows after substituting all the bilinear 
forms in (\ref{eq:64}, \ref{eq:19}) with their explicit expressions given
in (\ref{derniere1}, \ref{derniere2}) and Lem.~\ref{lem:4.6}.

Upon the identification \eqref{eq:spinors}, one finds that ${\cal
L}^f_\rho$  coincides with the Dirac lagrangian
	in lorentzian signature \eqref{L_M} (with the covariant
        derivative $\frak D_\mu$ to take into account the coupling
        with the electromagnetic field, but in the temporal
        Weyl gauge $\frak D_0=\partial _0$)
\begin{equation}
\label{eq:lagDir2}	
	{\cal L}_M 
	= i\Psi_l^\dag \big(\frak D_0 - \sigma_j\frak D_j \big)\Psi_l
	+ i\Psi_r^\dag \big(\frak D_0 + \sigma_j\partial_j \big)\Psi_r
	- m \big( \Psi_l^\dag\Psi_r + \Psi_r^\dag\Psi_l \big),
\end{equation}
        as soon as one imposes that
	$\partial_0\Psi = if_0\Psi$, i.e.\
\begin{equation}
\Psi(x_0, x_j) = \Psi(x_j) e^{if_0x_0}.
\label{eq:2}
\end{equation}

The mass terms also match up correctly if one imposes the parameter 
$d\in\mathbb{C}$ to be purely imaginary as $d :=- im$.
	This is in agreement with the non-twisted
	electrodynamics, cf.\ \cite[Rem.~4.4]{DS}.  $\qed$
\end{proof}

\bigskip

The above Prop.~\ref{prop:eq-Dirac} extends the
analysis done for the Weyl equation, in \S\ref{subsec:Weyl}, to the Dirac equation.
It confirms the interpretation of the zeroth component of the real field $f_\mu$, 
arising in the twisted fluctuation, as an energy. It also shows that the other 
field $g_\mu$ is well-identified with the electromagnetic gauge potential, as in the 
non-twisted case.   

\smallskip

But this does not say anything about the other components $f_i$ for
$i=1,2,3$ since they do not appear in the lagrangian \eqref{eq:lagDir1}.
It is tempting to identify them with the momenta. This, in fact, makes
sense if one implements a Lorentz transformation, as discussed in the next
section.

\smallskip

Another motivation to study the action of the Lorentz transformations on
the twisted fermionic action is that the temporal Weyl
gauge we ended with, is not Lorentz invariant. One must check
whether the interpretation of the twisted fermionic action provided by
Prop.~\ref{prop:eq-Dirac} is robust enough to survive Lorentz transformations. 


\clearpage
\newpage 
	\chapter{Open Questions}
\label{chap:5}

In this chapter, we touch upon some open issues that arise as a result of this 
thesis. We show in \S\ref{sec:Lorentzinv} that the $\rho$-inner product 
and, hence, the fermionic action associated to a minimally twisted manifold is 
invariant under Lorentz boosts. That being said, the origin of Lorentz 
transformations (or, equally, the Lorentz group) within the context of (twisted) 
noncommutative geometry is not yet fully understood.

\smallskip

In \S\ref{sec:5.2}, we work out the squared twisted-covariant Dirac 
operator $\DX^2$ for the minimally twisted manifold with curvature. It is the 
first step towards writing a heat-kernel expansion for spectral action associated
to this twisted spectral triple. Here, we give the explicit expression for the 
endomorphism term that accounts for the potential terms in the spectral action.
This formula gives the impression that there is a coupling between the curvature 
and the $X_\mu$ field that appears in the non-vanishing twisted fluctuation $\DX$ of the 
Dirac operator $\eth$.

\smallskip

These will be the subject of future works and a full exploration into these lines 
will appear elsewhere.

\section{Lorentz invariance of fermionic action}
\label{sec:Lorentzinv}

A Lorentz boost $S[\Lambda]$ in the Dirac spinor representation 
$\left( \frac{1}{2}, 0 \right) \+ \left( 0, \frac{1}{2} \right)$
is given by
\begin{equation}
\label{eq:slambda}
	S[\Lambda] = \begin{pmatrix} \Lambda_+ & 0_2 \\ 0_2 & \Lambda_- \end{pmatrix} 
	\qquad \text{ with } \quad 
	\Lambda_\pm = \exp(\pm\frac{\mathbf{b}\cdot\boldsymbol{\sigma}}{2}),
\end{equation}
where $\boldsymbol{\sigma} := (\sigma_1, \sigma_2, \sigma_3)$ is the
Pauli vector and $\mathbf{b}\in \mathbb R^3$ is the boost parameter. 
Under such a boost, a lorentzian spinor $\psi_M$ and the lorentzian Dirac operator 
$\eth_M$ transform as 
\begin{equation}
\label{eq:boost}
\begin{split}
	\psi_M & \to S[\Lambda]\psi_M,\\[5pt]
	\eth_M & \to S[\Lambda]\,\eth_M \,S[\Lambda]^{-1}.
\end{split}
\end{equation}
We define the action of a boost on the minimal twist $(C^\infty(\M)\otimes
\mathbb C^2,\, L^2(\M, {\cal S}),\, \eth)$ of a closed euclidean manifold $\M$
of dimension $4$, as follows
\begin{equation}
  \label{eq:67}
\begin{split}
  \psi \; &\to \; \psi_\Lambda = S[\Lambda]\psi, \qquad \forall \psi\in L^2(\M,{\cal S}), \\[5pt]
\eth \; &\to \; \eth_\Lambda := \rho(S[\Lambda]) \;\eth \; S[\Lambda],
\end{split}
\end{equation}
where $\rho(S[\Lambda]) = \R S[\Lambda]\R^\dag$ for $\R=\g^0$ given in 
\eqref{eq:definR}, i.e.
\begin{equation}
\label{3.7}
	\rho(S[\Lambda]) 
	=\g^0\begin{pmatrix} \Lambda_+ & 0_2 \\ 0_2 & \Lambda_- \end{pmatrix} \g^0
	= \begin{pmatrix} \Lambda_- & 0_2 \\ 0_2 & \Lambda_+ \end{pmatrix} 
	= S[\Lambda]^{-1}.
\end{equation}

The boost components $\Lambda_\pm$ can be decomposed into their
`even' and `odd' parts as shown in the following lemma.

\smallskip

\begin{lem}
	For the Lorentz boost components \eqref{eq:slambda}:
\begin{equation}
\label{Lambda}
	\Lambda_\pm := \exp(\pm{\bf a}\cdot\boldsymbol\sigma) \qquad \text{ with } \quad
	{\bf a} := (b/2){\bf n},
\end{equation}
	where $(b/2)$ is the rapidity and $\bf n$ is the direction of the boost,
  one has
  \begin{equation}
    \label{eq:68}
    \Lambda_\pm = \Lambda_e \pm \Lambda_o \qquad \text{with} \quad
\begin{cases}
	\Lambda_e & := \left(\cosh |{\bf a}|\right) \mathbb I_2 \\
    \Lambda_o & := \left(\sinh |{\bf a}|\right){\bf n}\cdot{\boldsymbol \sigma}
\end{cases}.
  \end{equation}
\end{lem}

\begin{proof}
Using $\{\sigma_i,\sigma_j\} = 2\delta_{ij}\mathbb{I}_2$, one gets
\begin{equation*}
\begin{split}
      (\pm{\bf a}\cdot\boldsymbol\sigma)^2 \enskip
      & = \enskip (a_1\sigma_1 + a_2\sigma_2 + a_3\sigma_3)^2 \\[2pt]
     & = \enskip (a_1^2 + a_2^2 + a_3^2)\,\mathbb{I}_2 \enskip
     = \enskip |{\bf a}|^2\,\mathbb{I}_2.
    \end{split}
  \end{equation*}
  Collecting the terms with even
  and odd powers in the expansion of~$\exp(\pm\bf a\cdot\boldsymbol\sigma)$, 
  one obtains
\begin{equation*}
\begin{split}
	\Lambda_\pm \enskip
& = \enskip \sum_{n=0}^\infty \frac{(\pm{\bf a}\cdot\boldsymbol\sigma)^n}{n!} \\[2pt]
& = \enskip \sum_{k=0}^\infty \frac{(\pm{\bf a}\cdot\boldsymbol\sigma)^{2k}}{(2k)!} \;
+ \; \sum_{k=0}^\infty \frac{(\pm{\bf a}\cdot\boldsymbol\sigma)^{2k+1}}{(2k+1)!} \\[2pt]
& = \enskip \sum_{k=0}^\infty \frac{|{\bf a}|^{2k}}{(2k)!}\;\mathbb{I}_2
\; \pm \; \sum_{k=0}^\infty \frac{|{\bf a}|^{2k}}{(2k+1)!}\;({\bf a}\cdot\boldsymbol\sigma) \\[2pt]
& = \enskip \sum_{k=0}^\infty \frac{|{\bf a}|^{2k}}{(2k)!}\;\mathbb{I}_2
\; \pm \; \sum_{k=0}^\infty \frac{|{\bf a}|^{2k+1}}{(2k+1)!}\;({\bf n}\cdot\boldsymbol\sigma) \\[2pt]
& = \enskip \left(\cosh|{\bf a}|\right)\mathbb I_2 
\;\pm \; \left(\sinh|{\bf a}|\right){\bf n}\cdot{\boldsymbol \sigma} \\[2pt]
&  =: \enskip \Lambda_e \;\pm\; \Lambda_o. \qed
      \label{Lambda_decomp}
    \end{split}
  \end{equation*}
\end{proof}

The above decomposition \eqref{eq:68} comes very handy in proving an important property of
the real structure $\J$ and the Lorentz boost $S[\Lambda]$ given below.

\bigskip

\begin{prop}
\label{prop:jboost}
	The Lorentz boosts twist-commute with the real structure $\J$ \eqref{eqn:3.9} 
	of a four-dimensional riemannian manifold, i.e.
\begin{equation}	
{\cal J}S[\Lambda]- \rho(S[\Lambda]) {\cal J}=0.
\end{equation}
\end{prop}	
		
\begin{proof}
	For $\J = \begin{pmatrix} 
		-\sigma_2 & 0 \\ 0 & \sigma_2
	\end{pmatrix}cc$ and $S[\Lambda]$ as in \eqref{eq:slambda}, we have
\begin{equation}
\begin{split}
\label{eq:jslambda}
	\J S[\Lambda] & = 
	\begin{pmatrix} 
		-\sigma_2\overline\Lambda_+ & 0 \\ 0 & \sigma_2\overline\Lambda_-
	\end{pmatrix}cc, \\[5pt] \text{and } \qquad S^{-1}[\Lambda]\J & =
	\begin{pmatrix} 
		-\Lambda_-\sigma_2 & 0 \\ 0 & \Lambda_+\sigma_2
	\end{pmatrix}cc,
\end{split}
\end{equation}
where we use $cc\Lambda_\pm = \overline\Lambda_\pm cc$, 
with the bar denoting the complex conjugation.
For the decomposition \eqref{eq:68}, we have 
\begin{equation}
	\overline\Lambda_\pm = \overline\Lambda_e \pm \overline\Lambda_o 
	= \Lambda_e \pm \overline\Lambda_o,
\end{equation}
where
\begin{equation}
	\overline\Lambda_o = 
	\sum_{k=0}^\infty\frac{|{\bf a}|^{2k}}{(2k+1)!}\;\overline{(\bf a\cdot\boldsymbol\sigma)},
\end{equation}
with
$$\overline{(\bf a.\boldsymbol\sigma)} 
= a_1\sigma_1 - a_2\sigma_2 + a_3\sigma_3.$$
Further, recalling that $$({\bf a}\cdot\boldsymbol\sigma)\sigma_2 
	= -\sigma_2\overline{({\bf a}\cdot\boldsymbol\sigma)}$$ due to 
	$\{\sigma_i,\sigma_j\} = 2\delta_{ij}\mathbb{I}_2$, we notice that
\begin{equation}
	\Lambda_o\sigma_2 = -\sigma_2\overline\Lambda_o,
\end{equation}
whence
\begin{equation*}
\begin{split}
	\Lambda_\pm\sigma_2 
&	= \Lambda_e\sigma_2 \pm \Lambda_o\sigma_2
	= \sigma_2\Lambda_e \mp \sigma_2\overline\Lambda_o \\[3pt]
&	= \sigma_2 (\Lambda_e \mp \overline\Lambda_o) 
	= \sigma_2\overline{(\Lambda_e \mp \Lambda_o)} 
	= \sigma_2\overline\Lambda_\mp.
\end{split}
\end{equation*} 
Then, \eqref{eq:jslambda} reads as
  $\J S[\Lambda] = S^{-1}[\Lambda]\J$,
and the result follows from \eqref{3.7}. $\qed$
\end{proof}	



\clearpage

As we recalled in the Introduction, the $\rho$-inner product
of euclidean spinors coincides with the Kre\u{\i}n product of lorentzian spinors. 
For the action \eqref{eq:67} of the boost to be coherent, it should leave this 
product invariant. This is indeed the case as we shall see in what follows.

\bigskip

\begin{lem}
\label{lemma:boostinvproduit}
	The canonical $\rho$-inner product \eqref{rho-p} of a minimally twisted 
	four-dimensional riemannian manifold is invariant under action~\eqref{eq:67}
	of the Lorentz boost $S[\Lambda]$.
\end{lem}

\smallskip

\begin{proof} 
	The product $\langle\psi,\phi\rangle_{\rho}$ is mapped to (omitting the 
	argument $\Lambda$ of $S$)
\begin{equation*}
\begin{split}
\langle \psi_\Lambda, \phi_\Lambda \rangle_{\rho}
& = \langle S\psi, S\phi \rangle_{\rho}
  = \langle \psi, S^+S\phi \rangle_{\rho} \\
&  = \langle \psi, \rho(S)^\dag S\phi \rangle_{\rho}
 = \langle \psi, (S^{-1})^\dag S\phi \rangle_{\rho} \\
&  = \langle \psi, S^{-1}S\phi \rangle_{\rho}
  = \langle \psi, \phi \rangle_{\rho},
\end{split}
\end{equation*}
using \eqref{3.7} and the fact that $S^{-1}$ is 
Hermitian. $\qed$
\end{proof}

\bigskip

\begin{cor}
\label{cor:boost}
	The fermionic action of a minimally twisted four-dimensional euclidean 
	manifold is boost invariant.
\end{cor}

\smallskip

\begin{proof}
Using  $\J S=S^{-1}\J$ from Prop.~\ref{prop:jboost} and \eqref{3.7}, one has that 
\begin{equation}
\begin{split}
  \label{eq:25}
	\J\psi_\Lambda & = \J S\psi = S^{-1}\J\psi, \\ 
\text{and } \qquad \eth_\Lambda \phi_\Lambda 
	& = \rho(S)\,\eth\,S^{-1} S\phi = S^{-1}\eth \phi. 
\end{split} 
\end{equation}
By Lem.~\ref{lemma:boostinvproduit}, one then gets
\begin{equation*}
\langle \J\psi_\Lambda ,\eth_\Lambda\phi_\Lambda\rangle_{\rho} = \langle S^{-1}\J\psi, S^{-1}\eth\phi\rangle_{\rho} = \langle \J\psi,  \eth\phi \rangle_\rho. \qed
\end{equation*}
\end{proof}

\clearpage

\section{Lichnerowicz formula for minimally twisted manifold}
\label{sec:5.2}

To compute the Seeley-de Witt coefficients in the asymptotic expansion 
\eqref{eq:SBexp} of the spectral action $S^b$ \eqref{eq:SB}, one uses the standard
local formula for the heat-kernel expansion \cite[\S4.8]{Gi} on  
the square $\D_\omega^2$ of the fluctuated Dirac operator $\D_\omega$.

\bigskip

Let us first recall here the statement of \cite[Lem.~4.8.1]{Gi}.

\medskip

\begin{mdframed}
\begin{theorem}
\label{thrm:Gilks}
Given a differential operator $D$ acting on the sections of a vector bundle
$\V$ on a compact riemannian manifold $(\M,g)$ with the leading symbol given by
the metric tensor. That is, $D$ has the local form
\begin{equation}
\label{eq:elliptic}
	D = -\big(\gMN\Id\dm\dn + A^\mu\dm + B\big),
\end{equation}
where $\gMN$ is the inverse metric, $\Id$ the identity matrix, and $A^\mu$ and $B$
are endomorphisms of $\V$. Then, $D$ can uniquely be written as 
\begin{equation}
\label{eq:Lichnerowicz}
	D = \nabla^*\nabla - E,
\end{equation}
where $\nabla$ is a connection on $\V$ with the associated laplacian 
$\nabla^*\nabla$ and $E$ is an endomorphism of $\V$. Explicitly, one has that
\begin{equation}
\label{eq:5.15}
	\nabla_\mu := \dm + \om, \qquad
	\om := \half g_{\mu\lambda} \big(\alpha^\lambda + \G^\lambda\cdot{\sf id}\big),
	\quad\text{with } \G^\lambda := \gMN\G^\lambda_{\mu\nu},
\end{equation}
where $\sf id$ is the identity endomorphism of $\V$ and $\G^\lambda_{\mu\nu}$ are 
the Christoffel symbols of the Levi-Civita connection of the metric $g$; and
\begin{equation}
\label{eq:Endmrphsm}
	E = B + \big(\G^\nu\cdot{\sf id} - \gMN\nabla_\mu \big)\on.
\end{equation} 
\end{theorem}
\end{mdframed}

\medskip

Now, we fix following the notation:
\begin{equation}
\label{eq:not1}
	\eth_{\cal S} := -i\gm\nsm, \qquad\text{where}\quad \nsm := \dm + \osm,
\end{equation}
and similarly
\begin{equation}
\label{eq:not2}
	\DX := -i\gm\nxm, \qquad\text{where}\quad 
	\nxm := \dm + \oxm,
\end{equation}
with
\begin{equation}
\label{eq:not3}
	\oxm := \osm + X_\mu \qquad \text{and} \qquad \Ox := -i\gm\oxm.
\end{equation}
Thus, we have that
\begin{equation}
\label{eq:not4}
	 \nxm = \nsm + X_\mu.
\end{equation}

\clearpage

With that under the belt, we now give an expression for $\DX^2$ as an
elliptic operator of the laplacian type \eqref{eq:elliptic}, in order to write 
down a generalized Lichnerowicz formula \eqref{eq:Lichnerowicz} for it, 
using Theorem~\ref{thrm:Gilks}.

\bigskip

\begin{prop}
	The squared twisted-covariant Dirac operator can be written as
\begin{equation}
	\DX^2 = -\big(\gMN\dm\dn + \alpha^\nu\dn + \beta\big),
\end{equation}
	where
\begin{equation}
\label{eq:5.7}
	\alpha^\nu = i\big(\eth\gn\big) + i\big\{\Ox,\gn\big\}, \qquad
	\beta  = -\big(\eth\Ox\big) -\Ox^2.
\end{equation}
\end{prop}

\smallskip

\begin{proof}
	We have
\begin{equation}
\label{eq:5.3}
	\DX^2 \; = \; \big(\eth + \Ox\big)^2 
\; = \; \eth^2 + \Ox\eth  +  \eth\Ox + \Ox^2.
\end{equation} 
	The first term of \eqref{eq:5.3} is
\begin{equation}
\begin{split}
\label{eq:5.4}
	\eth^2\; & =\; \big(-i\gm\dm\big)\big(-i\gn\dn\big) \\ 
	& =\; -\;\gm\dm\gn\dn \\
	& =\; -\;\gm\gn\dm\dn -\;\gm\big(\dm\gn\big)\dn \\ 
	& =\; -\;\gMN\dm\dn\; -\;i\big(\eth\gn\big)\dn,
\end{split}
\end{equation}
	where the last equality holds by using the identity $\gm\gn = 
	\half\big[\gm,\gn\big] + \gMN\Id$ and the symmetry $\dm\dn = \dn\dm$ 
	as following:
\begin{equation*}
\begin{split}
	\big[\gm,\gn\big]\dm\dn \; & = \; \gm\gn\dm\dn - \gn\gm\dm\dn \\
	& = \; \gm\gn\dm\dn  \;-\; \gm\gn\dn\dm \\ & =\; \gm\gn\big[\dm,\dn\big] \\ 
	& =\; 0.
\end{split}
\end{equation*}
	The second and third terms of \eqref{eq:5.3}, respectively, are
\begin{equation}
\begin{split}
\label{eq:5.6}
	\Ox\eth \; & = \; -\; i\Ox\gn\dn, \\[3pt]
	\eth\Ox \; & = \; -\; i\gn\dn\Ox \\ 
	& = \; - \; i\gn\Ox\dn \; - \; i\gn\big(\dn\Ox\big)  \\
	& = \; - \; i\gn\Ox\dn \; + \; \big(\eth\Ox\big),
\end{split}
\end{equation}
	Substituting \eqref{eq:5.4} and \eqref{eq:5.6} into \eqref{eq:5.3}, the result
	follows:
\begin{equation*}
	\DX^2 = -\gMN\dm\dn -\left( i\big(\eth\gn\big) + i\Ox\gn + i\gn\Ox \right)\dn 
			+ \big(\eth\Ox\big) + \Ox^2. \qed
\end{equation*}
\end{proof}

\clearpage

In accordance with the notation (\ref{eq:not1}--\ref{eq:not4}), we define the 
covariant derivatives associated with the adjoint action of the corresponding
connections as 
\begin{equation}
\label{eq:not5}
	\dx_\mu := \dm + [\oxm, \cdot] \qquad \text{and} \qquad
	\ds_\mu := \dm + [\osm, \cdot],
\end{equation}
using which we now give two lemmata that will be useful for the subsequent proofs.

\bigskip

\begin{lem}
\label{lem:5.1}
	One has
\begin{equation}
	\gm\dx_\mu\gn = -\G^\nu\Id -2\gm\gn X_\mu.
\end{equation}	
\end{lem}

\medskip

\begin{proof} 
Since the commutator $[\dm,\gl]$ acts on the Hilbert space $L^2(\M,\cal S)$ as the
bounded operator $(\dm\gl)$, Def.~\eqref{eq:not5} gives
\begin{equation}
	\ds_\mu\gl = (\dm\gl) + [\osm,\gl] = [\nsm,\gl].
\end{equation}
If $c$ denotes the Clifford action: $c(dx^\lambda) = \gl$, then by definition of 
the spin connection, 
\begin{equation}
\label{eq:5.29}
	[\nsm, c(dx^\lambda)] = c(\nabla_\mu^{LV}dx^\lambda) 
	= c(-\G^\lambda_{\mu\nu}dx^\nu) = -\G^\lambda_{\mu\nu}c(dx^\nu),
\end{equation}
where $\nabla^{LV}$ denotes the covariant derivative (on the cotangent bundle 
$\T^*\M$) associated to the Levi-Civita connection, and we used the linearity of 
the Clifford action. \eqref{eq:5.29} in terms of $\g$-matrices gives, for any $\mu,\nu$:
\begin{equation}
\label{eq:5.30}
	\ds_\mu\gl = -\G^\lambda_{\mu\nu}\gl.
\end{equation}
	From \eqref{eq:not3} and \eqref{eq:not5}, we have that 
	$\dx_\mu := \ds_\mu + [X_\mu, \cdot]$, 
	which acting on $\gn$ becomes
\begin{equation}
\label{eq:5.2}
	\dx_\mu\gn = \ds_\mu\gn + [X_\mu,\gn] 
	= -\G^\nu_{\mu\kappa}\gk -2\gn X_\mu,
\end{equation}
	where
\begin{equation*}
	\big[X_\mu,\gn\big] = X_\mu\gn -\gn X_\mu 
	= \gn\rho(X_\mu) -\gn X_\mu = -2\gn X_\mu,
\end{equation*}
	since $X_\mu\gn = \gn\rho(X_\mu)$ with $X_\mu = f_\mu\gf$, so 
	$\rho(X_\mu) = -X_\mu$. Further, multiplying \eqref{eq:5.2} by $\gm$, we get
\begin{equation}
\label{eq:5.4a}
	\gm\dx_\mu\gn = -\G^\nu_{\mu\kappa}\gm\gk -2\gm\gn X_\mu.
\end{equation}
	Using the identity $\gm\gk = g^{\mu\kappa}\Id + \half\big[\gm,\gk\big]$ and the 
	symmetry $\G^\nu_{\mu\kappa} = \G^\nu_{\kappa\mu}$,
\begin{equation}
\begin{split}
\label{eq:G5.34}
	\G^\nu_{\mu\kappa}\gm\gk 
	& = \G^\nu_{\mu\kappa}g^{\mu\kappa}\Id + \half\G^\nu_{\mu\kappa}\big[\gm,\gk\big] \\
	& = \G^\nu\Id + \half\big[\G^\nu_{\mu\kappa}, \G^\nu_{\kappa\mu}\big]\gm\gk
	= \G^\nu\Id,
\end{split}
\end{equation}
substituting which in \eqref{eq:5.4a} the result follows. $\qed$
\end{proof} 

\clearpage

\begin{cor}
\label{cor:5.6.1}
One has
\begin{equation}
\dx_\mu(\gl\gm) = -\G^\lambda\Id -\gl\gm\G^\mu_{\mu\nu}.
\end{equation}
\end{cor}

\begin{proof}
Using the Leibniz rule and \eqref{eq:5.2}, it follows that
\begin{equation*}
\begin{split}
	\dx_\mu(\gl\gm) & = \dx_\mu(\gl)\gm + \gl\dx_\mu(\gm) \\ 
& = \big( -\G^\lambda_{\mu\nu}\gn -2\gl X_\mu \big)\gm 
	+ \gl\big( -\G^\mu_{\mu\nu}\gn -2\gm X_\mu \big) \\
& = -\gn\gm\G^\lambda_{\mu\nu} +2\gl\gm X_\mu -\gl\gn\G^\mu_{\mu\nu} -2\gl\gm X_\mu \\
& = -\gn\gm\G^\lambda_{\mu\nu} -\gl\gn\G^\mu_{\mu\nu} \\
& = -\G^\lambda_{\mu\nu}\Id -\gl\gn\G^\mu_{\mu\nu},
\end{split}
\end{equation*}
where we used the fact that $X_\mu = f_\mu\g^5$ anticommutes with any $\g$-matrix
and the last equality follows from \eqref{eq:G5.34}. $\qed$
\end{proof}

\begin{lem} One has
\begin{equation}
\label{}
\gMN(\dx_\mu g_{\nu\kappa}\Id) = (\G_\kappa + \G^\mu_{\mu\kappa})\Id.
\end{equation}
\end{lem}

\begin{proof}
Since $g_{\nu\kappa}\Id$ is a multiple of the identity matrix; 
for any $\mu,\nu,\kappa$ one has 
\begin{equation*}
	\dx_\mu g_{\nu\kappa}\Id = \ds_\mu g_{\nu\kappa}\Id.
\end{equation*}
By Leibniz rule, one has
\begin{equation*}
	\ds_\mu(\gMN g_{\nu\kappa}\Id) =
\begin{cases}
	\ds_\mu(\delta^\mu_\kappa\Id) = 0 \\[3pt]
	\gMN \dx_\mu(g_{\nu\kappa}\Id) + \ds_\mu(\gMN\Id)g_{\nu\kappa} 
\end{cases}
\end{equation*}
Hence,
\begin{equation}
\label{eq:5.37}
	\gMN \dx_\mu(g_{\nu\kappa}\Id) = -g_{\nu\kappa}\ds_\mu(\gMN\Id)
	= -\half g_{\nu\kappa}\ds_\mu(\gm\gn + \gn\gm).
\end{equation} 
By \eqref{eq:5.30}, we have
\begin{align*}
	\ds_\mu(\gm\gn) 
& = -\gm\G^\nu_{\mu\lambda}\gl -\G^\mu_{\mu\lambda}\gl\gn 
  = -\G^\nu_{\mu\lambda}\gm\gl -\G^\mu_{\mu\lambda}\gl\gn, \\[5pt]
	\ds_\mu(\gn\gm) 
& = -\gn\G^\mu_{\mu\lambda}\gl -\G^\nu_{\mu\lambda}\gl\gm 
  = -\G^\mu_{\mu\lambda}\gn\gl -\G^\nu_{\mu\lambda}\gl\gm,
\end{align*}
so that
\begin{align*}
	\ds_\mu(\gm\gn + \gn\gm) & = \ds_\mu(\gm\gn) + \ds_\mu(\gn\gm) \\[3pt]
&	= -2\big( \G^\mu_{\mu\lambda}g^{\mu\lambda} + \G^\mu_{\mu\lambda}g^{\nu\lambda} \big)\Id 
	= -2\big( \G^\nu + g^{\nu\lambda} \G^\mu_{\mu\lambda} \big)\Id,
\end{align*}
and, therefore, \eqref{eq:5.37} give
\begin{equation*}
	\gMN \dx_\mu(g_{\nu\kappa}\Id) 
	= g_{\nu\kappa}\big( \G^\nu + g^{\nu\lambda} \G^\mu_{\mu\lambda} \big)\Id
	= (\G_\kappa + \G^\mu_{\mu\kappa})\Id. \qed
\end{equation*}
\end{proof}

Moving forward, we make use of the above Lem.~\ref{lem:5.1} to obtain explicit 
expressions for the endomorphism terms $\alpha^\nu$ and $\beta$ in \eqref{eq:5.7}.

\begin{prop}
\label{lem:5.3}
	One has that
\begin{equation}
\begin{split}
	\alpha^\nu & = 2\gMN\oxm -2\gm\gn X_\mu -\G^\nu\Id, \\[5pt]
	\beta & = \gm\gn \big( \nabla^{\cal S}_\mu-X_\mu \big) \oxn -\G^\nu\oxn.
\end{split}
\end{equation}
\end{prop}

\begin{proof}
	From \eqref{eq:5.7}, writing $\alpha^\nu$ as
\begin{equation*}
\begin{split}
	\alpha^\nu\; & =\; \big\{ \gn,\gm\oxm \big\} \;+\; \gm\big(\dm\gn\big) \\[3pt]
	& =\; \gn\gm\oxm \;+\; \gm\oxm\gn \;+\; \gm\big(\dm\gn\big) \\[3pt]
	& =\; \gn\gm\oxm \;+\; \gm\gn\oxm \;+\; \gm\big[\oxm,\gn\big] \;+\; \gm\big(\dm\gn\big) \\[3pt]
	& =\; \big\{\gm,\gn\big\}\oxm \;+\; \gm\dx_\mu\gn \\[3pt]
	& =\; 2\gMN\oxm \;-\; \G^\nu\Id \;-\; 2\gm\gn X_\mu,
\end{split}
\end{equation*}
	and the first result follows, using Lem.~\ref{lem:5.1} and the identity $\gm\gn = 
	\half\big[\gm,\gn\big] + \gMN\Id$. 

	Next, $\beta$ in \eqref{eq:5.7} can be
	expanded as
\begin{equation*}
\begin{split}
	\beta\; & =\; \gm\oxm\gn\oxn \;+\; \gm \big( \dm\gn\oxn \big) \\[3pt]
	& =\; \gm\big[\oxm,\gn\big]\oxn \;+\; \gm\gn\oxm\oxn \;+\; \gm\big(\dm\gn\big)\oxn \;+\; \gm\gn\big(\dm\oxn\big) \\[3pt]
	& =\; \gm\gn\big(\dm+\oxm\big)\oxn \;+\; \gm\left( \big(\dm\gn\big) + \big[\oxm,\gn\big] \right)\oxn \\[3pt]
	& =\; \gm\gn\big(\dm + \osm + X_\mu\big)\oxn \;+\; \gm\big(\dx_\mu\gn\big)\oxn \\[3pt]
	& =\; \gm\gn\big(\nsm + X_\mu\big)\oxn \;-\; \G^\nu\oxn \;-\; 2\gm\gn X_\mu\oxn
\end{split}
\end{equation*}
and the second result follows. $\qed$
\end{proof}

\bigskip

For the connection $\omega$ defined in \eqref{eq:5.15}, we denote 
\begin{equation}
	\bm\omega := -i\gm\om,
\end{equation}
recalling the notation ${\bf X} := -i\gm X_\mu$, and obtain the following relation 
between them.

\begin{lem}
\label{lem:5.7}
	One has
\begin{equation}
	\om = \oxm -\chi_\mu \qquad \text{and} \qquad 
	\bm\omega = \Ox + 2{\bf X},
\end{equation}
	where $\oxm$ and $\Ox$ are as in \eqref{eq:not3}, and we have defined
\begin{equation}
\label{eq:chinu}
	\chi_\mu := \gn\gamma_\mu X_\nu. 
\end{equation}
\end{lem}

\begin{proof}
	Substituting $\alpha^\nu$, from Prop.~\ref{lem:5.3}, in $\om$ given by 
	\eqref{eq:5.15}, we have
\begin{equation*}
\begin{split}
	\om & = \half\gmn \big( \alpha^\nu + \G^\nu\cdot{\sf id}\big) \\[3pt]
	& = \half\gmn \big( 2g^{\lambda\nu}\omega_\lambda^X -2\gl\gn X_\lambda \big) \\[3pt]
	& = \gmn g^{\lambda\nu}\omega_\lambda^X -\gmn\gl\gn X_\lambda \\[3pt]
	& = \oxm - \gl\gamma_\mu X_\lambda,
\end{split}
\end{equation*}
and the first result follows identifying $\gl\gamma_\mu X_\lambda =: \chi_\mu$. 
Further, multiplying the first result by $-i\gm$, the second result is obtained
\begin{equation*}
\begin{split}
	-i\gm\om \; & =\; -\;i\gm\oxm \;+\; i\gm\gl\gamma_\mu X_\lambda, \\[3pt]
\text{i.e. } \qquad \bm\omega \; & = \; \Ox \;-\;2i\gl X_\lambda \\ 
	& =\;\Ox \;+\; 2{\bf X},
\end{split}
\end{equation*}
where we used the identity $\gm\gl\gamma_\mu = -2\gl$. $\qed$
\end{proof}

\bigskip

We now give another lemma and use to compute the endomorphism $E$ that 
gives the potential terms in the spectral action of the minimally twisted manifold.

\begin{lem}
\label{lem:useful}
One has the following relations
\begin{equation} \gMN \dx_\mu\chi_\nu 
= \gl\gm\dx_\mu X_\lambda + \G_\kappa\gl\gk X_\lambda -\G^\lambda X_\lambda,
\end{equation} \begin{equation}
	\chi\cdot\chi = -2\gl\gk X_\lambda X_\kappa, \qquad \qquad
	\G^\mu \chi_\mu = \G_\kappa \gl\gk X_\lambda.
\end{equation}
\end{lem}

\begin{proof}
	Using \eqref{eq:chinu}, the Leibniz rule for $\dx_\mu$ as in Lem.~\ref{lem:5.1}
	and Cor.~\ref{cor:5.6.1}, we expand as following
\begin{align*}
	\gMN \dx_\mu\chi_\nu = \gMN \big( \dx_\mu \gl\g_\nu X_\lambda \big)
= \gMN \gl\g_\nu \dx_\mu X_\lambda + \gMN (\dx_\mu \gl\g_\nu) X_\lambda 
\end{align*}
where the first term is $\gl\gm\dx_\mu X_\lambda$ and the second term becomes
\begin{align*}
	\gMN (\dx_\mu \gl\g_\nu) X_\lambda
& = \gMN (\dx_\mu g_{\nu\kappa} \gl\gk) X_\lambda \\[3pt]
& = \gMN(\dx_\mu g_{\nu\kappa})\gl\gk X_\lambda + \gMN g_{\nu\kappa}\dx_\mu(\gl\gk) X_\lambda \\[3pt]
& = (\G_\kappa + \G^\mu_{\mu\kappa})\gl\gk X_\lambda + \dx_\mu(\gl\gm) X_\lambda \\[3pt]
& = (\G_\kappa + \G^\mu_{\mu\kappa})\gl\gk X_\lambda -(\G^\lambda\Id + \gl\gm\G^\mu_{\mu\nu}) X_\lambda \\[3pt]
& = \G_\kappa\gl\gk X_\lambda -\G^\lambda X_\lambda,
\end{align*}
and so the first result follows. In the same manner, we obtain
\begin{equation*}
	 \G^\nu \chi_\nu = \G^\mu\gl\g_\mu X_\lambda 
	 = \G_\kappa g^{\mu\kappa}\gl\g_\mu X_\lambda
	 = \G_\kappa \gl\gk X_\lambda,
\end{equation*}
and 
\begin{align*}
	\chi\cdot\chi = \gMN\chi_\mu\chi_\nu 
&	= \gMN(\gl\g_\mu X_\lambda) (\gk\g_\nu X_\kappa) \\[3pt]
&	= \gMN\gl\g_\mu\gk\g_\nu X_\lambda X_\kappa \\[3pt]
&	= \gmn\gl\gn\gk\gm X_\lambda X_\kappa = -2\gl\gk X_\lambda X_\kappa. \qquad \qed
\end{align*}.
\end{proof}

\begin{prop}
\label{prop:Lchnrwcz}
The endomorphism term \eqref{eq:Endmrphsm} for the Lichnerowicz formula 
\eqref{eq:Lichnerowicz} of the twisted-covariant Dirac operator $\DX$ is
\begin{equation}
	E = \half\gm\gn\big( F_{\mu\nu}^X + 2\,\dx_\nu X_\mu + 4\,X_\mu X_\nu \big) 
		- \G^\mu X_\mu,
\end{equation}
where
\begin{equation}
	F_{\mu\nu}^X := \nxm\oxn - \nxn\oxm
\end{equation}
is the field strength of the connection $\oxm$ and $\dx_\mu$ is the covariant 
derivative \eqref{eq:not5} of its adjoint action.
\end{prop}

\begin{proof}
Substituting $\beta$ from Prop.~\ref{lem:5.3} into $E$ \eqref{eq:Endmrphsm} and 
using $\om = \oxm -\chi_\mu$ of Lem.~\ref{lem:5.7} in the form 
$\nabla_\mu = \nxm -\chi_\mu$, one has
\begin{align}
	E & = \gm\gn \big( \nsm-X_\mu \big)\oxn -\gMN(\nxm - \chi_\mu)\on -\G^\nu(\oxm-\om) \nonumber \\[3pt]
	& = \gm\gn \big( \nsm-X_\mu \big)\oxn -\gMN\nxm\oxn +\gMN\nxm\chi_\nu +\gMN\chi_\nu\om -\G^\nu\chi_\nu \nonumber \\[3pt]
	& = \gm\gn \big( \nsm-X_\mu \big)\oxn -\gm\gn\nxm\oxn +\half[\gm,\gn]\nxm\oxn \nonumber \\ 
& \qquad\qquad\qquad\qquad + \gMN(\dx_\mu\chi_\nu + 2\chi_\nu\oxm + \chi_\nu\om) -\G^\nu\chi_\nu \label{eq:5.31} \\[3pt]
	& = \gm\gn \big( \nsm-X_\mu -\nxm \big)\oxn + \half\gm\gn(\nxm\oxn - \nxn\oxm) \nonumber \\
& \qquad\qquad\qquad\qquad + \gMN\dx_\mu\chi_\nu + 2\gMN\chi_\nu\oxm - \gMN\chi_\nu\chi_\mu -\G^\nu\chi_\nu \nonumber \\[3pt] 
 & = -2\gm\gn X_\mu\oxn + \half \gm\gn F^X_{\mu\nu} \nonumber \\ 
 & \qquad\qquad\qquad\qquad + \gMN\dx_\mu\chi_\nu 
 +2\gm\gn X_\mu\oxn - \chi\cdot\chi -\G^\nu\chi_\nu \label{eq:5.33} \\[3pt]
 & = \half \gm\gn F^X_{\mu\nu} + \gMN\dx_\mu\chi_\nu - \chi\cdot\chi -\G^\nu\chi_\nu, \label{eq:5.34}
\end{align}
where in \eqref{eq:5.31} we used the identity $\gMN\Id = \gm\gn -\half[\gm,\gn]$
and the relation from \eqref{eq:not3}
\begin{equation}
	\dx_\mu\chi_\nu = \dm\chi_\nu + [\oxm,\chi_\nu]
	= \nxm\chi_\nu - \chi_\nu\oxm;
\end{equation}
and in \eqref{eq:5.33} we used \eqref{eq:not4} and \eqref{eq:chinu}.
Finally, substituting in \eqref{eq:5.34} the expressions from the previous
Lem.~\ref{lem:useful} directly yields the result. $\qed$
\end{proof}

\bigskip

Prop.~\ref{prop:Lchnrwcz}, of course, reduces to the correct expression in the flat
case \cite[Prop.~5.3]{DM}. The only difference is that of the last term $\G^\mu 
X_\mu$, which asserts a coupling between the $X_\mu$ field and the curvature.
It is tempting to speculate that $X_\mu$ has something to do with torsion
due to its form $X_\mu = -i\gm f_\mu\gf$ \eqref{Xmu}, which
also appears as a modification of the spin connection (with curvature) in the 
above analysis. 
However, this is yet to be confirmed by fully computing the spectral action.

\bigskip

A related result in this context is \cite{HPS}, where the spectral 
action for pure gravity with torsion is calculated. The (skew-symmetric) torsion 
is incorporated into the \emph{twisted} Dirac operators, which are twisted in a 
different sense than what we mean in our context. A comparative study might
shed some light on the geometric understanding of the $X_\mu$ field.

\clearpage
\newpage 
	\addcontentsline{toc}{chapter}{Conclusions}

\chapter*{Conclusions}

\bigskip

Here we conclude the thesis highlighting the key results with some passing comments.

\bigskip

The fermionic actions associated to the minimal twist of the spectral triples of a
$U(1)$ gauge theory and electrodynamics respectively, yield the Weyl and the Dirac 
equations in lorentzian signature, although one starts with the euclidean signature
(Prop.\ \ref{Prop:Weyl} and \ref{prop:eq-Dirac}). 
That a similar transition of metric signature (from the riemannian to the 
pseudo-riemannian) at the level of fermionic action happens also for the minimal 
twist of the Standard Model should 
be checked. This will be the subject-matter of future works.

\bigskip

At any rate, these results we put forward here strengthen the suggestion of 
twisting noncommutative geometries as an alternative way to approach the problem of 
extending the theory of spectral triples to lorentzian manifolds. The fact that the
twist does not satisfactorily implement the Wick rotation -- it does so 
only for the Hilbert space -- is not so relevant after all. What is far more 
important and interesting from a physical point of view than giving a purely 
spectral characterization of pseudo-riemannian manifolds is to arrive at an action
that is meaningful in a lorentzian context. This thesis makes the case
that it occurs for minimally twisted spectral triples, at least at the level of the
fermionic action.

\bigskip

This reminds us of the results of \cite{Ba} where, by dissociating the 
$KO$-dimension from the metric dimension, one imposes the lorentzian signature for 
the internal spectral triple, and thus obtains a fermionic action allowing 
right-handed neutrinos.

\bigskip

Indeed, the question of a lorentzian spectral action or the spectral action 
associated to twisted spectral triples remains wide open. The interpretation of the
zeroth component of the real field $X_\mu$ as an energy (cf.\ discussions right
after Prop.\ \ref{Prop:Weyl} and \ref{prop:eq-Dirac}) should nevertheless play a 
role for the spectral action, where this field also appears (Prop.\
\ref{prop:Lchnrwcz}).
As shown in \cite{DM} for the twisted Standard Model that the 
contribution of this real field $X_\mu$ to the spectral action is minimized when 
$X_\mu$ vanishes, i.e.\ the case when no twisting occurs. Based on that and the 
results presented here regarding the Wick rotation of the fermionic action, one 
might wonder if the lorentzian (twisted) geometry is a vacuum excitation of the
(non-twisted) riemannian geometry or, in other words, the twist is indeed a spontaneous breaking of 
the symmetry from a riemannian geometry to a pseudo-riemannian one.

\bigskip

The regularity condition imposed in \cite[(3.4)]{CMo} (see also 
Rem.~\ref{rem:autmodul}) has its origin in Tomita-Takesaki theory (App.~\ref{appb}). 
Particularly, the automorphism $\rho$ defining a twisted spectral triple should be 
seen as the evaluation of a one-parameter modular group $\{\rho_t\}$ of 
automorphisms at some specific value $t$. 
For the minimal twist of spectral triples, the flip \eqref{eq:flip} turned out to 
be the only possible automorphism that makes the twisted commutator bounded 
\cite[Prop.~4.2]{LM1}. It is not yet determined what the modular group of 
automorphisms corresponding to this flip would be. Should it exist, this will 
indicate that the time evolution in the Standard Model has its origin in such a 
modular group. 
This is precisely the essence of the `thermal time hypothesis' 
proposed in \cite{CR}. So far, this hypothesis has
been applied to algebraic quantum field theory \cite{Ma,MR}, and for general
considerations in quantum gravity \cite{RS}. Its application to the Standard
Model would be a novelty.

\clearpage
\newpage

\appendix
	\addcontentsline{toc}{chapter}{Appendices}

\chapter{Gel'fand Duality}
\label{app:A}

Gel'fand duality is an algebraic characterization of topological spaces, providing 
one-to-one correspondence between compact Hausdorff topological spaces and 
commutative $C^*$-algebras.

The following definitions are from \cite[\S2.1,~\S4.3]{Su}.

\paragraph{Algebras.} An $\mathbb{F}$-{\it algebra} $A$ is a vector space over the 
field $\mathbb{F}$ with a bilinear associative product:
\begin{equation}
\label{def:alg}
	A \times A \to A, \quad (a,b) \mapsto ab, \qquad \forall a,b\in A.
\end{equation}
$A$ is said to be {\it unital} if there exists a unit $1 \in A$ satisfying
$1a = a1 = a\;(\forall a \in A)$.

\paragraph{*-algebras and their representations.} An algebra $\A$ is called a 
$*$-{\it algebra} (or, an involutive algebra), if there exists an involution (that 
is, a conjugate linear map) $* : \A \to \A$ such that
\begin{equation}
\label{def:*-alg}
	(ab)^* = b^*a^*, \quad (a^*)^* = a, \qquad \forall a,b\in\A.
\end{equation}
A {\it representation} $\pi$ of $\A$ on a Hilbert space $\HH$ is given by a 
$*$-algebra map
\begin{equation}
\label{eq:piAtoLH}
	\pi : \A \to L(\HH),
\end{equation}
where $L(\HH)$ denotes the $*$-algebra of operators on $\HH$ with the product given
by composition and the involution given by hermitian conjugation.

\paragraph{$C^*$-algebras and their representations.} A {\it $C^*$-algebra} $\fra$ 
is a complex norm-complete $*$-algebra that satisfies the $C^*$-property:
\begin{equation}
\label{eq:C*cond}
	\Vert a^*a\Vert = \Vert a\Vert^2, \qquad \forall a \in \fra.
\end{equation}
A {\it representation} $(\HH,\pi)$ of a $C^*$-algebra $\fra$ on a Hilbert space 
$\HH$ is given by a $*$-algebra map
\begin{equation}
	\pi : \fra \to \B(\HH),
\end{equation}
where $\B(\HH)$ denotes the $*$-algebra of bounded operators on $\HH$. 

\medskip

A representation $(\HH,\pi)$ is called {\it irreducible} if $\HH\ne0$ and the only 
closed subspaces in $\HH$ that are invariant under the action of $\fra$ are $\{0\}$ 
and $\HH$ itself. 

\medskip

Two representations $(\HH_1,\pi_1)$ and $(\HH_2,\pi_2)$ of $\fra$ are called {\it 
unitarily equivalent} if there exists a unitary map $\U : \HH_1 \to \HH_2$ such 
that
\begin{equation}
	\pi_1(a) = \U^*\pi_2(a)\U, \qquad \forall a \in \fra.
\end{equation}

\medskip

Define the {\it structure space} $\widehat\fra$ of $\fra$ as the set of all unitary 
equivalence classes of the irreducible representations of $\fra$. And, let $C(\X)$ 
denote the algebra of continuous $\mathbb{C}$-valued functions on a compact 
Hausdorff topological space $\X$. Then, Gel'fand duality asserts that

\medskip

\begin{mdframed}
\begin{enumerate}
\item The structure space $\widehat\fra$ of a commutative (non-)unital 
	$C^*$-algbera $\fra$ is a (locally) compact Hausdorff topological space, 
	and $\fra \simeq C(\widehat\fra)$ via the {\it Gel'fand transform}:
	\begin{equation}
		\fra \ni a \mapsto \widehat a \in \widehat\fra,
		\qquad \widehat a(\pi) = \pi(a).
	\end{equation}
\item For any compact Hausdorff topological space $\X$, we have 
	$\widehat{C(\X)}\simeq\X$. 
\end{enumerate}
\medskip
\end{mdframed}

\clearpage
\newpage

\chapter{Clifford Algebras}
\label{app:cliff}

The definitions and the notations here are primarily taken from \cite[\S4.1]{Su}.

\bigskip

A {\bf quadratic form} $\Q$ on a finite-dimensional $\mathbb{F}$-vector space $\V$ 
is a map $\Q:\V\to\mathbb{F}$ such that $\Q(\lambda v) = \lambda^2\Q(v)$ for all 
$\lambda\in\Q, v\in\V$ and the function $\Q(v+w)-\Q(v)-\Q(w)$ is bilinear for all 
$v,w\in\V$. 

\smallskip

Given a quadratic form $\Q$ on $\V$, the {\bf Clifford algebra} $Cl(\V,\Q)$ is a 
unital associative algebra generated (over $\mathbb{F}$) by $\V$
satisfying $v^2 = \Q(v)1$ for all $v\in\V$.

\medskip

\noindent {\bf \textsc{Property} 1.} Clifford algebras are 
{\bf $\mathbb{Z}_2$-graded algebras}, with 
{\bf grading} $\chi$ given by 
\begin{equation}
	\chi(v_1\cdots v_k) = (-1)^k(v_1\cdots v_k),
\end{equation}
and, thus, can be decomposed into even and odd parts, respectively, as follows:
\begin{equation}
	Cl(\V,\Q) = Cl^0(\V,\Q) \+ Cl^1(\V,\Q).
\end{equation}

\noindent {\bf \textsc{Property} 2.} For all $v,w\in\V$, one has $vw+wv=2g_\Q(v,w)$,
where the symmetric bilinear form $g_\Q:\V\times\V\to\mathbb{F}$ associated to $\Q$
is given by
\begin{equation}
	g_\Q(v,w) = \frac 1{2}\big(\Q(v+w)-\Q(v)-\Q(w)\big).
\end{equation}
 
\smallskip 

For the Clifford algebras generated by the vector spaces $\mathbb{R}^n$ and 
$\mathbb{C}^n$, respectively, we fix the following notation
\begin{equation}
	Cl^\pm_n := Cl(\mathbb{R}^n,\pm\Q_n), \qquad
	\mathbb{C}l_n := Cl(\mathbb{C}^n,\pm\Q_n),
\end{equation}
with the standard quadratic form $\Q_n(x_1,\ldots,x_n) = x_1^2 + \ldots + x_n^2$.
The algebras $Cl^\pm_n$ are generated over $\mathbb{R}$ by $\{e_1,\ldots,e_n\}$ 
subject to
\begin{equation}
\label{eq:eiejdelta}
	e_ie_j + e_je_i = \pm2\delta_{ij}, \qquad \forall i,j \in \{1,\ldots,n\}.
\end{equation}
The even part $(Cl^\pm_n)^0$ and the odd part $(Cl^\pm_n)^1$ of $Cl^\pm_n$ consists
of products, respectively, of an even and an odd number of $e_i$'s. 
$\mathbb{C}l_n$ is the complexification of the algebras $Cl^\pm_n$ and is, 
therefore, generated over $\mathbb{C}$ by the same $\{e_1,\ldots,e_n\}$ respecting 
\eqref{eq:eiejdelta}. 

\smallskip

Further, one checks that
\begin{equation}
	\Dim_\mathbb{R}(Cl^\pm_n) = \Dim_\mathbb{C}(\mathbb{C}l_n) = 2^n.
\end{equation}
The lower dimensional Clifford algebras (for $n=1,2$) are obtained explicitly as
\begin{equation}
\label{eq:Cl12}
	Cl_1^+ \simeq \mathbb{R}\+\mathbb{R}, \quad Cl_1^- \simeq \mathbb{C}; \qquad
	Cl_2^+ \simeq M_2(\mathbb{R}), \quad Cl_2^- \simeq \mathbb{H}.
\end{equation}
The map $\Phi(e_i) = e_{n+1}e_i$ on the generators extends to the following 
isomorphisms:
\begin{equation}
	Cl^-_n \simeq (Cl^\pm_{n+1})^0.
\end{equation}
Similarly, the map defined by
\begin{equation}
	\Psi(e_i) =
	\begin{cases}
		1\x e_1, & i = 1,2 \\
		e_{i-2}\x e_1e_2, & i = 3,\ldots,n
	\end{cases}
\end{equation}
extends to
\begin{equation}
\label{eq:recur}
	Cl^\pm_k \x_\mathbb{R} Cl^\mp_2 \simeq Cl^\mp_{k+2}, \qquad \forall k \ge 1,
\end{equation}
which, along with its base cases \eqref{eq:Cl12}, recursively generates the Table 
\ref{tab:Cliff}.
\begin{table}[h]
\centering
\begin{tabular}{|c|l|l|l|}
\hline
$n$ & $Cl^+_n$ 								& $Cl^-_n$ 								& $\mathbb{C}l_n$ \\
\hline
$1$ & $\mathbb{R} \+ \mathbb{R}$			& $\mathbb{C}$ 							& $\mathbb{C} \+ \mathbb{C}$ \\
$2$ & $M_2(\mathbb{R})$ 					& $\mathbb{H}$ 							& $M_2(\mathbb{C})$ \\
$3$ & $M_2(\mathbb{C})$ 					& $\mathbb{H} \+ \mathbb{H}$ 			& $M_2(\mathbb{C}) \+ M_2(\mathbb{C})$\\
$4$ & $M_2(\mathbb{H})$ 					& $M_2(\mathbb{H})$						& $M_4(\mathbb{C})$ \\
$5$ & $M_2(\mathbb{H}) \+ M_2(\mathbb{H})$	& $M_4(\mathbb{C})$						& $M_4(\mathbb{C}) \+ M_4(\mathbb{C})$ \\
$6$ & $M_4(\mathbb{H})$						& $M_8(\mathbb{R})$						& $M_8(\mathbb{C})$ \\
$7$ & $M_8(\mathbb{C})$						& $M_8(\mathbb{R}) \+ M_8(\mathbb{R})$	& $M_8(\mathbb{C}) \+ M_8(\mathbb{C})$ \\
$8$ & $M_{16}(\mathbb{R})$					& $M_{16}(\mathbb{R})$					& $M_{16}(\mathbb{C})$ \\
\hline
\end{tabular}
\caption{Clifford algebras $Cl^\pm_n$ and their complexifications $\mathbb{C}l_n$
for $n=1,\ldots,8$.}
\label{tab:Cliff}
\end{table}

For $k=n+2$, \eqref{eq:recur} gives
\begin{equation} \begin{split}
\label{eq:Clif11}
	Cl^\pm_{n+4} & \simeq Cl^\mp_{n+2} \x_\mathbb{R} Cl^\pm_2, \\
	& \simeq Cl^\pm_n \x_\mathbb{R} Cl^\mp_2 \x_\mathbb{R} Cl^\pm_2, \\
	& \simeq Cl^\pm_n \x_\mathbb{R} M_2(\mathbb{H}),
\end{split} \end{equation}
where in the second step we used \eqref{eq:recur} for $k=n$ and in the third step
$\mathbb{H} \x_\mathbb{R} M_2(\mathbb{R}) \simeq M_2(\mathbb{R}) \x_\mathbb{R}
\mathbb{H} \simeq M_2(\mathbb{H})$. Further, substituting $n\to n+4$ in 
\eqref{eq:Clif11}, we have
\begin{equation} \begin{split}
\label{eq:Clif12}
	Cl^\pm_{n+8} & \simeq Cl^\pm_{n+4} \x_\mathbb{R} M_2(\mathbb{H}), \\
	& \simeq Cl^\pm_n \x_\mathbb{R} M_2(\mathbb{H}) \x_\mathbb{R} M_2(\mathbb{H}), \\
	& \simeq Cl^\pm_n \x_\mathbb{R} M_{16}(\mathbb{R}),
\end{split} \end{equation}
where the second step uses \eqref{eq:Clif11} and the third 
$M_2(\mathbb{H}) \x_\mathbb{R} M_2(\mathbb{H}) \simeq M_{16}(\mathbb{R})$. 

\medskip

Thus, with \eqref{eq:Clif12}, one concludes that $Cl^\pm_{n+8}$ is Morita 
equivalent to $Cl^\pm_n$ and, therefore, one determines $Cl^\pm_n$ for all $n$. 

\medskip

In this sense, for the real Clifford algebras, Table \ref{tab:Cliff} has the 
periodicity of eight. Similarly, the complex Clifford algebras have periodicity two:
\begin{equation}
	\mathbb{C}l_{n+2} \simeq \mathbb{C}l_n \x_\mathbb{C} M_2(\mathbb{C}).
\end{equation}
and, thus, $\mathbb{C}l_{n+2}$ is Morita equivalent to $\mathbb{C}l_n$.

\medskip

\subsubsection{Clifford bundles}

The Clifford algebraic structure can be naturally imported to riemannian manifolds,
thanks to the metric structure on them.

\medskip

A {\bf riemannian metric} $g$ on a manifold $\M$ is a symmetric bilinear form
$g:\G(\T\M)\times\G(\T\M)\to C(\M)$ satisfying the following conditions:
\begin{enumerate}[label=(\roman*)] 
\setlength\itemsep{-0.25em}
	\item $g(X,Y)$ is a real function if $X$ and $Y$ are real vector fields;
	\item $g$ is $C(\M)$-bilinear, that is
\begin{equation*}
	g(fX,Y) = g(X,fY) = fg(X,Y), \qquad \forall f\in C(\M),\;\forall X,Y\in\G(\T\M);
\end{equation*}
	\item 
	$g(X,X)\ge0$ for all real vector fields $X$ and equality holds iff $X=0$.
\end{enumerate}

\medskip

On the fibers $\T_x\M$ of the tangent bundle $\T\M$ over a riemannian manifold
$(\M,g)$, the inner product defined by the metric:
\begin{equation}
	g_x(X_x,Y_x) := g(X,Y)|_x
\end{equation}
associates the following quadratic form on the tangent space $\T_x\M$:
\begin{equation}
	\Q_g(X_x) = g_x(X_x,X_x).
\end{equation}
Then, at every $x\in\M$, one has the Clifford algebra $Cl(\T_x\M,\Q_g)$ and its
corresponding complexification $\mathbb{C}l(\T_x\M,\Q_g)$.

\clearpage

The {\bf Clifford bundle} $Cl^\pm(\T\M)$ is the bundle of algebras 
$Cl(\T_x\M,\pm\Q_g)$ along with the transition functions $\tau$ inherited from the 
tangent bundle $\T\M$ (i.e.\ for open sets $U,V\subset\M$, 
$\tau_{UV}:U\cap V\to SO(n)$, where $n=\Dim(\M)$) and their action on each fiber 
$\T_x\M$ extended to $Cl(\T_x\M,\pm\Q_g)$ by
\begin{equation}
	v_1v_2\cdots v_k \mapsto \tau_{UV}(v_1)\cdots\tau_{UV}(v_k), \qquad
	v_1,\ldots,v_k \in \T_x\M.
\end{equation}
Similarly, complexified algebras $\mathbb{C}l(\T_x\M,\pm\Q_g)$ define the Clifford 
bundle $\mathbb{C}l(\T\M)$.

\bigskip

The algebra of continuous real-valued sections of $Cl^\pm(\T\M)$ is denoted by
\begin{equation}
\label{alg-Cliff}
	Cliff^\pm(\M) := \G(Cl^\pm(\T\M)),
\end{equation}
and the algebra of continuous sections of $\mathbb{C}l(\T\M)$ by 
\begin{equation}
	\mathbb{C}liff(\M) := Cliff^\pm(\M) \x_\mathbb{R} \mathbb{C}.
\end{equation} 

\clearpage
\newpage

	\chapter{Tomita-Takesaki Modular Theory}
\label{appb}

Modular theory first appeared in two unpublished lecture notes of Minoru 
Tomita~\cite{To1,To2} and a more accessible version was later presented by 
Masamichi Takesaki~\cite{Ta1}. It provides a way to construct `modular 
automorphisms' of von Neumann algebras via polar decomposition of an involution. 
For a more involved account, see~\cite{Ta2}.

\bigskip

A $C^*$-algebra \eqref{eq:C*cond} is a $*$-algebra of bounded operators on a 
Hilbert space $\HH$ that is closed in the \emph{operator norm topology}. In 
particular, a {\bf von Neumann algebra} $\frm$ is a unital $C^*$-algebra closed in 
the \emph{weak operator topology}. The {\bf commutant} $\frm'$ of $\frm$ is defined
as

\begin{equation}
	\frm' := \{ m' \in \frm : \; m'm=mm',\, \forall m \in \frm \}.
\end{equation}

\bigskip

For a von Neumann algebra $\frm$, let a unit vector $\varpi \in \HH$ be {\bf 
separating} (that is, the map $\frm \to \frm\varpi$ is injective) and {\bf cyclic} 
(that is, $\frm\varpi$ is dense in $\HH$). Then, there exist two unique canonical 
operators, namely the {\bf modular operator} $\Delta$ and the {\bf modular 
conjugation} or {\bf modular involution} $J$, such that

\begin{itemize}
	\item $\Delta^* = \Delta$ is positive and invertible (but not bounded),
	\item the set $\{\Delta^{it} : t \in \mathbb{R} \}$ of unitaries induces a 
	strongly continuous one-parameter group $\{\alpha_t\}$ of {\bf modular 
	automorphisms} $\alpha_t:\frm\to\frm$ (with respect to $\varpi$) defined by
\begin{equation}
	\alpha_t(m) = \Ad(\Delta^{it})m = \Delta^{it}m\Delta^{-it}, 
	\qquad \forall m \in \frm,\; \forall t \in \mathbb{R},
\end{equation}
	\item $J = J^* = J^{-1}$ is {\bf antilinear} (i.e.\ $\langle J\psi, J\phi
		\rangle = \overline{\langle\psi,\phi\rangle} = \langle\phi,\psi\rangle,\, 
		\forall\psi,\phi\in\HH$) and it commutes with $\Delta^{it}$, implying
\begin{equation}
	\Ad(J)\Delta := J\Delta J^{-1} = \Delta^{-1},
\end{equation}
	\item $J : \frm \to \frm'$, defined by $J\frm J = \frm'$. Thus, $\frm$ is 
	anti-isomorphic to its commutant $\frm'$ and the anti-isomorphism is given by 
	the $\mathbb{C}$-linear map
\begin{equation}
\label{eq:b4}
	\frm \ni m \; \mapsto \; J m^* J^{-1} \in \frm',
\end{equation}
	\item $\varpi$ is a $+1$-eigenvector of both the operators, that is,
	\begin{equation}
		\Delta\varpi = \varpi = J\varpi,
	\end{equation}	
	\item The unbounded antilinear operators $S_0$ and $F_0$ defined on $\HH$ with 
		domains $\frm\varpi$ and $\frm'\varpi$, respectively, by setting
\begin{equation}
\begin{split}
	S_0(m\varpi) & := m^*\varpi, \qquad \forall m \in \frm \\[5pt] 
	F_0(m'\varpi) & := m'^*\varpi, \qquad \forall m' \in \frm';
\end{split}
\end{equation}
extend to their respective closures -- antilinear operators $S$ and $F=S^*$, 
defined on a dense subset of $\HH$ -- which have the following {\bf polar 
decomposition}:
\begin{equation}
\begin{split}
	S & = J|S| = J\Delta^\half = \Delta^{-\half}J \\[5pt]
	F & = J\vert F\vert = J\Delta^{-\half} = \Delta^\half J,
\end{split}
\end{equation}
implying 
\begin{equation}
\begin{split}
	\Delta & = S^*S = FS \\[5pt]
	\Delta^{-1} & = SF = SS^*.
\end{split}
\end{equation}
\end{itemize}



\clearpage
\newpage
	
\chapter{The Dirac Equation}
\label{GammaMatrices}

\section{$\g$-matrices in chiral representation}

In four-dimensional euclidean space, the gamma matrices are
\begin{equation}
\label{EDirac}
	\gamma^\mu = 
	\begin{pmatrix} 0 & \sigma^\mu \\ \tilde\sigma^\mu & 0 \end{pmatrix}, \qquad
	\gamma^5 := \gamma^1\gamma^2\gamma^3\gamma^0 =
	\begin{pmatrix} \mathbb{I}_2 & 0 \\ 0 & -\mathbb{I}_2 \end{pmatrix},
\end{equation}
where, for $\mu = 0,j$, we define
\begin{equation}
	\sigma^\mu := \big\{ \mathbb{I}_2, -i\sigma_j \big\}, \qquad
	\tilde\sigma^\mu := \big\{ \mathbb{I}_2, i\sigma_j \big\},
\end{equation}
with $\sigma_j$, for $j = 1,2,3$, being the Pauli matrices:
\begin{equation}
\label{Pauli}
		\sigma_1 = \begin{pmatrix} 0 & 1 \\ 1 & 0 \end{pmatrix},
\qquad	\sigma_2 = \begin{pmatrix} 0 & -i \\ i & 0 \end{pmatrix},
\qquad	\sigma_3 = \begin{pmatrix} 1 & 0 \\ 0 & -1 \end{pmatrix}.
\end{equation}

In $(3+1)$-dimensional minkowski spacetime, the gamma matrices are
\begin{equation}
\label{MDirac}
	\gamma^\mu_M =
	\begin{pmatrix} 0 & \sigma^\mu_M \\ \bar\sigma^\mu_M & 0 \end{pmatrix}, \qquad
	\gamma^5_M := \gamma^1_M\gamma^2_M\gamma^3_M\gamma^0_M = -i\gamma^5,
\end{equation}
where, for $\mu = 0,j$, we define
\begin{equation}
	\sigma^{\mu}_M := \big\{ \mathbb{I}_2, \sigma_j \big\}, \qquad
	\bar{\sigma}^{\mu}_M := \big\{ \mathbb{I}_2, -\sigma_j \big\},
\end{equation}
with $\sigma_j$, for $j = 1,2,3$, being the Pauli matrices~\eqref{Pauli}.

\section{Dirac lagrangian and Weyl equations}
\label{App:Dirac}

\bigskip

The Dirac lagrangian in euclidean space and minkowski spacetime, respectively, is 

\begin{equation}
\begin{array}{llll}
	{\cal L}	& := \chi^\dag(\eth + m) \psi 	& \qquad
	\eth 	& := -i\gamma^\mu \partial_\mu \\[5pt]
	{\cal L}_M 	& := -\bar\Psi(\eth_M  +m)\Psi 	& \qquad
	\eth_M 	& := -i\gamma^\mu _M\partial_\mu
\end{array}
\end{equation}
where $\chi, \psi$ are independent Dirac spinors, while the Dirac spinors
$\Psi, \bar\Psi$ are related by: $\bar\Psi := \Psi^\dag\gamma^0$. 
And $\gamma^\mu , \gamma^\mu _M$ are, respectively,
the euclidean~\eqref{EDirac} and minkowskian~\eqref{MDirac} gamma 
matrices. 

\bigskip

The Dirac spinor (or, the spin-$\frac{1}{2}$) representation of (the double cover 
of) the Lorentz group $SL(2,\mathbb{C})$ is reducible into two irreducible 
representations: 
$\left(\frac{1}{2}\,,\,0\right) \oplus \left(0\,,\,\frac{1}{2}\right)\!,$ 
which act only on the two-component~\emph{Weyl spinors} $\Psi_l$ and $\Psi_r$ of a
Dirac spinor $\Psi$, defined, in the chiral 
representation~(see~\S\ref{GammaMatrices}), by
\begin{equation}
	\Psi = \begin{pmatrix} \Psi_l \\ \Psi_r \end{pmatrix} \in L^2({\cal M, S}),
	\qquad \begin{array}{l} \Psi_l \in L^2({\cal M, S})_+ \\[3pt] 
	\Psi_r \in L^2({\cal M, S})_- \end{array}.
\end{equation}

\medskip

Under such decomposition into Weyl spinors, the Dirac lagrangian ${\cal L}_M$ 
becomes
\begin{equation}
\label{L_M}
\begin{split}
	{\cal L}_M
& = \begin{pmatrix} \Psi_l^\dag & \Psi_r^\dag \end{pmatrix} \!
	\begin{pmatrix} 0 & \mathbb{I}_2 \\ \mathbb{I}_2 & 0 \end{pmatrix} \!
	\left[ \begin{pmatrix}
		0 & i\sigma_M^\mu\partial_\mu \\ i\tilde\sigma_M^\mu\partial_\mu & 0
	\end{pmatrix}  -m \right] \!
	\begin{pmatrix} \Psi_l \\ \Psi_r \end{pmatrix} \\[5pt]
& = i\Psi^\dag_l \tilde\sigma^\mu_M \partial_\mu\Psi_l + 
	i\Psi^\dag_r \sigma^\mu_M \partial_\mu\Psi_r - m
	\left( \Psi^\dag_l\Psi_r + \Psi^\dag_r\Psi_l \right)\!,
\end{split}
\end{equation}
which, for $m=0$, describes the \emph{Weyl fermions} (massless spin-$\frac{1}{2}$ 
particles) in quantum field theory. The corresponding \emph{Weyl equations} of 
motion:
\begin{equation}
\label{eq:Weyl}
\begin{array}{lllll}
	{\cal L}_M^l
	& := i\Psi^\dag_l \tilde\sigma^\mu_M \partial_\mu\Psi_l 
	& \quad \longrightarrow \quad \tilde\sigma^\mu_M \partial_\mu\Psi_l
	& = \big(\mathbb{I}_2\partial_0 - \sigma_j\partial_j\big)\Psi_l & = 0, \\[5pt]
	{\cal L}_M^r
	& := i\Psi^\dag_r \sigma^\mu_M \partial_\mu\Psi_r
	& \quad \longrightarrow \quad \sigma^\mu_M \partial_\mu\Psi_r 
	& = \big(\mathbb{I}_2\partial_0 + \sigma_j\partial_j\big)\Psi_r & = 0,
\end{array}
\end{equation}
are derived from the relevant lagrangian density, by treating the Weyl spinor
$\Psi_{l/r}$ and its Hermitian conjugate $\Psi^\dag_{l/r}$ as independent 
variables in the Euler-Lagrange equation:
\begin{equation}
\label{eq:Weyl}
\begin{array}{lllll}
	{\cal L}_M^l
	& := i\Psi^\dag_l \tilde\sigma^\mu_M \partial_\mu\Psi_l 
	& \quad \longrightarrow \quad \tilde\sigma^\mu_M \partial_\mu\Psi_l
	& = (\mathbb{I}_2\partial_0 - \sum_{j=1}^3\sigma_j\partial_j)\Psi_l & = 0, \\[5pt]
	{\cal L}_M^r
	& := i\Psi^\dag_r \sigma^\mu_M \partial_\mu\Psi_r
	& \quad \longrightarrow \quad \sigma^\mu_M \partial_\mu\Psi_r 
	& = (\mathbb{I}_2\partial_0 +\sum_{j=1}^3 \sigma_j\partial_j)\Psi_r & = 0,
\end{array}
\end{equation}

\clearpage

The plane-wave solutions of these equations, with $x^0$ identified to
the time $t$ and $x^{j=1,2,3}$ the space coordinates, are
\begin{align}
  \label{eq:50}
  \Psi_{l/r}(x^0, x^j) = \Psi_0 e^{-i(p_jx^j- Et)}
\end{align}
where $(E, p_j)$ is the energy momentum $4$-vector and $\Psi_0$ is a constant spinor, solution of
\begin{align}
  \label{eq:52}
 (E\mathbb I_2 -\tilde\sigma^j p_j) \Psi_0 =0,& \text{ for the left handed
  solution } \Psi_l ,\\
 (E\mathbb I_2 -\sigma^j p_j )\Psi_0 =0,& \text{ for the right handed
  solution } \Psi_r. 
\end{align}


\newcommand{\noopsort}[1]{} \newcommand{\printfirst}[2]{#1}
  \newcommand{\singleletter}[1]{#1} \newcommand{\switchargs}[2]{#2#1}
\begin{thebibliography}{}

\end{thebibliography}


\begin{thebibliography}{99}


\bibitem[AMST]{AMST}
U.~Aydemir, D.~Minic, C.~Sun, T.~Takeuchi.
\emph{Pati-Salam unification from noncommutative geometry and the TeV-scale $W_R$ boson}.
\emph{Int.\ J.\ Mod.\ Phys.\ A} {\bf 31}(01): 1550223 (2016).

\bibitem[Ba]{Ba}
J.W.~Barrett.
\emph{Lorentzian version of the noncommutative geometry of the standard model of particle physics}.
\emph{J.\ Math.\ Phys.} {\bf 48}: 012303 (2007).

\bibitem[BB]{BB}
F.~Besnard, N.~Bizi.
\emph{On the definition of spacetimes in noncommutative geometry: Part I}.
\emph{J.\ Geom.\ Phys.} {\bf 123}: 292--309 (2016).

\bibitem[BBB]{BBB}
N.~Bizi, C.~Brouder, F.~Besnard.
\emph{Space and time dimensions of algebras with applications to Lorentzian noncommutative geometry and quantum electrodynamics}.
\emph{J.\ Math.\ Phys.} {\bf 59}: 062303 (2018).

\bibitem[BS]{BS}
J.~Boeijink, W.D.v.~Suijlekom.
\emph{The Noncommutative Geometry of Yang-Mills Fields}.
\emph{J.\ Geom.\ Phys.} {\bf 61}: 1122--1134 (2011).


\bibitem[BMS]{BMS}
S.~Brain, B.~Mesland, W.D.v.~Suijlekom. 
\emph{Gauge theory for spectral triples and the unbounded Kasparov product}.
\emph{J.\ Noncom.\ Geom.} {\bf 10}: 131--202 (2016).


\bibitem[BCDS]{BCDS}
T.~Brzezi\'nski, N.~Ciccoli, L.~D\k{a}browski, A.~Sitarz.
\emph{Twisted Reality Condition for Dirac Operators}.
\emph{Math.\ Phys.\ Anal.\ Geom.} {\bf 19}(3): 16 (2016).

\bibitem[BDS]{BDS}
T.~Brzezi\'nski, L.~D\k{a}browski, A.~Sitarz.
\emph{On twisted reality conditions}.
\emph{Lett.\ Math.\ Phys.} {\bf 109}(3): 643--659 (2019).


\bibitem[Ch94]{Ch94}
A.H.~Chamseddine.
\emph{Connection between Space-Time Supersymmetry and Non-Commutative Geometry}.
\emph{Phys.\ Lett.} {\bf 332}: 349--357 (1994).

\bibitem[CC96]{CC96} 
A.H.~Chamseddine, A.~Connes.
\emph{Universal Formula for Noncommutative Geometry Actions: Unification of Gravity and the Standard Model}.
\emph{Phys.\ Rev.\ Lett.} {\bf 77}(24): 4868--4871 (1996).

\bibitem[CC97]{CC97}
A.H.~Chamseddine, A.~Connes.
\emph{The Spectral Action Principle}.
\emph{Commun.\ Math.\ Phys.} {\bf 186}(3): 731--750 (1997).

\bibitem[CC06a]{CC06a}
A.H.~Chamseddine, A.~Connes.
\emph{Scale invariance in the spectral action}.
\emph{J.\ Math.\ Phys.} {\bf 47}(6): 063504(19) (2006).

\bibitem[CC06b]{CC06b}
A.H.~Chamseddine, A.~Connes.
\emph{Inner fluctuations of the spectral action}.
\emph{J.\ Geom.\ Phys.} {\bf 57}(1): 1--21 (2006).

\bibitem[CC08]{CC08}
A.H.~Chamseddine, A.~Connes.
\emph{Why the Standard Model}.
\emph{J.\ Geom.\ Phys.} {\bf 58}(1): 38--47 (2008).

\bibitem[CC10]{CC10}
A.H.~Chamseddine, A.~Connes.
\emph{Noncommutative geometry as a framework for unification of all fundamental interactions including gravity. Part I}.
\emph{Fortsch.\ Phys.} {\bf 58}(6): 553--600 (2010).

\bibitem[CC12]{CC12}
A.H.~Chamseddine, A.~Connes.
\emph{Resilience of the spectral standard model}.
\emph{JHEP} {\bf 09}: 104 (2012).

\bibitem[CCM]{CCM}
A.H.~Chamseddine, A.~Connes, M.~Marcolli.
\emph{Gravity and the standard model with neutrino mixing}.
\emph{Adv.\ Theor.\ Math.\ Phys.} {\bf 11}(6): 991--1089 (2007).

\bibitem[CCS1]{CCS1}
A.H.~Chamseddine, A.~Connes, W.D.v.~Suijlekom.
\emph{Inner fluctuations in noncommutative geometry without the first order condition}.
\emph{J.\ Geom.\ Phys.} {\bf 73}: 222--234 (2013).

\bibitem[CCS2]{CCS2}
A.H.~Chamseddine, A.~Connes, W.D.v.~Suijlekom.
\emph{Beyond the spectral standard model: emergence of Pati-Salam unification}.
\emph{JHEP} {\bf 11}: 132 (2013).

\bibitem[CCS3]{CCS3}
A.H.~Chamseddine, A.~Connes, W.D.v.~Suijlekom.
\emph{Grand unification in the spectral Pati-Salam model}.
\emph{JHEP} {\bf 11}: 011 (2015).

\bibitem[CF1]{CF1}
A.H.~Chamseddine, J.~Fr\"ohlich.
\emph{Particle Physics Models, Grand Unification, and Gravity in Non-Commutative Geometry}
(1993) [\href{https://arxiv.org/abs/hep-th/9311068}{hep-th/9311068}].

\bibitem[CF2]{CF2}
A.H.~Chamseddine, J.~Fr\"ohlich.
\emph{SO(10) Unification in Non-Commutative Geometry}.
\emph{Phys.\ Rev.\ D} {\bf 50}: 2893--2907 (1994).

\bibitem[CS]{CS}
A.H.~Chamseddine, W.D.v.~Suijlekom.
\emph{A survey of spectral models of gravity coupled to matter} (2019) 
[\href{https://arxiv.org/abs/1904.12392v2}{1904.12392}].



\bibitem[Col]{Col} 
S.~Coleman.
\emph{Aspects of Symmetry: Selected Erice Lectures}.
Cambridge University Press, Cambridge, U.K.~(1985).

\bibitem[C94]{C94} 
A.~Connes.
\href{https://www.alainconnes.org/docs/book94bigpdf.pdf}{\emph{Noncommutative Geometry}}.
Academic Press, San Diego (1994).

\bibitem[C95]{C95} 
A.~Connes.
\href{http://www.alainconnes.org/docs/reality.pdf}{\emph{Noncommutative geometry and reality}}.
\emph{J.\ Math.\ Phys.} {\bf 36}(11): 6194--6231 (1995).

\bibitem[C96]{C96}
A.~Connes.
\emph{Gravity coupled with matter and the foundations of noncommutative geometry}.
\emph{Commun.\ Math.\ Phys.} {\bf 182}(1): 155--176 (1996).

\bibitem[C13]{C13}
A.~Connes.
\emph{On the spectral characterization of manifolds}.
\emph{J.\ Noncom.\ Geom.} {\bf 7}(1): 1--82 (2013).

\bibitem[CMa]{CMa}
A.~Connes, M.~Marcolli. \href{https://www.its.caltech.edu/~matilde/coll-55.pdf}{\emph{Noncommutative Geometry, Quantum Fields and Motives}}. American Mathematical Society (2008).

\bibitem[CMo]{CMo}
A.~Connes, H.~Moscovici.
\emph{Type III and spectral triples}.
\emph{Traces in number theory, geometry and quantum fields,
Aspects Math., Friedr. Vieweg, Wiesbaden} {\bf E38}: 57--71 (2008).

\bibitem[CR]{CR}
A.~Connes, C.~Rovelli.
\emph{Von Neumann algebra automorphisms and time-thermodynamics relation in general covariant quantum theories}.
\emph{Class.\ Quant.\ Grav.} {\bf 11}(12): 2899--2918 (1994).


\bibitem[D'\!AKL]{D'AKL}
F.~D'\!Andrea, M.A.~Kurkov, F.~Lizzi.
\emph{Wick rotation and fermion doubling in noncommutative geometry}.
\emph{Phys.\ Rev.\ D} {\bf 94}(2): 025030 (2016).


%

\bibitem[DFLM]{DFLM} 
A.~Devastato, S.~Farnsworth, F.~Lizzi, P.~Martinetti.
\emph{Lorentz signature and twisted spectral triples}.
\emph{JHEP} {\bf 03}: 089 (2018).

\bibitem[DKL]{DKL}
A.~Devastato, M.~Kurkov, F.~Lizzi.
\emph{Spectral Noncommutative Geometry, Standard Model and all that} (2019)
[\href{https://arxiv.org/abs/1906.09583v2}{1906.09583}].

\bibitem[DLM1]{DLM1}
A.~Devastato, F.~Lizzi, P.~Martinetti.
\emph{Grand symmetry, spectral action and the Higgs mass}.
\emph{JHEP} {\bf 01}: 042 (2014).
  
\bibitem[DLM2]{DLM2}
A.~Devastato, F.~Lizzi, P.~Martinetti.
\emph{Higgs mass in noncommutative geometry}.
\emph{Fortsch.\ Phys.} {\bf 62}(9-10): 863--868 (2014).

\bibitem[DM]{DM}
A.~Devastato, P.~Martinetti.
\emph{Twisted spectral triple for the Standard Model and spontaneous breaking of the Grand Symmetry}.
\emph{Math.\ Phys.\ Anal.\ Geom.} {\bf 20}(1): 2 (2017).


\bibitem[DPR]{DPR}
K.v.d.~Dungen, M.~Paschke, A.~Rennie.
\emph{Pseudo-Riemannian spectral triples and the harmonic oscillator}.
\emph{J.\ Geom.\ Phys.} {\bf 73}: 37--55 (2013).


\bibitem[DS]{DS} 
K.v.d.~Dungen, W.D.v.~Suijlekom.
\emph{Electrodynamics from noncommutative geometry}.
\emph{J.\ Noncom.\ Geom.} {\bf 7}(2): 433--456 (2013).





\bibitem[Fr]{Fr}
N.~Franco.
\emph{Temporal Lorentzian spectral triples}.
\emph{Rev.\ Math.\ Phys.} {\bf 26}(08): 1430007 (2014).

\bibitem[FE]{FE}
N.~Franco, M.~Eckstein.
\emph{An algebraic formulation of causality for noncommutative geometry}.
\emph{Class.\ Quant.\ Grav.} {\bf 30}(13): 135007 (2013).


\bibitem[Gi]{Gi}
P.B.~Gilkey. \emph{Invariance theory, the heat equation and the atiya-singer theorem}. 
\emph{Publish or Perish} (1984).

\bibitem[HPS]{HPS}
F.~Hanisch, F.~Pfaeffle, C.A.~Stephan.
\emph{The Spectral Action for Dirac Operators with skew-symmetric Torsion}.
\emph{Commun.\ Math.\ Phys.} {\bf 300}: 877--888 (2010).


\bibitem[LM1]{LM1}
G.~Landi, P.~Martinetti.
\emph{On twisting real spectral triples by algebra automorphisms}.
\emph{Lett.\ Math.\ Phys.} {\bf 106}(11): 1499--1530 (2016).

\bibitem[LM2]{LM2}
G.~Landi, P.~Martinetti.
\emph{Gauge transformations for twisted spectral triples}.
\emph{Lett.\ Math.\ Phys.} {\bf 108}(12): 2589--2626 (2018).

\bibitem[LM]{LM}
B.~Lawson, M.L.~Michelson.
\emph{Spin Geometry}. 
Princeton University Press, Princeton, NJ (1989).
 
\bibitem[LMMS]{LMMS} 
F.~Lizzi, G.~Mangano, G.~Miele, G.~Sparano.
\emph{Fermion Hilbert space and fermion doubling in the noncommutative geometry approach to gauge theories}.
\emph{Phys.\ Rev.\ D} {\bf 55}(10): 6357--6366 (1997).

\bibitem[Ma]{Ma}
P.~Martinetti.
\emph{Conformal mapping of Unruh temperature}.
\emph{Mod.\ Phys.\ Lett.\ A} {\bf 24}(19): 1473--1483 (2009).

\bibitem[MR]{MR}
P.~Martinetti, C.~Rovelli.
\emph{Diamond's temperature: Unruh effect for bounded trajectories and thermal time hypothesis}.
\emph{Class.\ Quant.\ Grav.} {\bf 20}: 4919--4931 (2003).
 
\bibitem[MS]{MS}
P.~Martinetti, D.~Singh.
\emph{Lorentzian fermionic action by twisting euclidean spectral triples} 
(2019) [\href{https://arxiv.org/abs/1907.02485v1}{1907.02485}].
 


\bibitem[Ri]{Ri}
M.A.~Rieffel. 
\emph{Vector bundles and Gromov-Hausdorff distance.} 
\emph{J. K-Theory} {\bf 5}: 39--103 (2010).

\bibitem[RS]{RS}
C.~Rovelli, M.~Smerlak.
\emph{Thermal time and the Tolman-Ehrenfest effect: `temperature as the speed of time'}.
\emph{Class.\ Quant.\ Grav.} {\bf 28}: 075007 (2010).



\bibitem[Ta1]{Ta1}
M.~Takesaki.
\emph{Tomita's Theory of Modular Hilbert Algebras and Its Applications}.
\emph{Lecture Notes in Math.} {\bf 128}, 
Springer-Verlag Berlin Heidelberg (1970).

\bibitem[Ta2]{Ta2}
M.~Takesaki.
\emph{Theory of Operator Algebras II}.
\emph{Encyclopaedia of Math.\ Sci.} {\bf 125},
Springer-Verlag Berlin Heidelberg (2003).

\bibitem[To1]{To1}
M.~Tomita.
\emph{On canonical forms of von Neumann algebras}.
\emph{Fifth Func.\ Anal.\ Sympos.} (in Japanese).
\emph{T\^ohoku Univ., Sendai: Math.\ Inst.} 101--102 (1967).

\bibitem[To2]{To2}
M.~Tomita.
\emph{Quasi-standard von Neumann algebras}.
Mimeographed note, unpublished (1967).

\bibitem[Su]{Su}
W.D.v.~Suijlekom.
\emph{Noncommutative Geometry and Particle Physics}. 
Springer Netherlands (2015).

\bibitem[Va]{Va}
D.V.~Vassilevich. 
\emph{Heat Kernel Expansion: User's Manual}. 
\emph{Phys.\ Rep.} {\bf 388}: 279--360 (2003).

\bibitem[W]{W}
R.~Wulkenhaar.
\emph{SO(10) unification in noncommutative geometry revisited}.
\emph{Int.\ J.\ Mod.\ Phys.\ A} {\bf 14}: 559--588 (1999).

\end{thebibliography}
\end{document}